\documentclass[12pt]{puthesis}

\title{Geometric Aspects and Neutral Excitations in the Fractional Quantum Hall Effect}

\submitted{September 2013}  
\copyrightyear{2013}  
\author{Bo Yang}
\adviser{Professor F.D.M. Haldane}  
\department{Physics}

\usepackage{amsfonts}
\usepackage{amsmath}
\usepackage{tipa}
\usepackage{wasysym}
\usepackage{esint}
\usepackage{psfrag}
\usepackage{subfigure}
\usepackage{float}

\usepackage{graphicx,bm,units}

\usepackage{verbatim}

\usepackage{multirow}
\usepackage{longtable}

\usepackage{booktabs}

\ifdefined\printmode

\usepackage{url}

\else

\ifdefined\proquestmode

\usepackage{hyperref}
\hypersetup{bookmarksnumbered}

\makeatletter
\hypersetup{pdftitle=\@title,pdfauthor=\@author}
\makeatother

\else

\usepackage{hyperref}
\hypersetup{bookmarksnumbered}

\makeatletter
\hypersetup{pdftitle=\@title,pdfauthor=\@author}
\makeatother

\fi 
\fi 

\ifodd 0
\renewcommand{\maketitlepage}{}
\renewcommand*{\makecopyrightpage}{}
\renewcommand*{\makeabstract}{}

\else

\abstract{
In this thesis, I will present studies on the collective modes of the fractional quantum Hall states, which are bulk neutral excitations reflecting the incompressibility that defines the topological nature of these states. It was first pointed out by Haldane that the non-commutative geometry of the fractional quantum Hall effects (FQHE) plays an important role in the intra-Landau-level dynamics. The geometrical aspects of the FQHE will be illustrated by calculating the linear response to a spatially varying electromagnetic field, and by a numerical scheme for constructing model wavefunctions for the neutral bulk excitations. Compared to early studies of the magneto-roton modes with single mode approximation (SMA), the scheme presented in this thesis is good not only in the long wavelength limit, but also for large momenta where the neutral excitations evolve into quasihole-quasiparticle pair. It is also shown that in the long wavelength limit, the SMA scheme produces exact model wavefunctions describing a quadrupole excitation. The same scheme can also extend to describe the neutral fermion mode in the Moore-Read state, reflecting its non-Abelian nature.

The numerically generated model wavefunctions are then identified with a family of analytic wavefunctions that describe both the magneto-roton modes and the neutral fermion modes. Like the ground state wavefunction of the Laughlin and Moore-Read state, the family of the analytic wavefunctions do not have any variational parameters. This set of analytic wavefunctions unifies previous numerical works on neutral excitations of single-component FQH states, both from the Jack polynomial point of view presented in this thesis, and from the composite fermion picture developed by Jain and collaborators. The compact analytic forms also lend much insight into the nature of the neutral excitations from the plasma analogy. In particular, the quadrupole excitation gap is related to the free energy cost of the fusion of charged particles in a two-dimensional plasma with a neutralizing background.
}

\acknowledgements{
I cannot imagine what my past five years at Princeton would be like without so many helps I received along the way. I would like to first thank my advisor, Prof. Duncan Haldane, who is always there to discuss about physics with much patience and enthusiasm. I consider myself very fortunate to have the opportunity to learn many aspects of condensed matter physics from one of the defining scholars in the field. Prof. Haldane exemplifies how physical intuition and creativity should be combined with mathematical rigor, and he will always remind me of the level of dedication it takes to venture into the world of physics. Most importantly, Prof. Haldane initiated me into the field of fractional quantum Hall effect, a fascinating subject with which I have had a love-hate relationship for the past few years, which will probably last for many years to come.

I would also like to thank my pre-thesis advisor, Prof. Shivaji Sondhi. Not only was his guidance and advice (including a very useful reading list for condensed matter physics) invaluable at the early stage of my life in Princeton, he also helped me in many ways throughout my graduate career, letting me tag along with his group discussions, being on the committee of my pre-thesis presentations, and writing recommendation letters for my summer school applications, just to name a few. I am also very thankful for his careful reading of my final thesis and many constructive feedbacks.

Princeton University offers me a unique intellectual environment where I can meet a lot of talented and passionate physicists. I would like to especially thank Prof. Andrei Bernevig for his guidance and comments on my works about the collective modes in FQHE. His earlier works on entanglement spectrum and FQHE model wavefunctions have always been inspirations for me; I am very glad to see the field of the topological insulators having its first pedagogical textbook, evolving from Andrei's series of lectures at Princeton I was fortunate to have the opportunity to attend.

I am also fortunate to work with Hu Zi-xiang and Zlatko Papic, two talented postdocs who know everything about the numerics of the quantum Hall systems. I thank Zi-xiang for teaching me the physics of FQHE on the disk, and for tolerating many of my stupid questions on coding while we were working together. I thank Zlatko Papic for teaching me the phase transition of the FQHE in higher Landau levels, and how to make sense out of seemingly messy numerical data. The manuscripts we write always read better after his final touch, and I thank him for a very careful reading of my thesis draft, when it was still full of typos and missing articles.

I am also grateful for many fellow graduate students around me, who made my life at Princeton stimulating both academically and socially. During my early days here I learnt a lot of physics from discussions with Sid Parameswaran and Chris Laumann. It was also a pleasant experience to learn DMRG from PCTS postdoc Bryan Clark. Discussions with Yeje Park in between long conversations with our common advisor often times made my thoughts and understandings much clearer. I also enjoyed lively conversations with my officemates Anushya Chandran and Wu Yang-le, and Arijeet Pal, who in addition to being a fellow explorer in physics, was also responsible for a lot of my improvements on the tennis court. I thank Sonika Johri, who helped me plot the FQH spectrum on the cylinder in this thesis. I am also thankful for my fellow students on pu2012 mailing list, and especially my friend Ming, who made a lot of my weekends and holidays memorable moments of my graduate life.

Finally, I would like to thank Agency for Science, Technology and Research in Singapore for the scholarship over the past five years, and Prof. Jason Petta for letting me work in his lab for my experimental project. I also thank Prof. Herman Verlinde and Prof. Nai Phuan Ong for being on the committee of my thesis defense. Needless to say I owe everything to my parents. The pursuit of PhD has been a uniquely rewarding and humbling experience, I am very thankful to have it as an episode of my life.

}

\dedication{To my parents.}

\fi  

\begin{document}

\makefrontmatter



\chapter{Introduction\label{ch:intro}}

\begin{figure}[ttt]
 \centerline{\includegraphics[width=0.95\linewidth,angle=0]{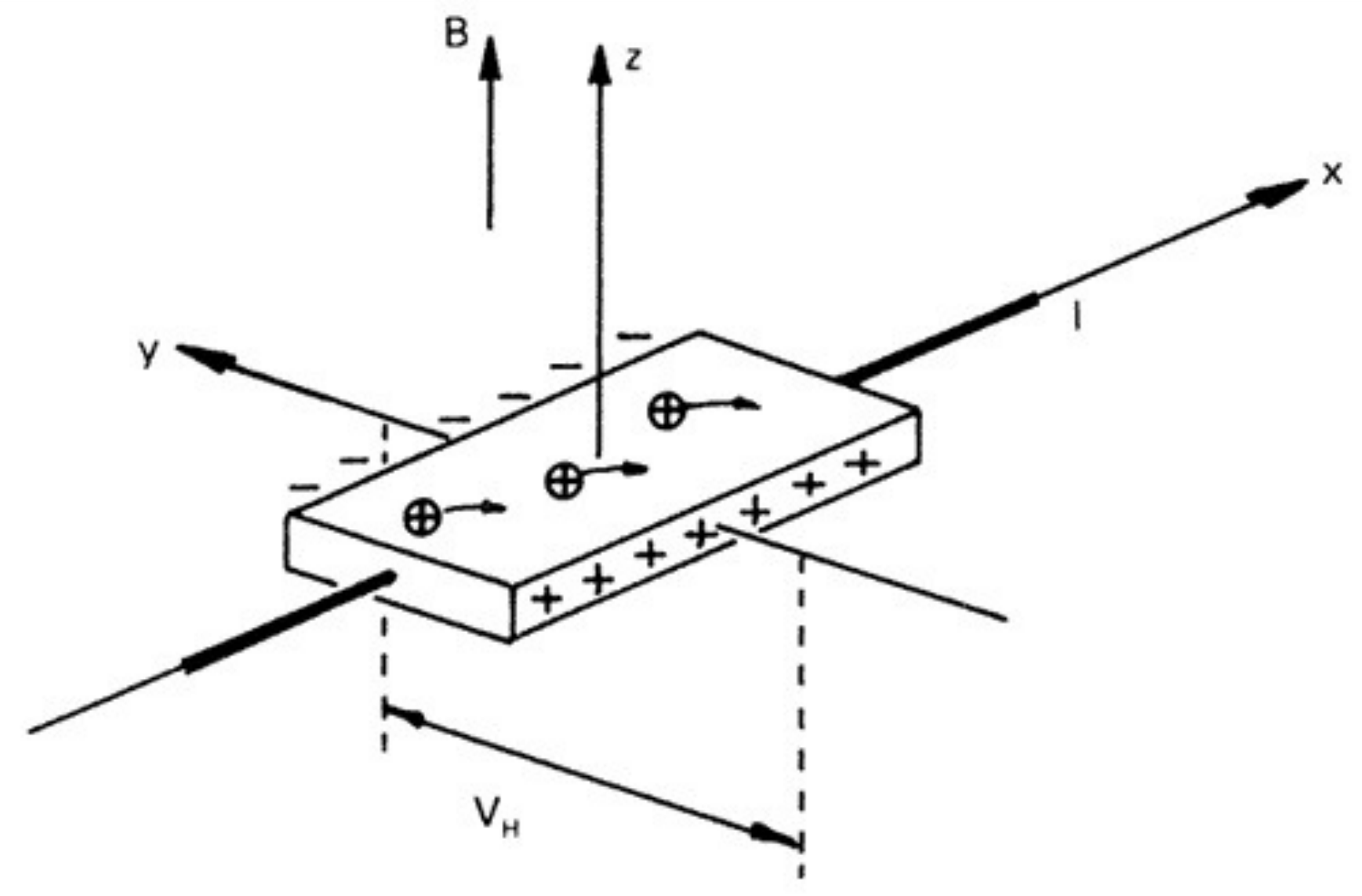}}
\caption{Schematic drawing of the experimental setup of quantum Hall sample (Kosmos, 1986)}
\label{2deg}
\end{figure}

The quantum Hall effect (QHE)\cite{qhe1} is one of the great discoveries in the history of condensed matter physics. It leads to many exciting physical concepts in $(2+1)$ dimensional spacetime, including fractional\cite{wilczek} and non-Abelian statistics\cite{mr}, classification of matter with topological phases\cite{wen1}, bulk-edge correspondence\cite{wen2,lh} and the framework of topological quantum computing\cite{cn}, just to name a few. Quantum Hall systems are experimentally realized by confining an electron gas to a two-dimensional manifold with a strong perpendicular magnetic field which breaks time reversal symmetry (see Fig.(\ref{2deg})). Experimental discovery of the integer quantum Hall effect (IQHE) dates back to 1980, when Klaus von Klitzing\cite{klitzing} found the quantization of the Hall conductivity at integer multiples of $e^2/h$, where $e$ is the elementary charge and $h$ is the Planck's constant. The formation of plateaus and the vanishing of dissipative longitudinal resistivity are hallmarks of the quantum Hall effect, suggesting a gapped phase with non-trivial attributes very robust against disorder. The integer coefficients multiplying $e^2/h$ at these plateaus are accurate up to $10^{-8}$. These integers are equal to the ratio of the number of electrons to the number of flux quanta $h/e$ at the special incompressible points (which are typically in the middle of the plateau). We call this ratio the filling factor $\nu$. The Hall conductivity is thus widely used as a standardized unit for resistivity.

\begin{figure}[ttt]
\centerline{\includegraphics[width=0.8\linewidth,angle=0]{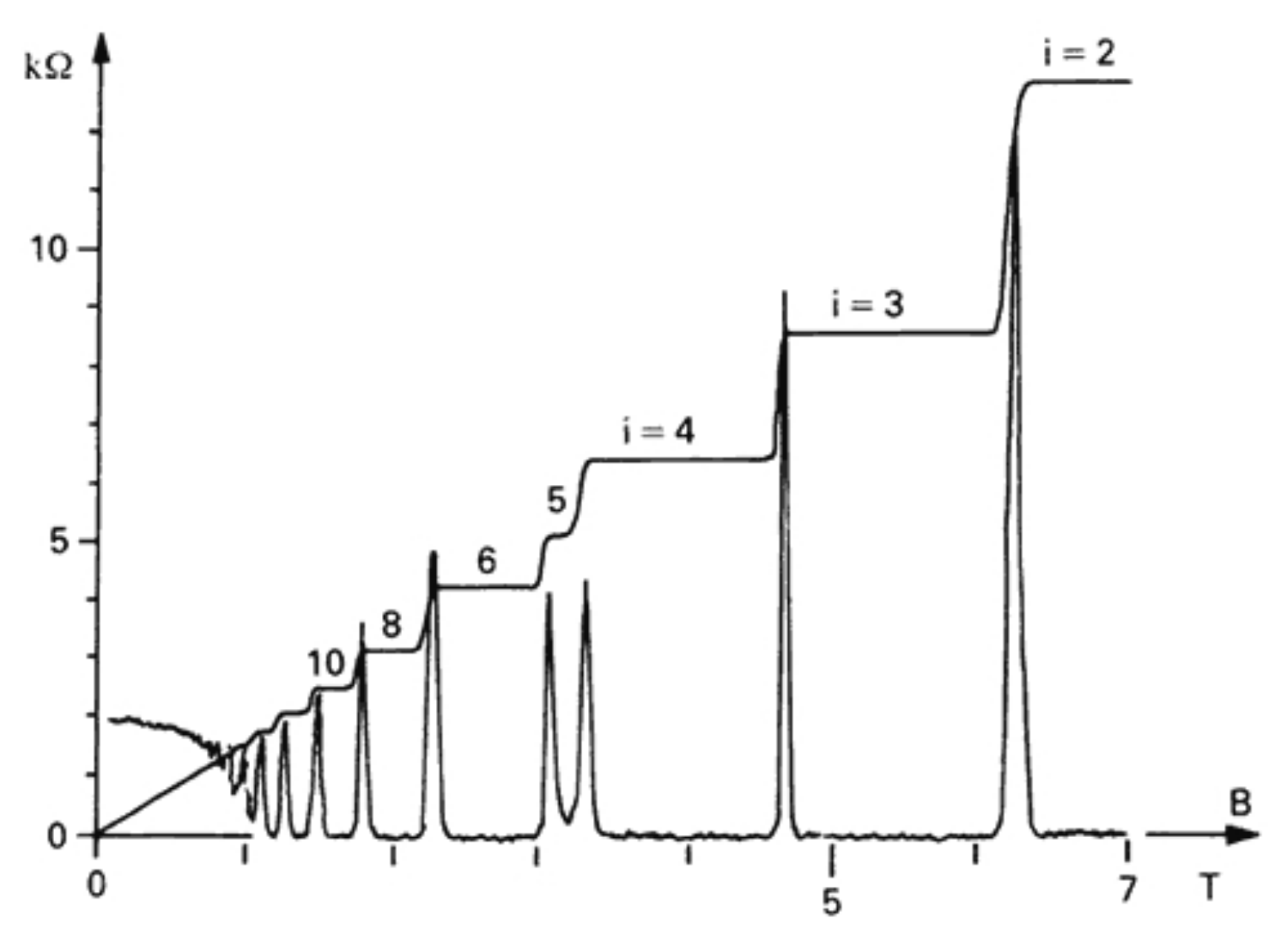}}
\caption{Integer quantum Hall effect (Kosmos, 1986)}
\label{iqhe}
\end{figure}  

\begin{figure}[ttt]
\centerline{\includegraphics[width=0.8\linewidth,angle=0]{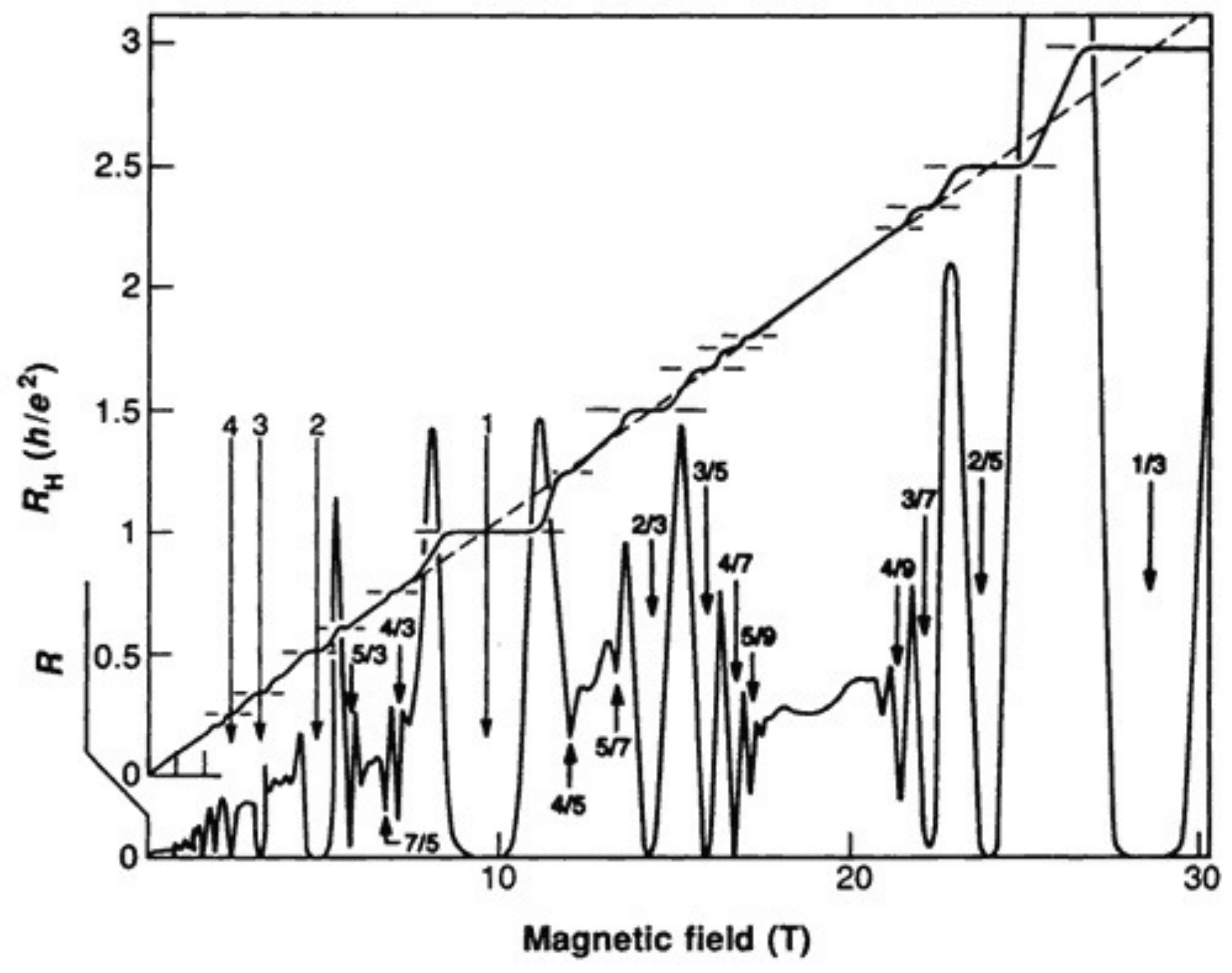}}
\caption{Fractional quantum Hall effect\cite{stormer}}
\label{fqhe}
\end{figure}  

The fractional quantum Hall effect (FQHE) was discovered in 1982 by Tsui, Stormer and Gossard\cite{tsg}, where the plateau in the Hall conductivity was found in the lowest Landau level (LLL) at fractional filling factors (notably at $\nu=1/3$). Unlike the IQHE, which can be primarily explained by single particle physics, the FQHE is a result of strong interactions between electrons within a single Landau level, in which the single particle kinetic energy is a trivial constant. Theoretical understandings of the FQHE was initiated by R. Laughlin\cite{laughlin1}; in his seminal paper the wavefunctions for the ground state and charged excitations were proposed for filling factors $\nu=1/m$, where $m$ is an odd integer for fermions. Soon after that Haldane\cite{haldane1} and later Trugman and Kivelson\cite{tk} constructed the model Hamiltonians in the form of pseudopotentials where Laughlin-like model wavefunctions are exact zero energy states gapped from the rest of the energy spectrum. These elegant model Hamiltonians are intuitively appealing, and are believed to be adiabatically connected to the realistic physical interactions in the thermodynamic limit. Though general arguments of gauge invariance and recognition of the non-trivial band topology were first inspired by IQHE\cite{laughlin2}, the idea that topological phases can exist beyond Landau's paradigm of spontaneous symmetry breaking became widespread after people start to understand the FQHE. Till this day, the FQHE is one of the very few experimental examples (and probably the only reasonably well-understood theoretical model) where strongly correlated topological phases are realized without the protection of any symmetry (also see ref.\cite{hos} for a relatively modern understanding of the superconducting phases).

A more comprehensive review of the quantum Hall effect can be found in the Les Houches lecture notes prepared by Steve Girvin\cite{girvin1}. In this chapter I will give a brief overview of some of the important aspects of the QHE. A formal approach with the microscopic model will be presented. Though the effect of disorder is crucial in the experimental realization of the QHE, in this thesis the disorder is ignored unless otherwise stated.

\section{Aspects of Quantum Hall Effects}

Most features of the integer quantum Hall effect (IQHE) can be understood in the framework of single particle physics. The energy levels of a two-dimensional electron gas (2DEG) subject to a perpendicular magnetic field form Landau levels (LL), each with macroscopic degeneracy $N_\phi=BA\cdot\frac{e}{h}$, which is proportional to the system size. Here $B$ is the strength of the magnetic field, $A$ is the area of the 2DEG and $h/e$ is the flux quantum; thus $N_\phi$ is the number of flux quanta piercing through the Hall surface. The magnetic length is given by $l_B=\sqrt{1/eB}$, where we set $\hbar=1$, and $e$ is the charge of the particle. For Galilean invariant systems the energy spacing of LLs is given by the characteristic cyclotron frequency $\omega_c=eB/mc$, where $m$ is the effective mass of the particle and $c$ is the speed of light. It is easy to see that the ground state of completely filled LLs are gapped. On the other hand the partially filled LLs are compressible due to the macroscopic degeneracy. For a translationally invariant system the Hall conductivity is always equal to $\nu\frac{e^2}{h}$, where the filling factor $\nu$ can be any real number. In this ideal situation experimental measurement of the Hall conductivity \emph{cannot} distinguish between compressible and incompressible phases.

In real samples, the presense of disorder, however weak it is, is expected to localize the state and suppress the Hall conductivity\cite{localization1,localization2}. The surprising fact is at integer filling factor when the system is incompressible, the Hall conductivity is \emph{unaffected} by disorder. On the contrary, any small deviation from these integer filling factors creates particle or hole charge carriers that are localized by disorder, forming the plateau around these integer filling factors\cite{localization2}. The presense of disorder is not necessary for the physics of the QHE, but is essential to experimentally expose these special filling factors, where the ground state is gapped and dissipationless.

It is by now understood that any quantity that is robust against small perturbation is likely to have a topological origin. The famous gedanken experiment by Laughlin\cite{laughlin2} shows that if the 2-D Hall manifold has a cylinder geometry, one can thread magnetic fluxes through the cylinder along the longitudinal axis. By gauge invariance the system should be the same before and after adiabatic threading of a single flux quantum. One can follow the spectral flow of the single particle orbitals during this adiabatic process. At the end of the process the spectrum returns back to the intial configuration, implying an integer number of particles pumped from one edge to the other, leading to the integer Hall conductivity. Even in the presense of disorder that affects the details of the spectral flow, the initial and final configuration has to be identical by gauge invariance. By the argument that one cannot make small changes to quantities that are integers, the integer Hall conductivity has to be robust against disorder.

 The topological nature of the QHE was further illustrated by the work of TKNN\cite{tknn}, and later substantiated by Niu et.al\cite{niu}, with the Kubo formula developed by the linear response theory on a lattice. The single particle wavefunctions on a lattice can be viewed as a section of a $U(1)$ fiber bundle, where the periodic momentum space is the base space. It was shown for a fully occupied conduction band, the Hall conductivity is the first Chern number of the $U(1)$ fiber bundle over the Brillioun zone. This was later generalized to Chern insulators first introduced by Haldane\cite{haldane2}, whereby the filled valence band structure has none zero Chern number even when the net magnetic field per unit cell is zero. This observation has ignited a flurry of research in symmetry protected topological insulators both in two and three dimensions\cite{kanehasan}. 

In contrast to the IQHE, the fractional quantum Hall effect (FQHE) is primarily due to strong interactions between electrons, where the single particle dynamics is ignored in the limit of a strong magnetic field. The incompressibility of the FQH fluid at certain fractional filling factors results from intricate interplay between interactions and the truncated Hilbert space defined by the filling factor. Theoretical understanding of the FQHE was initiated by Laughlin's many-body trial wavefunctions\cite{laughlin1} for the ground states of odd-denominated fractional filling factors, followed by the model Hamiltonians\cite{haldane1,tk} with a gapped spectrum, such that Laughlin's trial wavefunctions are exact ground states. 

An interesting fact about these trial wavefunctions is that as long as rotational invariance is assumed, no variational parameters are typically necessary to optimize these trial wavefunctions within the lowest Landau level (LLL): they are intrinsically good model wavefunctions. Following this line, the study of the FQHE via model wavefunctions and model Hamiltonians has been a very fruitful endeavor. At filling factors $\nu=\frac{1}{m}$, where $m$ is odd, wavefunctions of ground states and charged (quasielectrons and quasiholes) excitations can be written down in nice analytic forms, and the model Hamiltonians are just two-body intereactions with judiciously selected Haldane pseudopotentials\cite{haldane1}. While at even $m$ the FQHE state is generally forbidden because of the fermionic statistics of electrons, at $m=2$ a pairing mechanism is introduced to explain the experimental observation of the $\nu=5/2$ plateaux\cite{willet}, or $\nu=1/2$ state in the $1^{\text{st}}$ Landau Level (1LL). The ground state can be written down analytically as a Pfaffian\cite{mr} multiplying the Jastrow factor. Charged excitations can also be written down following the similar procedure of flux insertion introduced by Laughlin. The model Hamiltonian is the three-body interaction that allows pairing of electrons but penalizes the congregation of three electrons. This is physically possible in higher LLs, where the effective two-body interaction has nodes, allowing particles to stay close to each other.

It is thus natural to formally generalize to $(k+1)$-body interactions that allow clustering of $k$ electrons but penalizes congregation of $k+1$ electrons. This leads to the Read-Rezayi (RR) series\cite{rr} of the single component FQH states. The $k=1$ case of the RR series is the Laughlin state, while the $k=2$ case is the MR state. The set of the many-body wavefunctions from the FQHE are themselves quite fascinating objects. Even with explicit analytic forms, they are quite complex. The simplest case of the Laughlin wavefunctions has the following form:
\begin{eqnarray}\label{laughlin}
\psi_{l}=\prod_{i<j}\left(z_i-z_j\right)^me^{-\frac{1}{2}\sum_i|z_i|^2}
\end{eqnarray}
where the holomorphic variables are used with $z_i=\frac{1}{\sqrt{2}}(x_i+iy_i)$, and $i$ is the particle index. Though it has a compact analytic form, there is no closed expression for its normalization constant as a function of the number of particles. The Moore-Read ground state at half filling is a bit more complicated\cite{pfaffian}:
\begin{eqnarray}\label{pfaffianw}
\psi_{\text{mr}}=\text{Pfaff}\left(\frac{1}{z_i-z_j}\right)\prod_{i<j}\left(z_i-z_j\right)^qe^{-\frac{1}{2}\sum_i|z_i|^2}
\end{eqnarray}
The Pfaffian is defined by
\begin{eqnarray}\label{pfaffian}
\text{Pfaff}\left(M_{ij}\right)=\frac{1}{2^{N/2}(N/2)!}\sum_{\sigma\in S_N}\text{sgn}\sigma\prod_{k=1}^{N/2}M_{\sigma(2k-1),\sigma(2k)}
\end{eqnarray}
for an $N\times N$ antisymmetric matrix $M_{ij}$, and $S_n$ is the permutation group on $n$ indices. In Eq.(\ref{pfaffianw}), $q$ is even for fermions, and $q=2$ at $\nu=1/2$. To tackle these wavefunctions analytically, there are efforts to reinterpret them in more revealing ways. It was first noticed by Laughlin\cite{laughlin1} that the norm of Eq.(\ref{laughlin}) describes a system of a two-dimensional one-component plasma (2DOCP) with logarithmic Coulomb interactions and a neutralizing background. The physical picture of 2DOCP, which is well studied in plasma physics, lends insight on charged excitations in the FQHE, as well as possible ground state phase transitions from a fluid state to a symmetry breaking Wigner crystal state. It also allows effective use of Monte-Carlo techniques in calculating wavefunction overlaps and correlation functions\cite{montecarlo}. In Chapter 4 and 5 a set of model wavefunctions for the neutral collective excitations in the FQHE will be introduced both from a numerical perspective\cite{yb1} and an analytical perspective\cite{yb2}. The latter extends the way we understand the FQHE via the plasma analogy to include the neutral bulk excitations. It seems the analogy between the FQHE and the 2DOCP is not only limited to the Hilbert space, but also includes the energy spectrum as well.

Almost parallel to the development of the Pfaffian wavefunctions, it was realized that many trial wavefunctions in the FQHE can be written as correlators in 2-D conformal field theory (CFT)\cite{mr2}. On first sight, one would be surprised to think that CFT, which describes quantum critical systems with gapless excitations, would play a role in gapped systems like the FQHE. On the other hand, the FQHE is gapless when an edge is present. It was shown by Wen\cite{wen2} that while the bulk of the FQHE can be described by an effective Chern-Simons theory, the requirement of the gauge invariance for a system with a boundary predicts gapless neutral edge excitations that can be described by CFT.  From a more formal perspective, the connection between CFT and the topological field theory (TFT) was previously established by Witten\cite{witten}. A microscopic interpretation of the edge and bulk excitations of the FQHE in the framework of $W_\infty$ algebra will be presented in Chapter 5, where the analogy between edge excitations and CFT is made explicit.

The conformal block description of the FQHE wavefunctions has the practical use of calculating the statistics of quasiparticle excitations with much ease. In principle, the anyonic statistics of quasiparticles in the Laughlin FQHE, and the non-abelian braiding statistics of those in the MR states are entirely encoded within the explicit first quantized wavefunctions. While it is relatively straightforward to show the anyonic statistics of the quasiparticle excitations of the Laughlin state\cite{asw}, the non-abelian statistics of the RR series with $k>1$ is much harder to prove. It is only until recently a rigorous proof was presented in \cite{bonderson} for the Moore-Read state at $k=2$.

The plasma analogy and the CFT connections are very limited in describing the \emph{dynamics} of FQHE, since in both cases the Hamiltonians are not explicitly involved. For many physical systems, the ground state \emph{does} contain information about the low-lying excitations in the spectrum (e.g. the Goldstone modes of the symmetry-breaking ground state). It is thus hopeful that the FQHE ground state will yield information about some part of the excitation spectrum, even though there is no symmetry-breaking for the quantum Hall fluid, and the bulk is gapped. The entanglement spectrum of the bulk ground state has been found to yield information on the gapless edge modes\cite{lh}, and the connection to CFT plays a significant role here\cite{qi1}. The bulk neutral excitation is known to depend on the guiding center Hall viscosity and the ground state structure factor at least in the long wavelength limit\cite{gmp}, and recently the relationship between the entire branch of the neutral excitations and the ground state has been made much clearer with an explicit set of analytic wavefunctions\cite{yb2}.

The single component FQHE given by the RR series (including Laughlin and MR states) cannot explain all the experimentally observed plateau at fractional filling factors. Haldane and Halperin\cite{haldane3,halperin} introduced a hierarchy picture where additional plateau can be explained as incompressible QH states of the quasiholes/quasiparticles. Later on Jain introduced the composite fermion picture whereby the ``elementary particles" in the FQHE are taken as electron-vortex composite, instead of bare electrons\cite{jain1}. These composite particles obey fermionic statistics. It is conjectured that the FQHE can be mapped into the IQHE of composite fermions forming its own ``Landau levels" (also refered to as the ``$\Lambda$ levels) in an effective magnetic field, leading to the ``Jain hierarchy" that is very successful in explaining most plateau observed experimentally. The composite fermion picture is also very useful in numerically generating model wavefunctions for these hierarchical states, including both charged and neutral excitations.

The hierarchical states can be described as multi-component FQH states, where different types of Hall fluids coexist. This is the place where TFT becomes very efficient in characterizing various types of FQH states. For the FQHE descended from the Abelian Laughlin states, both single component and multicomponent Abelian states can be expressed in a unified way by the K-matrix formulation\cite{wen3}. For the RR series with $k>1$, which are non-abelian FQHE due to the braiding statistics of the quasiparticle excitations, there are efforts in formulating effective field theory by introducing Majorana fermion fields\cite{hansson}, and it is still a field of active research.

Numerical analysis has been an indispensible tool in studying the FQHE, given the inherent difficulty in characterizing strongly correlated systems analytically. Historically Laughlin justified the validity of his model wavefunctions by their large overlap with the ground state of Coulomb interaction found by exact diagonalization. It is a remarkable fact that even for system sizes as small as a few electrons, exact diagonalization can reveal the physics of the FQHE quite clearly. Haldane developed the numerical formalism for the FQH systems on the sphere and torus geometry\cite{haldane3, haldane4}. These compact geometries do not have boundaries, making them especially convenient for studying the bulk properties of the finite FQH fluid. Other common geometries include disk\cite{disk} and cylinder geometry\cite{cylinder}, where the edge physics of FQH can be explored.

Recently, many model wavefunctions of the FQHE are identified with Jack polynomials\cite{haldane5}, which substantially enhances the capability of numerically generating wavefunctions at various FQH filling factor. While model wavefunctions have compact analytic forms, most finite-size calculations require explicit knowledge of the coefficients of expansions in terms of the orbital occupation basis. These coefficients are geometry dependent. With model Hamiltonians this information can be obtained via exact diagonalization, an expensive numerical procedure that grows exponentially with the system size. In comparison, the Jack polynomials have rich algebraic structures\cite{bernevig} and can be generated numerically via a recursive procedure, and one can adapt them onto different geometries just by proper single particle normalization, as long as these geometries have genus zero. A brief discussion about the Jack polynomials can be found in Chapter 4.

There has been a recent effort in understanding the FQHE, and QHE in general, from a geometric point of view\cite{haldane6}. Formally, a magnetic field perpendicular to a 2D Hall manifold maps a four dimensional phase space for each electron onto two sets of 2D real space coordinates - the cyclotron coordinates and the guiding center coordinates. In the limit of strong magnetic field, the incompressibility of the IQHE is governed by the dynamics of the cyclotron coordinates, which depends on the single particle kinetic energy\cite{note}. On the other hand, the FQHE is governed by the dynamics of the guiding center coordinates only, from the many-body interaction. Thus the IQHE and the FQHE exist in two different Hilbert spaces; in each of the Hilbert space, the spatial coordinates do not commute with each other, leading to quantum fluctuations of their respective metric. The fluctuation of the cyclotron metric is suppresed by strong magnetic field, while the fluctuation of the guiding center metric plays an important role on the bulk neutral excitations in the long wavelength limit.

Closely related to the geometric aspects of the FQHE is a topological quantity called the Hall viscosity\cite{asz,rr2}. The formal definition of the Hall viscosity will be presented in Chapter 3. From a heuristic hydrodynamic point of view, the Hall viscosity induces a force in the fluid proportional to the gradient of the velocity field; unlike the common dissipative viscosity, this force is perpendicular to the velocity field, thus it does not lead to any energy dissipation. It is only present in systems where time reversal symmetry is broken, such as in the QH system. In fact the Hall viscosity is related to the average angular momentum per particle in the fluid; for a rotationally invariant system it is defined with a metric, and is quantized just like the angular momentum. 

The guiding center Hall viscosity is an important quantity in the FQHE, because the filling factor as a topological index does not fully characterize the FQHE\cite{wen3}. The Hall viscosity, or the average angular momentum per particle, is another topological index differentiating between different phases, and is stable against perturbations that do not close the gap, as long as rotational invariance is preserved\cite{asz,rr2}. 

Phenomenologically, the FQHE can be viewed as consisting of fluids of particle-flux composites with a finite areal extension on the order of the square of the magnetic length. Different types of composite particles define different topological orders, and each composite particle carries a charge. Since both the particle and the flux carries angular momentum, each composite particle also carries a ``spin" relating to the Hall viscosity -thus the composite particles are ``topological" objects. On the sphere where the Hall manifold is curved, the spin of the composite particles will couple to the curvature of the manifold, resulting in a shift - an $O(1)$ correction to the number of states available due to the Berry phase of the coupling\cite{wen4}. This shift is also quantized by the Gauss-Bonnet theorem and is basically the same quantity as the Hall viscosity. 

The finite areal extension of the composite particles requires a metric to define its shape (as well as its spin). Thus even on a flat Hall manifold, the adiabatic deformation of the shape will couple to the guiding center spin in a non-trivial way. This interesting interplay between the topology and geometry in the FQHE was first emphasized by Haldane\cite{haldane6}, who conjectured the quantum fluctuation of the metric of the composite particle and its coupling to the guiding center spin captures the dynamics of the FQHE, or its collective mode, at least in the long wavelength limit.

The collective modes in the FQHE are neutral excitations completely dictated by the dynamics of the guiding center degrees of freedoms, which defines the incompressibility of the topological phase. It is the less well-known part of the FQHE spectrum, as compared to charged excitations like quasiparticles and quasiholes. The neutral excitations were first studied by Girvin, Macdonald and Platzman\cite{gmp} using single mode approximation (SMA) within the LLL. The collective mode is similar to the roton-modes in the Helium-4 superfluid\cite{feynman}. It has a roton minimum at momentum around the inverse of the magnetic length, hence the name ``magneto-roton mode". Unlike the collective mode in the Helium-4 superfluid, in the long wavelength limit the magneto-roton mode is gapped. Thus the roton-minimum defines the gap of the FQHE. Experimental realization of the FQH phases requires a much better understanding of these neutral excitations. With the recent discovery of fractional Chern insulators on the lattice  and their apparent connections to the FQHE\cite{fti1,fti2,fti3}, there is a pressing need to understand the collective modes better. Since the collective modes are neutral, the term "neutral excitations" will also be used in this thesis to refer to the collective modes in the FQHE in Chapter 4 and 5.

\section{Formalism of the Quantum Hall Problems}

A magnetic field perpendicular to the two-dimensional Hall surface leads to the minimal coupling of the kinetic momentum with the in-plane vector potential $\vec A$ with $\nabla\times\vec A=B$, where $B$ is the strength of the magnetic field. The length scale is thus defined by the magnetic length $l_B=\sqrt{1/eB}$, where $e$ is the effective charge of the particles and we set $\hbar=1$. For a Hall surface of an area $\mathcal A$ the total number of the magnetic flux is given by $N_\phi=\frac{\mathcal A}{2\pi l_B^2}$. Normally, we pick a gauge for the vector potential and solve the single particle Hamiltonian to get the wavefunctions for the eigenstates. For a rotationally invariant system the convenient gauge is the symmetric gauge, and the single particle wavefunctions are coherent states of electrons undergoing cyclotron motion about the origin. For translationally invariant systems the Landau gauge is often used, where the single particle wavefunctions are plane waves in one direction, and confined Gaussian packages in the other direction.

This chapter aims to give a very general treatment of the formalism of the QHE, without recourse to explicitly picking a guage for the external vector potential. In this way, the algebraic structure of the Hilbert space of the two-dimensional Hall surface is fully exploited with explicit gauge invariance. Unlike most previous literature, the geometric aspect of the quantum Hall problem is emphasized by requiring real space coordinates to have the upper indices and the covariant momentum vectors to have lower indices. Einstein summation convention is assumed unless otherwise stated. Metric dependence of various quantities are shown explicitly, without the assumption of rotational invariance, allowing the existence of several metrics with different physical origins. The notations of this thesis will also be fixed in this section.

\subsection{Algebra and Hilbert space}

\begin{figure}[ttt]
\centerline{\includegraphics[width=0.4\linewidth,angle=0]{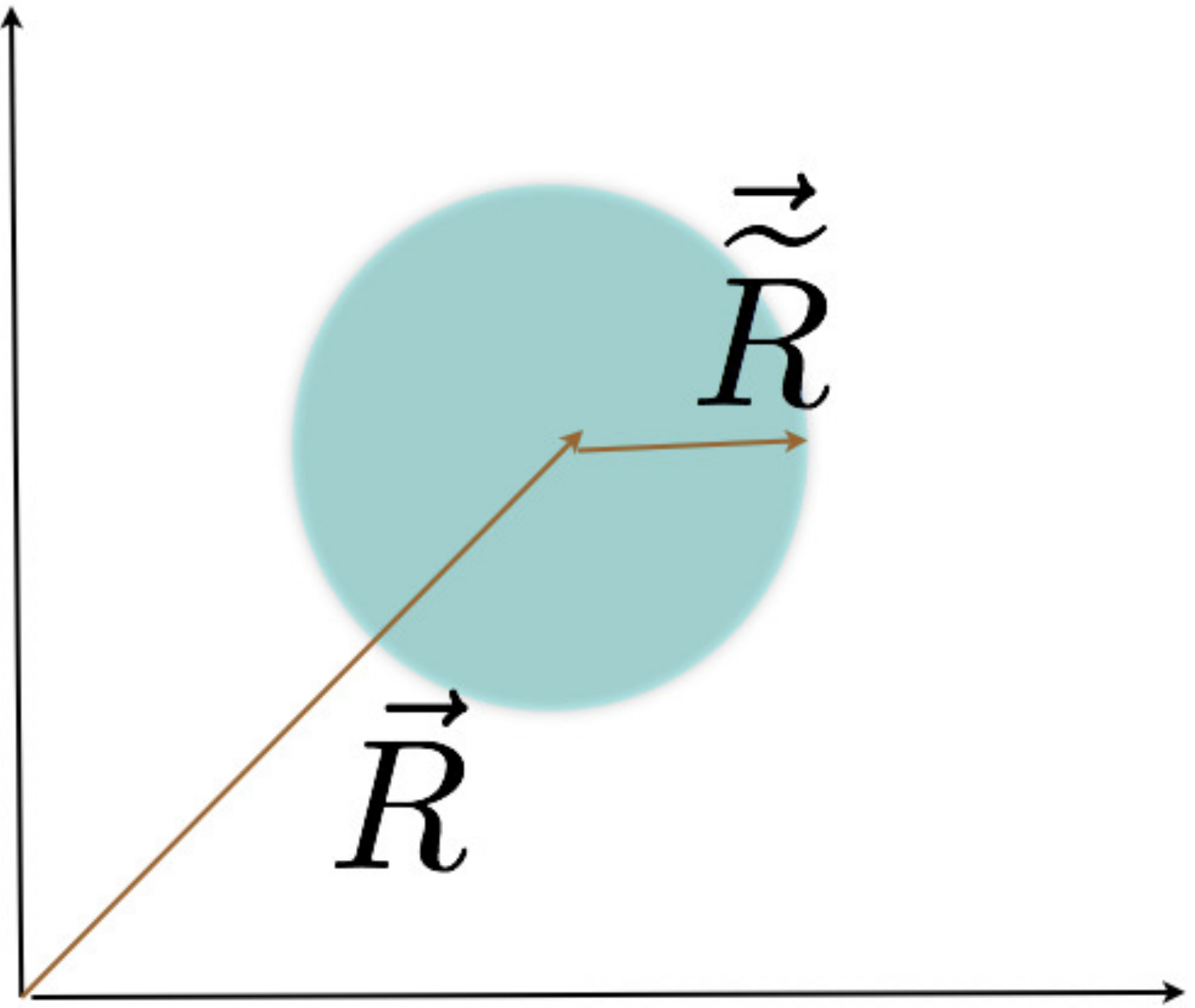}}
\caption{The density of an electron coherent state is smeared out within a single LL. The real coordinates of the electron can be separated into the cyclotron coordinates $\vec{\widetilde R}$ and the guiding center coordinates $\vec R$. The shaded area represents the Gaussian package where most of the electron density is concentrated.}
\label{cood}
\end{figure}

The phase space of the 2D Hall surface is four-dimensional for each particle, with spatial coordinates $r^a$ and momentum coordinates $P_a=-i\partial_a$ satisfying commutation relations $[r^a,P_b]=i\delta ^a_b$. With a perpendicular uniform magnetic field, the covariant momentum is given by $\pi_a=P_a-eA_a$. We choose a new basis for the four-dimensional phase space by writing $r^a=\widetilde R^a+R^a$ with $\widetilde R^a=-l_B^2\epsilon^{ab}\pi_b$ and the following algebra :
\begin{eqnarray}\label{newbasis}
[\widetilde R^a,\widetilde R^b]=il_B^2\epsilon^{ab}\qquad [R^a,R^b]=-il_B^2\epsilon^{ab},\qquad [\widetilde R^a,R^b]=0
\end{eqnarray}
Physically, while $r^a$ gives the location of the particle in the real space, we can separate $r^a$ into the cyclotron coordinates $\widetilde R^a$ and the guiding center coordinates $R^a$. Now the phase space is mapped onto two copies of 2D real spaces, with transparent physical meanings in the two-dimensional Hall manifold (See Fig.(\ref{cood})).

The generator of translations can be separately defined for the cyclotron and guiding center coordinates as follows:
\begin{eqnarray}\label{translation}
&&\widetilde T=e^{iq_a\widetilde R^a},\quad \bar T=e^{iq_aR^a}\\
&&\widetilde T\widetilde R^a\widetilde T^\dagger=\widetilde R^a+l_B^2\epsilon^{ab}q_b,\quad \bar TR^a\bar T^\dagger=R^a-l_B^2\epsilon^{ab}q_b
\end{eqnarray}
The generator of rotation can be defined similarly. Note the definition of angular momentum operator requires a unimodular metric $g_{ab}$ with $\det g=1$. Taking $\hat z=\hat r^1\times \hat r^2$ we have $L_z=\vec r\times\vec P=\epsilon_{ab}r^ag^{bc}P_c$. To separate it into the cyclotron and guiding center parts, we define $\widetilde\Lambda^{ab}=\frac{1}{2}\{\widetilde R^a,\widetilde R^b\},\Lambda^{ab}=\frac{1}{2}\{R^a,R^b\}$; the cyclotron and the guiding center angular momentum operators are
\begin{eqnarray}\label{rotation}
\widetilde L&=&\frac{1}{2}\widetilde g_{ab}\widetilde\Lambda^{ab},\quad L=\frac{1}{2}\bar g_{ab}\Lambda^{ab}\label{rotation}\\
e^{i\theta\widetilde L}\widetilde R^ae^{-i\theta\widetilde L}&=&\cos\theta \widetilde R^a+\sin\theta\epsilon^{ab}\widetilde g_{bc}\widetilde R^c\nonumber\\
e^{i\theta L}R^ae^{-i\theta L}&=&\cos\theta R^a-\sin\theta\epsilon^{ab}\bar g_{bc}R^c
\end{eqnarray}
Now the cyclotron and guiding center angular momentum operators are defined with their respective metric: the cyclotron metric $\widetilde g_{ab}$, and the guiding center metric $\bar g_{ab}$. This is possible because the two angular momentum operators act on different Hilbert spaces. Rotational invariance in the real space asserts $\widetilde g_{ab}=\bar g_{ab}$, which is a special case generally adopted in the literature for technical convenience.

For systems with more than one particle, the density operator is given by $\rho_q=\sum_ie^{iq_ar_i^a}=\sum_ie^{iq_a\widetilde R_i^a}e^{iq_aR_i^a}$, where $i$ is the particle index. The cyclotron density operators can be defined as:
\begin{eqnarray}\label{cdensity}
\widetilde\rho_q=\sum_ie^{iq_a\widetilde R_i^a},\quad [\widetilde\rho_{q_1},\widetilde\rho_{q_2}]=-2i\sin\left(\frac{l_B^2}{2}\vec q_1\times\vec q_2\right)\widetilde\rho_{q_1+q_2}
\end{eqnarray}
while the guiding center density operators can be defined as
\begin{eqnarray}\label{gdensity}
\bar\rho_q=\sum_ie^{iq_aR_i^a}, \quad [\bar\rho_{q_1},\bar\rho_{q_2}]=2i\sin\left(\frac{l_B^2}{2}\vec q_1\times\vec q_2\right)\bar\rho_{q_1+q_2}
\end{eqnarray}
The algebra in Eq. (\ref{cdensity}) and Eq.(\ref{gdensity}) is also called the Girvin-Macdonald-Platzman (GMP) algebra in their respective Hilbert space, which is isomorphic to the $W_\infty$ algebra. To show that, let us factorize the unimodular metric tensor by a set of complex vectors $\omega^a$ satisfying the constraint $i\epsilon^{ab}\omega_a^*\omega_b=-1$. Explicitly we have
\begin{eqnarray}\label{omega}
g_{ab}=\omega_a\omega_b^*+\omega_a^*\omega_b
\end{eqnarray}
These complex vectors are useful in constructing the ladder operators from non-commuting coordinates. As an example, from the guiding center coordinates we can define $b^\dagger=\omega_a^*R^a,b=\omega_aR^a$ such that $[b,b^\dagger]=1$. Taking $W_{m,n}=\left(b^\dagger\right)^{m+1}b^{n+1}$ with $m,n\ge -1$, the $W_\infty$ algebra\cite{winfinity} is given by
\footnotesize
\begin{eqnarray}\label{winfinity}
[W_{m,n},W_{k,l}]=\sum_{s=0}^{\text{Min}\left(n,k\right)}\frac{\left(n+1\right)!\left(k+1\right)!}{\left(n-s\right)!\left(k-s\right)!\left(s+1\right)!}W_{m+k-s,n+l-s}-\left(m\leftrightarrow k, n\leftrightarrow l\right).
\end{eqnarray}
\normalsize
That the $W_\infty$ algebra is isomorphic to the density operator algebra can be seen with the wavelength expansion of the density operators in their respective coordinates. Let us illustrate this by the expansion of the guiding center density operators. The procedure for the cyclotron density operators is exactly the same.

The regularized guiding center density operator is given by 
\begin{eqnarray}\label{rgdensity}
\delta\bar\rho_q=\bar\rho_q-\langle\bar\rho_q\rangle_0
\end{eqnarray}
where $\langle\cdots\rangle_0$ is the ground state expectation value. We thus have $\langle\delta\bar\rho_q\rangle_0=0$. The regularized guiding center density operator still obeys the algebra 
\begin{eqnarray}\label{reducedalgebra}
[\delta\bar\rho_q,\delta\bar\rho_{q'}]=2i\sin\frac{q\times q'l_B^2}{2}\delta\bar\rho_{q+q'}.
\end{eqnarray}
Before expansion, we formally define
\begin{eqnarray}\label{expancomponent}
\bar\Lambda^{a_1,a_2,\cdots,a_n}_i&=&(-i)^ne^{-\frac{i}{2}q_a\bar R_i^a}\left(\partial_{q_{a_1}}\partial_{q_{a_2}}\cdots \partial_{q_{a_n}}e^{iq_a\bar R_i^a}\right)e^{-\frac{i}{2}q_a\bar R_i^a}\nonumber\\
&=&\lim_{q\rightarrow 0}(-i)^n\left(\partial_{q_{a_1}}\partial_{q_{a_2}}\cdots \partial_{q_{a_n}}e^{iq_a\bar R_i^a}\right)
\end{eqnarray}
The expansion of the guiding center density operator is thus given by
\begin{eqnarray}\label{gdensityexpansion}
\delta\bar\rho_q=\sum_{n=1}\frac{i^n}{n!}q_{a_1}q_{a_2}\cdots q_{a_n}\sum_i\bar\Lambda_i^{a_1a_2\cdots a_n}
\end{eqnarray}
The reason for this notably elaborate definition is that in general the operators $R_i^a$ are not bounded in the thermodyamic limit. On the other hand, $\delta \bar\rho_q$ is well-defined when the periodic boundary condition is chosen. In this case only discrete values of $q$ are allowed, but in the thermodynamic limit the partial differential is properly defined. Less formally we can write 
\begin{eqnarray}\label{rewrite}
\Lambda_i^{a_1a_2\cdots a_n}=\frac{1}{n!}\mathcal A[R_i^{a_1}R_i^{a_2}\cdots R_i^{a_n}]
\end{eqnarray}
where $\mathcal A[\cdots]$ anti-symmetrizes over the upper indices. Since in the FQHE the cyclotron coordinates are bounded, the cyclotron counter-part defined in the form of Eq.(\ref{rewrite}) has no problem at all.

The expansion in Eq.(\ref{gdensityexpansion}) reveals two useful sub-algebras\cite{haldane8}. Writing $P_a=\epsilon_{ab}\sum_i\Lambda_i^b$, we have
\begin{eqnarray}\label{subalgebra}
&&[P^a,P^b]=0\label{translation}\\
&&[\Lambda^{ab},\Lambda^{cd}]=-i\left(\epsilon^{ac}\Lambda^{bd}+\epsilon^{ad}\Lambda^{bc}+\epsilon^{bd}\Lambda^{ac}+\epsilon^{bc}\Lambda^{ad}\right)\label{apd}
\end{eqnarray}
Note in Eq.(\ref{translation}) the extensive part proportional to the number of particles is regularized, and $P^a$ is none other than the generator of center-of-mass translation. $\bar\Lambda^{ab}$ is the generator of area-preserving deformation. Explicitly for any symmetric tensor $\alpha_{ab}$, we can define a unitary operator $U(\alpha)=e^{i\alpha_{ab}\Lambda^{ab}}$, which gives us
\begin{eqnarray}\label{deformation}
\bar R'^a_i=U(\alpha)R_i^aU(-\alpha)=e^{-2\epsilon^{ac}\alpha_{cb}}\bar R_i^b
\end{eqnarray}
with $[\bar R'^a_i,\bar R'^b_j]=-i\epsilon^{ab}\delta_{ij}$. The deformation in Eq.(\ref{deformation}) is equivalent to a Bogoliubov transformation of the guiding center ladder operators $b,b^\dagger$, where $\alpha_{ab}$ can be reparametrized as \begin{eqnarray}\label{repa}
g^{ab}\alpha_{ab}=k\cosh 2\theta, \quad\alpha_{ab}\omega^a\omega^b=\frac{1}{2}k\sinh 2\theta e^{2i\phi}
\end{eqnarray}
here $k$ parametrizes the overall phase of the Bogoliubov transformation, and is physically irrelevant.

\subsection{Hamiltonian and Dynamics}

For simplicity, only the QHE of a single species of (spin polarized) fermions is considered here. The full Hamiltonian of the many-body QH system is given by
\begin{eqnarray}\label{fullh}
\mathcal H_0=h_0+V
\end{eqnarray}
where $h_0$ is the single particle Hamiltonian and $V$ contains the many-body interactions. In principle $V$ can contain $k$- body interactions for any integer $k>1$. Here only the physical case of the two-body Coulomb interaction is considered. However, this is in no way reducing the generality of the effective Hamiltonian for the FQHE, since the $k>2$ interactions physically result from LL mixing, as we shall see later.

The single particle Hamiltonian is special because the minimal coupling of the external magnetic field to the kinetic momentum implies that $h_0$ is a function of \emph {only} the cyclotron coordinates. Assuming inversion symmetry, the most general form of $h_0$ is given by
\begin{eqnarray}\label{generalh0}
h_0=\sum_i\sum_{n=1}^\infty\frac{1}{(2n)!l_B^{2n}}\omega_{a_1,a_2,\cdots a_{2n}}\widetilde R_i^{a1}\widetilde R_i^{a2}\cdots \widetilde R_i^{a_{2n}}
\end{eqnarray}
where $\omega_{a_1,a_2,\cdots a_{2n}}$ are the fully symmetric tensors. In general $h_0$ does not have Galilean or rotational invariance. Physically the QHE is realized on a lattice system, where Galilean and rotational invariance can only emerge in the weak field limit $l_B\gg a$, where $a$ is the lattice constant. The energy spectrum of $h_0$ are generalized Landau levels, each with macroscopic degeneracy generated by the guiding center coordinates, because $h_0$ \emph{commutes} with $R_i^a$. If we do have rotational invariance for $h_0$, every $\omega_{a_1,a_2,\cdots a_{2n}}$ can be expressed as a function of a single metric - the cyclotron metric $\widetilde g_{ab}$. The general form of $h_0$ will be simplified to 
\begin{eqnarray}\label{rotationalh0}
h_0^r=\sum_i\sum_{n=1}^\infty\frac{c_n}{(2n)!l_B^{2n}}\left(\frac{1}{2}\widetilde g_{ab}\widetilde R_i^a\widetilde R_i^b\right)^n=\sum_i\sum_{n=1}^\infty\frac{c_n}{(2n)!l_B^{2n}}\widetilde L^n
\end{eqnarray}

In this case, the eigenstates are labeled by the cyclotron angular momentum. A familiar example is the massless Dirac fermions with the single particle Hamiltonian $h_0^r=\sqrt{1+\widetilde L}$. For free electrons confined in two-dimensions, we have Galilean invariance and the cyclotron metric is given by the effective mass tensor. In this case the LLs are equally spaced, and we can also define a cyclotron frequency $\omega_c=eB/m$ and the single particle Hamiltonian reduces to:
\begin{eqnarray}
h_0^g=\frac{\omega_c}{2l_B^2}\sum_i\widetilde R_i^a\widetilde R_i^b
\end{eqnarray}
The Galilean term is the leading term of the expansion in Eq.(\ref{generalh0}), and in most cases it defines the energy scale of the single particle Hamiltonian $\epsilon_s\sim\hbar\omega_c\sim B$.

The interaction term $V$ depends on the real space coordinates of particles $r_i^a$. Even though the Coulomb interaction is universal, the details of effective interaction between electrons confined in a two-dimensional manifold depends on the LL form factor and the experimental conditions, such as the single particle wavefunction in z direction (perpendicular to the Hall manifold), which depends on the thickness of the sample and the profile of the confinement potential. Denoting $V_q$ the Fourier component of the effective two-body interaction potential we have
\begin{eqnarray}\label{bareinteraction}
V=\int \frac{d^2ql_B^2}{(2\pi)^2}V_q\rho_q\rho_{-q}
\end{eqnarray}
The only length scale is given by the magnetic length $l_B$, thus the typical energy scale of the interaction is given by $\epsilon_{\text{int}}\sim e^2/l_B\sim\sqrt B$, which is subleading to $\epsilon_s$. In the limit of strong magnetic field, one is allowed to treat $V$ as a small perturbation, and we use this to organize the many-body Hilbert space. Formally, we write 
\begin{eqnarray}\label{lambdah}
H(\lambda_0)=h_0+\lambda_0V
\end{eqnarray}
In the limit of $\lambda_0\rightarrow 0$, there is a subspace $\mathcal H_{\lambda_0}$ spanned by eigenstates of $H(\lambda_0)$ that are degenerate with the ground state. If the filling factor $\nu$ is an integer, this subspace only contains the ground state, and the non-degenerate perturbation theory can be applied straightforwardly. This is the way we understand the IQHE. When partially filled LLs are present in the ground state, one has to apply the degenerate pertubation theory, which becomes intractable when the degeneracy is macroscopic in the thermodynamic limit. This is the case of the FQHE, which is formally treated by defining a projection operator $\mathcal P=\sum_n\left(|\bar\psi_n\rangle\langle\bar\psi_n|\right)$ for all $|\bar\psi_n\rangle\in\mathcal H_{\lambda_0\rightarrow 0}$. The interaction Hamiltonian can thus be written as
\begin{eqnarray}\label{interactionh}
V=\mathcal PV\mathcal P+O(\epsilon)
\end{eqnarray}
where $\epsilon\sim\epsilon_{\text{int}}/\epsilon_s\sim B^{-\frac{1}{2}}$. The projected interaction Hamiltonian $\bar V=\mathcal P V\mathcal P$ has the spectrum
\begin{eqnarray}\label{projected}
\bar V|\bar\psi_n\rangle=\bar\epsilon_n|\bar\psi_n\rangle
\end{eqnarray} 
In this projected Hilbert space, the kinetic energy of each particle is just a constant, the dynamics is dictated by the interaction alone. 

The leading order of Eq.(\ref{interactionh}) is the two-body interaction within a single partially filled LL (the case with more than one partially filled LLs is technically more cumbersome but conceptually the same). Terms of $O(\epsilon)$ contain LL mixing induced by the interaction, and can be calculated perturbatively\cite{bn}. The perturbation does not just renormalize the effective two-body interaction; the first order perturbation also gives effective three-body interactions, while higher-order perturbations lead to four-body interactions and more.

Formally, Eq.(\ref{interactionh}) can be written as a general effective Hamiltonian including $k-$body interactions for $k\ge 2$. The Hilbert space is still within a single LL, but the coefficients of every term in the Hamiltonian can be expanded in powers of $\epsilon$. Thus the FQHE can be completely described by the physics within a single LL even when LL mixing is included. Theorists can tune the coefficients of $k-$body interactions at will to realize different models of the FQHE; this makes numerical analysis a very powerful tool.  Perturbative calculations from realistic physical interactions, on the other hand, suggest that the effect of LL mixing is quite small, even though $\epsilon$ itself is not very small  ($\sim 0.3$) under most experimental conditions\cite{bn}. Thus a model of two-body interaction is sufficient in realizing most FQH states.

\section{Organization of the Thesis}

In Chapter 2, an overview of the numerical techniques in the FQHE is presented, focusing on the fact that the Hamiltonian matrix can be numerically constructed in a purely algebraic way that is manifestedly gauge invariant. The chapter also contains three examples that illustrate the power of the numerical analysis. The first example presents the entanglement spectrum of the FQH ground state on the sphere, with a notably new partition of the Hilbert space leading to a clear entanglement energy separation between the topological and the non-universal part of the entanglement spectrum. This new partition can be potentially useful for DMRG application. In the second example a numerical definition of the guiding center metric for the FQH fluid without rotational invariance is presented. In the last example, possible transitions from the incompressible FQH phase to compressible bubble/stripe phases are studied in the higher Landau levels, especially when the rotational invariance is broken.

In Chapter 3, the geometric aspect of the QHE is illustrated with microscopic calculations of the linear response to spatially varying electromagnetic fields. In particular, the term ``Hall viscosity" will be introduced in this chapter, which is an important quantity in the electromagnetic response, and is universal with rotational invariance. The Hall viscosity bridges the geometry of the QHE with its topological aspect, and also determines the gap of the neutral excitations in the long wavelength limit, as will be shown in Chapter 4 and 5. 

In Chapter 4 a numerical scheme for the construction of neutral excitation model wavefunctions in the Laughlin and Moore-Read state is presented. These model wavefunctions are compared with both the exact diagonalization and the single mode approximation. The dynamics of the long wavelength part of the magnetoroton mode is revealed to be both dependent on the Hall viscosity and the energy cost of the shear deformation of the ground state guiding center metric. 

With numerical results from Chapter 4 at hand, the analytic wavefunctions for the neutral excitations are presented in Chapter 5. These analytic wavefunctions are shown to be a generalization of the Laughlin and Moore-Read ground state wavefunctions, with no tuning parameters and transparent physical interpretations. The analytic calculations of the long wavelength neutral excitation gap in the thermodynamic limit reveals interesting connections to the dynamics of the two-dimensional plasma picture, where the energy gap of the quadrupole excitation is related to the free energy cost of the fusion of charges in the plasma. A lattice diagramatic representation of the model wavefunctions for the neutral excitations is also presented in this chapter, leading to a fresh point of view of the nature of quantum Hall many-body wavefunctions.

\chapter{Numerical Studies of the FQHE\label{ch:numerics}}

Numerical calculation is an indispensible tool in studying the FQHE. It is remarkable that in many cases the physics of the FQHE in the thermodynamic limit can be revealed with the numerical calculation of systems containing only a few particles. The Hilbert space of the FQHE is tractable once the system is projected into a single Landau level, with the proper boundary conditions. Effects of different cyclotron form factors in different LLs, modifications of the interaction by finite thickness, etc. can be modeled with a suitable choice of a set of the Haldane pseudopotentials for the two-body interactions. The effects of LL mixing, on the other hand, can be modeled by adding three (or even more) body interactions within a single LL.

In contrast to common practices in the FQHE numerical calculations, where one has to pick a gauge to specify the single particle wavefunctions, the numerical method presented in this chapter is based on the algebra of the FQH Hilbert space, and is manifestedly gauge invariant. This is both conceptually and technically advantageous over the use of real space wavefunctions of the single particle orbitals. While this chapter does not give a detailed guide for implementing numerical calculations and optimizations based on symmetry, it emphasizes the universal features of the numerical analysis in different geometries, from which the Hamiltonian matrix is built for exact diagonalization or DMRG analysis. For simplicity only spin polarized FQH systems are considered. The three examples in the chapter illustrate how numerical techniques can be implemented in spherical and torus geometry, to analyze both the ground state properties and the dynamics involving the entire energy spectrum.

\section{Landau Level Projection}

It is convenient to introduce the second-quantized formalism when doing numerical calculations.  The density operator is given by
\begin{eqnarray}\label{2nddensity}
\rho_q^s=\sum_{MN,mn}\langle M,m|e^{iq_ar^a}|N,n\rangle\xi^\dagger_{Mm}\xi_{Nn}
\end{eqnarray}
where the upper-case indices are LL indices, and the lower-case indices are the guiding center orbital indices. $\xi^\dagger_{Mm}$ creates a particle in the $\text{M}^{\text{th}}$ LL and the $\text{m}^{\text{th}}$ intra-LL orbital, and $\{\xi^\dagger_{Mm},\xi_{Nn}\}=\delta_{MN}\delta_{mn}$. Projection into the $\text{N}^{\text{th}}$ LL means only particles with LL index N are included in the Hilbert space. Writing $e^{iq_ar^a}=e^{iq_aR^a}e^{iq_a\widetilde R^a}$, we have
\begin{eqnarray}\label{formfactor}
\langle N,m|e^{iq_ar^a}|N,n\rangle=F_N(q)\langle m|e^{iq_aR^a}|n\rangle
\end{eqnarray}
where $F_N(q)=\langle N|e^{iq_a\widetilde R^a}|N\rangle$ is the LL form factor, completely determined by the single particle kinetic energy Hamiltonian $h_0$. If $h_0$ is rotationally invariant, i.e. containing only a single metric $\widetilde g_{ab}$, we can define the spectrum generating LL ladder operators $a=\omega_a\widetilde R^a, a^\dagger=\omega_a^*\widetilde R^a$ such that $a^\dagger a|N\rangle=N|N\rangle$ (where the complex vectors $\omega_a$ are defined in Eq.(\ref{omega})). An explicit calculation gives $F_N(q)=\mathcal L_N(q^2l_B^2/2)e^{-\frac{1}{4}q^2l_B^2}$, where $\mathcal L_N(x)$ is the $N^{\text{th}}$ Laguerre polynomial. For the purpose of numerical calculations, we are only going to deal with a rotationally invariant $h_0$, and the density operator is written as 
\begin{eqnarray}\label{2ndpdensity}
\rho_q=F_N(q)\sum_{mn}\langle m|e^{iq_aR^a}|n\rangle\xi^\dagger_m\xi_n
\end{eqnarray}
The LL index for the creation and annihilation operators are omitted without ambiguity. For numerical calculations on a flat surface, Eq.(\ref{2ndpdensity}) is the general implementation for the LL projection.

\section{Two-Body and Three-Body Interactions}

The general Hamiltonian for the two-body interaction is given by
\begin{eqnarray}\label{2body}
\mathcal H_{\text{2bdy}}=\int \frac{d^2q}{(2\pi)^2}V_q\bar\rho_q\bar\rho_{-q}=\int \frac{d^2q}{(2\pi)^2}V_q\sum_{i<j}e^{iq_a\left(R_i^a-R_j^a\right)}
\end{eqnarray}
where $\bar\rho_q=\sum_ie^{iq_aR_i^a}=\sum_{mn}\langle m|e^{iq_aR^a}|n\rangle\xi^\dagger_m\xi_n$ is the guiding center density operator and $i$ is the particle index. Comparing to Eq.(\ref{2nddensity}), the form factor is absorbed into $V_q$, the Fourier component of the two-body interaction. Different ways of organizing the single particle orbitals within a single LL is analogous to picking a gauge. Choosing an arbitrary complex vector $u_a$ we can define $R=u_aR^a$. Coherent states with $R|m\rangle=m|m\rangle$ is one way of labeling the single particle orbitals. If rotational invariance exists with metric $g_{ab}=\omega_a\omega_b^*+\omega_a^*\omega_b$, where $\omega_a$ is defined in Eq.(\ref{omega}), we can let $u_a=\omega_a$ so $R=b$ is the ladder operator. In this case a more natural way is to label the single particle orbital by its guiding center angular momentum. The single particle orbital is labeled by $|m\rangle$ with $b^\dagger b|m\rangle=m|m\rangle$. Writing $z=\omega_ar^a$, and in an infinite plane such states are given by
\begin{eqnarray}\label{rotationcoherentstate}
\langle z|m\rangle\sim z^me^{-\frac{1}{2}|z|^2}
\end{eqnarray}
which is analogous to the case when an explicit symmetric gauge is picked. The coherent state for a general $u_a$ is given by
\begin{eqnarray}\label{singleparticle}
\langle z|m\rangle\sim e^{\frac{1}{u_a\omega^a}\left(mz-\frac{1}{2}u_a\omega^{a*}z^2\right)}e^{-\frac{1}{2}zz^*}
\end{eqnarray}
Choosing $u_a$ to be a real vector is analogous to picking a ``Landau gauge", where Eq.(\ref{singleparticle}) is extended in the direction perpendicular to the vector $u_a$, and confined in the direction parallel to $u_a$. If the single particle wavefunction needs to satisfy certain boundary condition (e.g. periodic boundary condition on torus), Eq.(\ref{singleparticle}) is mathematically more complicated, and one is forced to pick a gauge for the vector potential. On the other hand, numerical calculations do not require us to deal with wavefunctions explicitly; they can be done algebraically with various different boundary conditions, as we shall see in the next section.

Since for two-body interactions only the relative coordinates are involved in the Hamiltonian, we write $R_{ij}=\frac{1}{\sqrt{2}}u_a\left(R_i^a-R_j^a\right), \bar R_{ij}=\frac{1}{\sqrt{2}}u_a\left(R_i^a+R_j^a\right)$. The two-body eigenstates are given by $|M,m\rangle=\xi_{M,m}^\dagger|\text{vac}\rangle$, where $M$ is the index for $\bar R_{ij}$, and $m$ is the index for $R_{ij}$.

The Hamiltonian is thus given by
\begin{eqnarray}\label{2ndqham}
\mathcal H_{\text{2bdy}}=\int \frac{d^2q}{(2\pi)^2}V_q\sum_{M,m,m'}\langle M,m|e^{iq_a\left(R_1^a-R^a_2\right)}|M,m'\rangle\xi^\dagger_{M,m}\xi_{M,m'}
\end{eqnarray}
If $V_q$ is rotationally invariant, we can expand it in the basis of Laguerre polynomials with $V_q=\sum_nc_nV_n$ and:
\begin{eqnarray}\label{2laguerre}
V_n=\mathcal L_n(q^2)e^{-\frac{1}{2}q^2}
\end{eqnarray}
This is because the Laguerre polynomials $\mathcal L_n(q^2)$ are orthogonal when integrated with the measure $e^{-\frac{1}{2}q^2}$. $V_n$ is called the $n^{\text{th}}$ Haldane pseudopotential. If we label the single particle orbitals by its guiding center angular momentum, i.e. $u_a=\omega_a^*$. This leads to $\langle M,m|e^{iq_a\left(R_1^a-R^a_2\right)}|M,m'\rangle=\delta_{m,m'}e^{-\frac{1}{2}l_B^2q^2}\mathcal L_m(q^2l_B^2)$. Due to the orthogonality between the Laguerre polynomials, Eq.(\ref{2ndqham}) is simplified to
\begin{eqnarray}\label{r2ndqham}
\mathcal H_{\text{2bdy}}=\sum_{M,m}c_m\xi^{(2)\dagger}_{M,m}\xi^{(2)}_{M,m}
\end{eqnarray}
Thus $c_m$ is the energy cost of a pair of particles having relative angular momentum $m$. The generalization to $k-$body interactions with $k>2$ is presented in \cite{kbodyint}. Here we are going to use the Jacobi coordinates to generalize the Haldane pseudopotentials. For three-body interactions the Jacobi coordinates are given by:
\small
\begin{eqnarray}\label{jacobi}
R^a_{ij}=\frac{1}{\sqrt{2}}\left(R^a_i-R^a_j\right),\quad R^a_{ij,k}=\frac{1}{\sqrt{6}}\left(R^a_i+R^a_j-2R^a_k\right),\quad R^a_{ijk}=\frac{1}{\sqrt{3}}\left(R^a_i+R^a_j+R^a_k\right)
\end{eqnarray}
\normalsize
Defining $R_{ij}=u_aR^a_{ij}, R_{ij,k}=u_aR^a_{ij,k}, R_{ijk}=u_aR_{ijk}^a$, the three-body Hilbert space is given by $|M,m,m'\rangle=\xi^\dagger_{M,m,m'}|\text{vac}\rangle$, where $M$ is the eigenstate index for $R_{ijk}$, $m$ is the eigenstate index for $R_{ij,k}$ and $m'$ is the eigenstate index for $R_{ij}$. The most general three-body Hamiltonian is given by
\footnotesize
\begin{eqnarray}\label{3body2ndham}
\mathcal H_{\text{3bdy}}=\int \frac{d^2q_1d^2q_2}{(2\pi)^4}V_{q_1,q_2}\sum_{Mm_1m_2m'_1m'_2}\langle Mm_1m'_1|e^{iq_{1a}R^a_{12}}e^{iq_{2a}R^a_{12,3}}|Mm_2m'_2\rangle\xi^{(3)\dagger}_{Mm_1m_1'}\xi^{(3)}_{Mm_2m_2'}
\end{eqnarray}
\normalsize
With rotational invariance we can do a similar expansion of the interaction with the basis of the Laguerre polynomials:
\begin{eqnarray}\label{3laguerre}
V_{q_1,q_2}=\sum_{nn'}c_{nn'}\mathcal L_n(q^2_1l_B^2)\mathcal L_{n'}(q^2_2l_B^2)e^{-\frac{1}{2}\left(q_1^2+q_2^2\right)l_B^2}
\end{eqnarray}
Again labeling the single particle orbitals by their guiding center angular momentum we have
\begin{eqnarray}\label{r3body2ndham}
\mathcal H_{\text{3bdy}}=\sum_{Mm_1m_1'}c_{m_1m_1'}\xi^{(3)\dagger}_{Mm_1m_1'}\xi^{(3)}_{Mm_1m_1'}
\end{eqnarray}
This is how the Haldane pseudopotentials for two-body interactions are generalized. The scheme can be naturally extended to $k-body$ interactions for $k>3$ with the set of coefficients $c_{m_1,m_2\cdots m_{k-1}}$; Eq.(\ref{2laguerre}) and Eq.(\ref{3laguerre}), and the physics of the coefficients of expansions, are completely general and independent of the geometry or topology of the Hall surface. The numerical study of the FQHE is tantamount to exploring the family of model Hamiltonians in the parameter space $\{c_{n_1},c_{n_1n_2},\cdots\}$.

\section{Disk, Cylinder and Torus}

From an experimental point of view, the quantum Hall droplet is realized on a two-dimensional plane of a finite size with electrons confined by an external potential. The details of the confinement potential perpendicular to the two-dimensional plane modifies the single particle wavefunction in the perpendicular direction, which in turn modifies the effective two-body interaction. Numerically, this can be modeled by tuning the different components of the pseudopotentials. The simplest geometry is the disk geometry with open boundary conditions; in this geometry both the bulk and edge excitations can be explored numerically\cite{disk,disk1}. We can also make the boundary condition open in one direction and periodic in the other; this gives a cylinder geometry with two chiral edges\cite{cylinder,cylinder1}. If we make both directions periodic, we have the torus geometry\cite{haldane4}. Torus geometry has no edge, which is convenient for exploring the bulk excitations. It also has a different topology (with genus 1), which is essential in studying the ground state degeneracy of different topological phases\cite{wendegeneracy}.

On the disk, rotational invariance is present and we can use Eq.(\ref{r2ndqham}) and Eq.(\ref{r3body2ndham}) directly. The two-particle creation operators can be expanded in terms of the single-particle creation operators:
\small
\begin{eqnarray}\label{2ndexpansion}
\xi^\dagger_{M,m}=2^{-\frac{1}{2}(M+m)}\sum_{a,b=0}^{a,b=M,m}(-1)^{a+b}\sqrt{\frac{\left(M+m-a-b\right)!(a+b)!}{M!m!}}\left(\begin{array}{ccc}
M\\
a\end{array}\right)\left(\begin{array}{ccc}
m\\
b\end{array}\right)\xi^\dagger_{M+m-a-b}\xi^\dagger_{a+b}\nonumber\\
\end{eqnarray}
\normalsize

 \begin{figure}[h!]
 \includegraphics[width=1\linewidth]{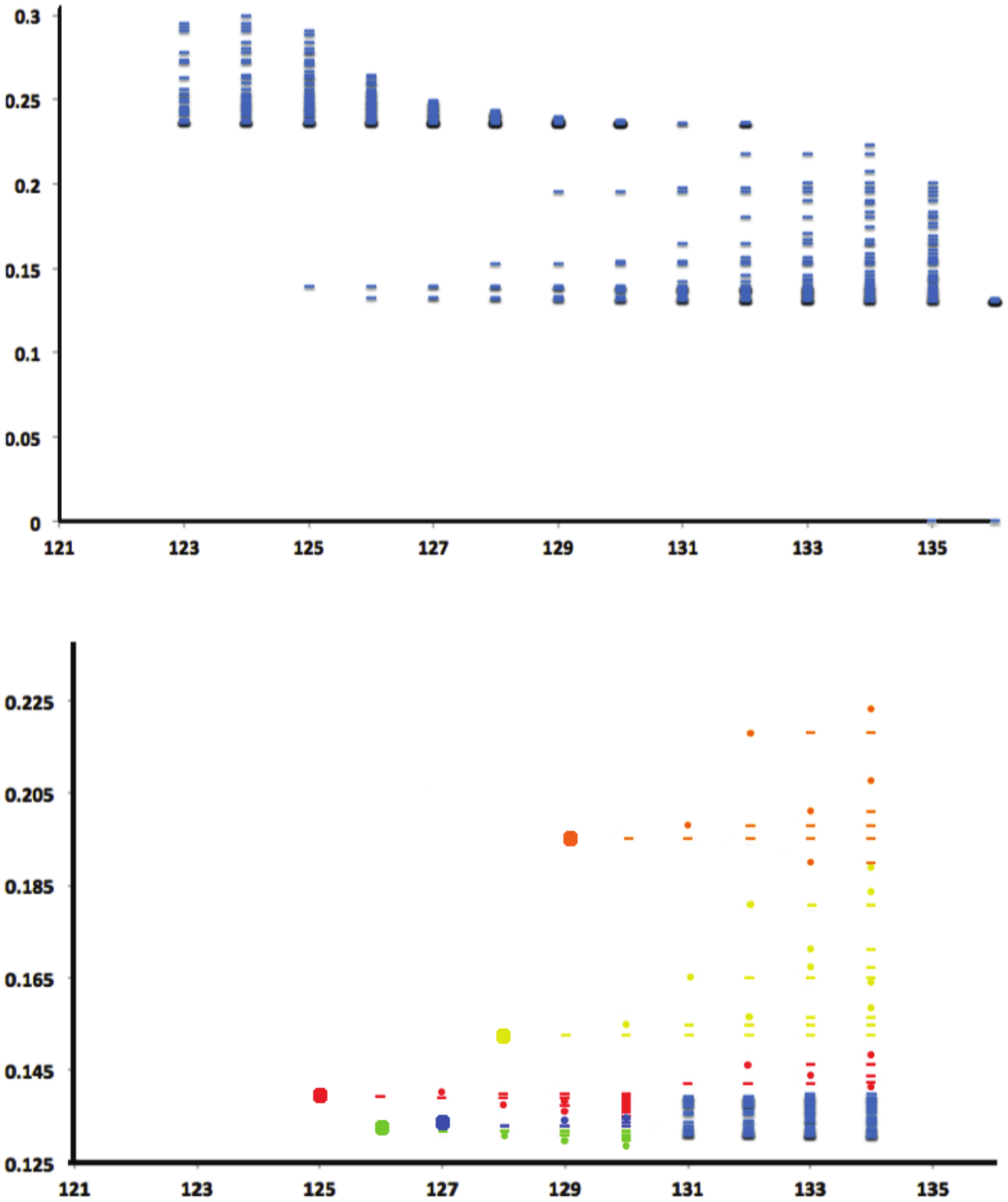}
\caption{The top graph is the energy spectrum of ten electrons on the disk at filling factor $\nu=1/3$. There is no restriction in the number of orbitals for exact diagonalization. The states with the lowest fifty energies in each angular momentum sector is plotted. The bottom graph is the zoom into the bulk excitation part of the disk spectrum (One should compare that with Fig.(\ref{fig_torus}) and Fig.(\ref{fig_sphere}), where only bulk neutral excitations are present.)}
\label{fig_disk}
\end{figure}
The numerical implementation is thus straightfoward, and the Hamiltonian will be block-diagonal with total angular momentum $M+m$ as a good quantum number. The finite size system consists of a finite number of particles $N_e$, and in principle there is no restriction of the number of orbitals, or the size of the disk. If a confining potential from the background positive charges is present, additional on-site single particle potential term will be added to the Hamiltonian\cite{disk,disk1}. One can also impose a sharp cut-off by restricting the number of orbitals, and thus truncating the Hilbert space in which the diagonalization is performed.

In Fig.(\ref{fig_disk}) the energy spectrum of a typical Laughlin state is plotted. What is interesting is the bottom half of the diagram, where we zoom into the bulk excitation part of the spectrum below the multi-roton continuum. Except for the big circular plot, each many-body state contains both bulk and edge excitations. The five different colors represent five different branches of the neutral excitations below the multi-roton continuum (except for the blue color at the lower right corner, where different branches mix and there is not enough resolution of the plot to differentiate between them). In each branch, the state with the big circular plot is the highest weight bulk state (with no edge excitations) corresponding to those on the sphere or the torus. In each branch the counting of the states follows the Virasoro algebra (see Chapter 5), and the small circular plots are the highest weight states. Just like the ground state, each bulk neutral excitation is the highest weight primary field where the Kac-Moody edge modes are generated. 

The cylinder geometry does not have rotational invariance. Let the cylinder to be periodic along the $y$-axis with the circumference $L$ and open along the $x$-axis, we can pick $u_x=1,u_y=0$ and the single particle orbitals are given by $R^x|m\rangle=\frac{2\pi l_B^2m}{L}|m\rangle$. We thus have
\begin{eqnarray}\label{cylinder}
e^{iq_xR^x}|m\rangle=e^{i\frac{2\pi l_B^2q_xm}{L}}|m\rangle,\quad e^{iq_yR^y}|m\rangle=|m+k\rangle
\end{eqnarray}
where $q_y=\frac{2\pi}{L}k$ is discrete for integer $k$, and $q_x$ is a continuous variable. We can thus pick $m$ to be an integer as well. In the second quantized form the guiding center density operator is given by
\begin{eqnarray}\label{cylinderdensity}
\bar\rho_q=\sum_me^{iq_x\left(m+\frac{1}{2}k\right)}\xi^\dagger_{m+k}\xi_m
\end{eqnarray}

 \begin{figure}[h!]
 \includegraphics[width=1\linewidth]{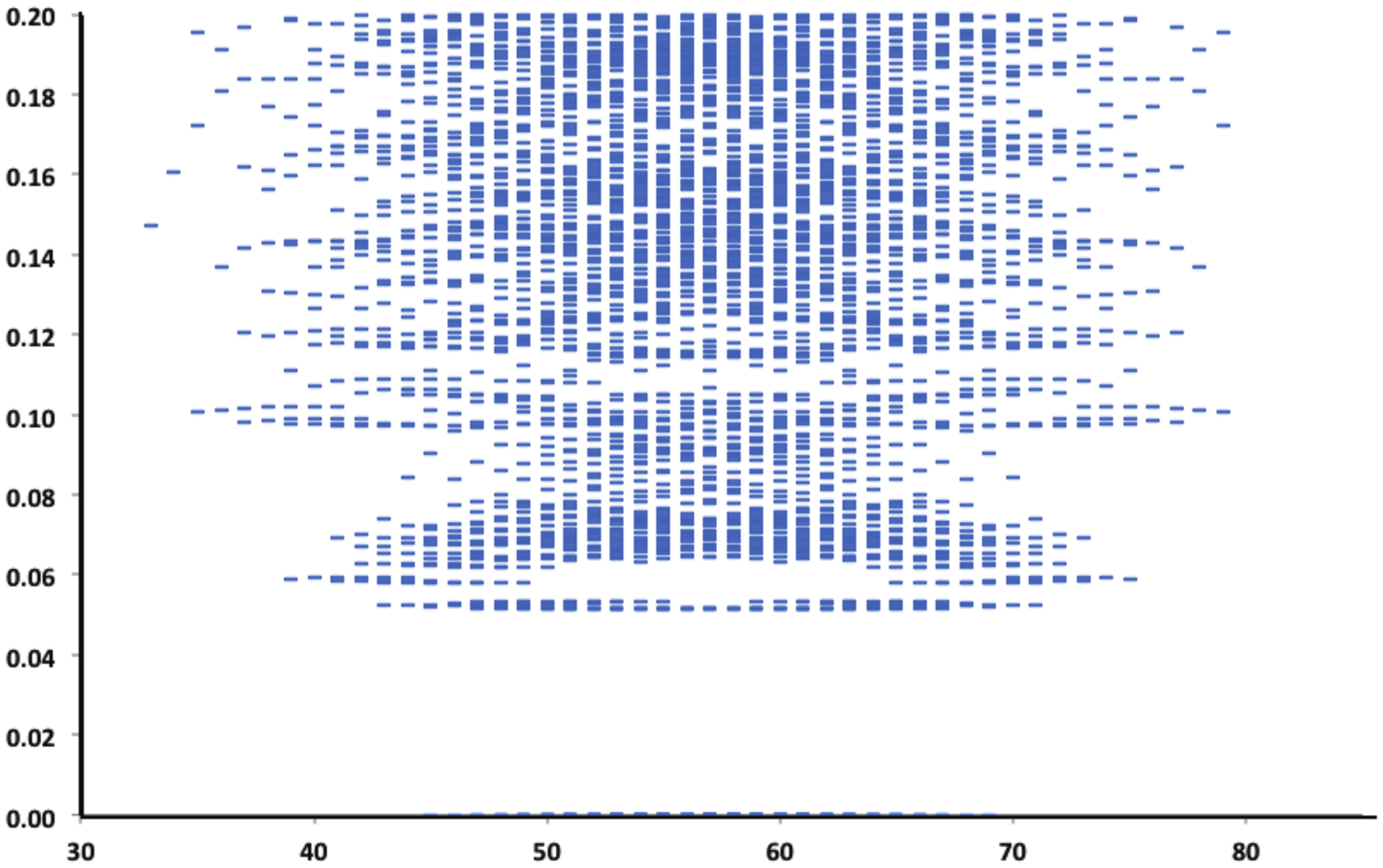}
\caption{The entire energy spectrum of six electrons on the cylinder at filling factor $\nu=1/3$ and aspect ratio equal to one, with no cut-off of orbitals. The degenerate manifold of zero energy states can be seen across all momentums due to translational invariance along the cylinder. The gapped neutral excitations are more complicated because of the presense of two chiral edges.(Plot courtesy: Sonika Johri).}
\label{fig_cylinder}
\end{figure}
Thus the two-body and three-body interaction Hamiltonians are given by
\begin{eqnarray}\label{cylinderham}
\mathcal H_{\text{2bdy}}&=&\int \frac{dq_xl_B}{2\pi}\sum_{kmn}V_qe^{i\frac{2\pi l_B^2q_x}{L}(m-n)}\xi^\dagger_{m+k}\xi^\dagger_{n-k}\xi_m\xi_n\label{c2bdy}\\
\mathcal H_{\text{3bdy}}&=&\int \frac{dq_{1x}dq_{2x}l_B^2}{4\pi^2}\sum_{k_1k_2m_1m_2m_3}V_{q_aq_b}e^{\frac{i\pi l_B^2}{L}\left(q_{1x}k_2+q_{2x}k_1\right)}\cdot\nonumber\\
&&e^{\frac{2\pi iq_{1x}}{L}\left(m_1-m_3\right)}e^{\frac{2\pi iq_{2x}}{L}\left(m_2-m_3\right)}\xi^\dagger_{m_1+k_1}\xi^\dagger_{m_2+k_2}\xi^\dagger_{m_3-k_1-k_2}\xi_{m_1}\xi_{m_2}\xi_{m_3}\label{c3bdy}
\end{eqnarray}
where in Eq.(\ref{c3bdy}) we have $q_a=\frac{1}{\sqrt{2}}\left(q_1+\frac{1}{\sqrt{2}}q_2\right), q_b=\sqrt{\frac{3}{2}}q_2$. 

The torus geometry is periodic in both directions. Unlike the disk and cylinder geometry, the torus is a compact manifold. To respect the boundary conditions, the total number of fluxes going through the surface has to be an integer\cite{haldane4}. This implies the total area of the torus to be $2\pi l_B^2N_\phi$, where the integer $N_\phi$ is the number of fluxes. Let the torus be defined by two principal vectors $\vec L_1=(L_1,0), \vec L_2=(L_2\cos\theta,L_2\sin\theta)$, so that $\vec L_1,\vec L_2$ forms a parallelgram, and the periodic boundary conditions require opposite sides of the parallelgram to be identified. The flux quantization condition is given by
\begin{eqnarray}\label{fluxquant}
L_1L_2\sin\theta=2\pi l_B^2N_\phi
\end{eqnarray}
 \begin{figure}[h!]
 \includegraphics[width=1\linewidth]{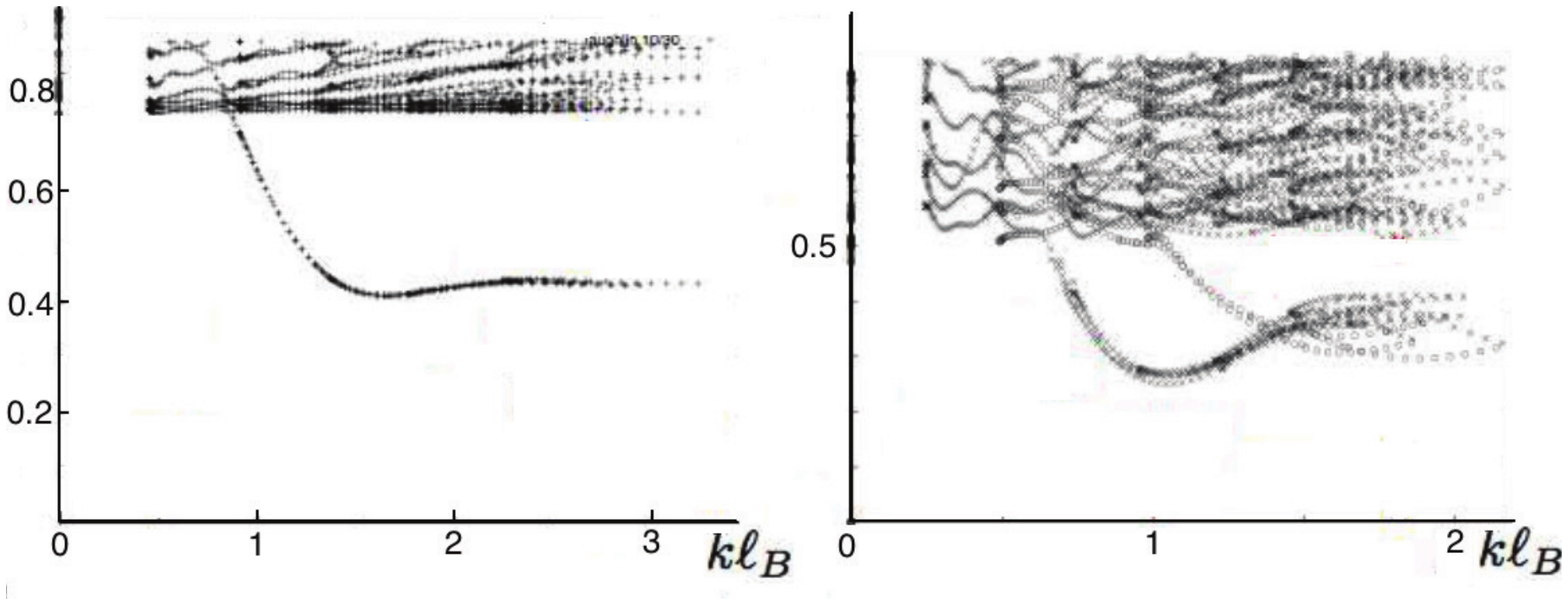}
\caption{This is the entire energy spectrum obtained on the torus geometry with exact diagonalization. On the left, the filling factor is $\nu=1/3$, with $V_1$ pseudopotential interactions. On the right, the filling factor is $\nu=1/2$, with the three-body model Hamiltonian and quartered Brillouin zone\cite{quartered}. The ground state with zero energy is set to have zero momentum, and a branch of magneto-roton modes can be seen clearly both for the Laughlin and the Moore-Read state, reaching the roton-minimum at $kl_B\sim 1.7$ for both cases, and merges into the multi-roton continuum in the long wavelength limit. For the Moore-Read state, the neutral fermion modes can also be seen to reach minimum gap at $kl_B\sim 1$. (Plot courtesy: F.D.M. Haldane).}
\label{fig_torus}
\end{figure}
The periodic boundary conditions fix a discrete set of allowed momentum vectors forming the reciprocal lattice, with primitive vectors $\vec q_1=\frac{L_2}{l_B^2N_\phi}\left(\sin\theta,-\cos\theta,0\right),\vec q_2=\frac{L_1}{l_B^2N_\phi}\left(0,1,0\right)$. There are in total $N_\phi^2$ allowed vectors within the Brillouin zone, with $\vec q=k_1\vec q_1+k_2\vec q_2$ for $k_1,k_2=0,1,\cdots, N_\phi-1$. 

The single particle orbitals can be defined in the same way as the case for the cylinder in Eq.(\ref{cylinder}), with additional periodic boundary conditions. Writing $R^x|m\rangle=\frac{2\pi l_B^2m}{L_2\sin\theta}|m\rangle$, with $q_x=\frac{L_2\sin\theta k_1}{l_B^2N_\phi},q_y=\frac{L_1k_2}{l_B^2N_\phi}$ for integers $k_1,k_2$, we have
\begin{eqnarray}\label{singleparticletorus}
e^{iq_xR^x}|m\rangle=e^{i\frac{2\pi k_1m}{N_\phi}}|m\rangle,\quad e^{iq_yR^y}|m\rangle=|m+k_2\rangle,\quad |m+N_\phi\rangle=|m\rangle
\end{eqnarray}
It is thus straightforward to rewrite Eq.(\ref{c2bdy}) and Eq.(\ref{c3bdy}) in terms of discrete sums over the reciprocal lattice momentum vectors:
\begin{eqnarray}\label{torusham}
\mathcal H_{\text{2bdy}}&=&\sum_{k_1k_2mn}V_{k_1k_2}e^{\frac{2\pi ik_1}{N_\phi}(m-n)}\xi^\dagger_{m+k_2}\xi^\dagger_{n-k_2}\xi_m\xi_n\label{t2bdy}\\
\mathcal H_{\text{3bdy}}&=&\sum_{k_1k_2k_3k_4m_1m_2m_3}V_{k_1k_2k_3k_4}e^{\frac{2\pi i}{N_\phi}\left(k_3k_2+k_4k_1\right)}\cdot\nonumber\\
&&e^{\frac{4\pi ik_3}{N_\phi}\left(m_1-m_3\right)}e^{\frac{4\pi ik_4}{N_\phi}\left(m_2-m_3\right)}\xi^\dagger_{m_1+k_1}\xi^\dagger_{m_2+k_2}\xi^\dagger_{m_3-k_1-k_2}\xi_{m_1}\xi_{m_2}\xi_{m_3}\label{t3bdy}
\end{eqnarray}
where in Eq,(\ref{t2bdy}) we have $V_{k_1k_2}=V_q,\vec q=\left(\frac{L_2\sin\theta k_1}{l_B^2N_\phi},\frac{L_1k_2}{l_B^2N_\phi}\right)$; in Eq.(\ref{t3bdy}) we have $V_{k_1k_2k_3k_4}=V_{q_aq_b}, \vec q_a=\left(\frac{L_2\sin\theta}{l_B^2N_\phi}\left(\frac{k_3}{\sqrt{2}}+\frac{k_4}{2}\right),\frac{L_1}{l_B^2N_\phi}\left(\frac{k_3}{\sqrt{2}}+\frac{k_4}{2}\right)\right),\vec q_b=\sqrt{\frac{3}{2}}\left(\frac{L_2\sin\theta k_4}{l_B^2N_\phi},\frac{L_1k_2}{l_B^2N_\phi}\right)$.

\subsection{Spherical Geometry}

Another geometry with no boundary is the spherical geometry, with a magnetic monopole sitting at the center of the sphere so that a total of $2S$ fluxes radiate through the surface\cite{haldane3}. By Dirac's quantization condition, $2S$ has to be an integer, and the single particle orbitals are spinors of total spin $S+N$, where $N$ is the Landau level index. Thus at $N^{\text{th}}$ LL the total number of states/orbitals is $2S+2N+1$.
 \begin{figure}[h!]
 \includegraphics[width=1\linewidth]{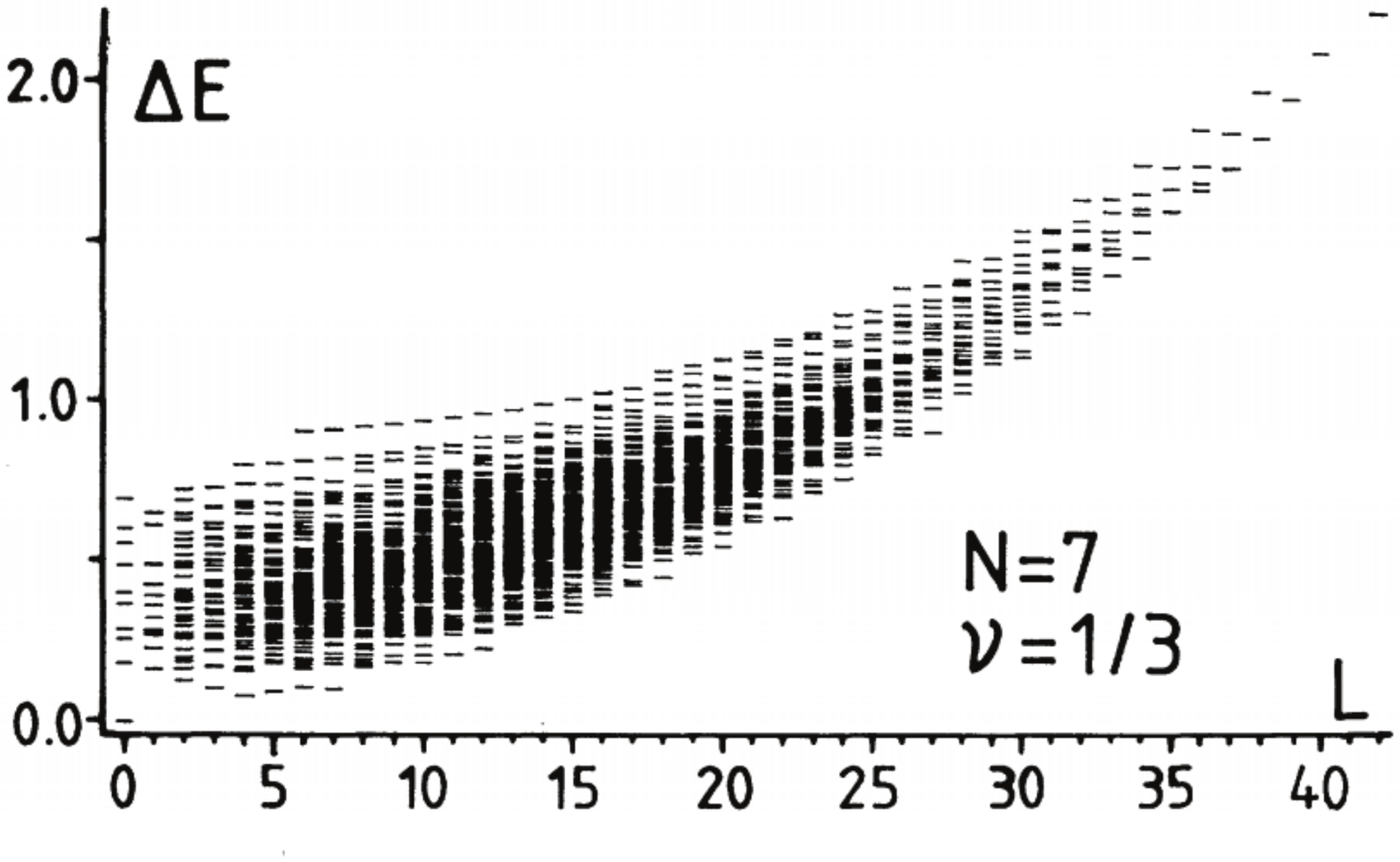}
\caption{The spectrum of 1656 multiplets (50388 states) of the $N_e=7$ electrons, $2S=18$ flux quanta system with Coulomb interactions, grouped by total angular momentum $L$. Energies (in units of $e^2/4\pi\epsilon l_B$) are shown relative to the incompressible ($\nu=\frac{1}{3}$) isotropic ($L=0$) ground state. Note the neutral excitations at the bottom of the spectrum, where the roton minimum occurs at around $L=4$.\cite{haldanerezayi}}
\label{fig_sphere}
\end{figure}
One can explicitly pick a gauge $\vec A=\left(S/eR\right)\times \vec\phi\cot\theta$, where $R$ is the radius of the sphere, $\phi$ is the azimuth angle and $\theta$ is the polar angle. In the LLL the $2S+1$ states have the wavefunctions expressed in terms of spinor coordinates $u=\cos\left(\frac{1}{2}\theta\right) e^{\frac{i}{2}\phi},v=\sin\left(\frac{1}{2}\theta\right) e^{-\frac{i}{2}\phi}$:
\begin{eqnarray}\label{spherewf}
\psi_m=u^{S+m}v^{S-m}
\end{eqnarray}
where $m$ is the $L_z$ component of the spinor which ranges from $-S$ to $S$. This directly corresponds to single particle orbitals on the disk with the guiding center angular momenta ranging from $0$ to $2S$. We can explicitly use the stereographic mapping by taking $z=2Rv/u$, where $R$ is the radius of the sphere. The single particle wavefunction of Eq.(\ref{spherewf}) can be written as $\psi_m=z^m/\left(1+zz^*/4R^2\right)^{1+N_\phi/2}$. Thus the Hamiltonian on the disk from Eq.(\ref{r2ndqham}) can be directly transcribed onto the sphere, where $\xi^\dagger_{M,m}$ is the creation operator of a pair of particles on the sphere with relative $L_z=m$ and total $L_z=M$. For particles in the $N^{\text{th}}$ LL, the states are given by spinors $|S+N,m\rangle$. Let the total spin of two particles be $L$ and the total azimuthal spin be $M$, the two-particle state is given by $|L,M\rangle$. The change of basis is given by
\begin{eqnarray}\label{cg}
|L,M\rangle=\sum_{m_1,m_2}C^{S+N,L,M}_{m_1,m_2}|S+N,S+N,m_1,m_2\rangle
\end{eqnarray}
where $C^{S+N,L,M}_{m_1,m_2}$ are the familiar Clebsch-Gordan coefficients. On the sphere, the $l^{\text{th}}$ pseudopotential is equivalent to the projection into a two-particle state with relative total angular momentum $2(S+N-l)$. Thus the form of Eq.(\ref{r2ndqham}) on the sphere is given by
\begin{eqnarray}\label{s2ndqham}
\mathcal H_{\text{2bdy}}=\sum_{l,m}c_l\xi^\dagger_{S+N-l,m}\xi_{S+N-l,m}
\end{eqnarray}
and Eq.(\ref{2ndexpansion}) on the sphere is given by
\begin{eqnarray}\label{s2ndexpansion}
\xi^\dagger_{l,m}=\sum_{m_1,m_2}C^{S+N,l,m}_{m_1,m_2}\xi^\dagger_{m_1}\xi^\dagger_{m_2}
\end{eqnarray}
The Hamiltonian matrix thus can be built up numerically in exactly the same way as the case for the disk geometry.

\section{Example A: Entanglement Spectrum of Spherical FQHE Ground State}

A major reason why we need to use numerics to solve FQH problems is because the many-body wavefunctions of the FQHE is intrinsically not simple. One way to quantify the complexity of a many-body state is to look at its entanglement. To do that, the Hilbert space of a many-body system $\bf{\mathcal H}$ is partitioned into two sub-Hilbert spaces
\begin{eqnarray}\label{cut}
\mathcal H=\mathcal H_A\otimes \mathcal H_B
\end{eqnarray}
The partition can be done in real space, momentum space, particle space, or in any other more abstract ways, and a wavefunction $|\Psi\rangle\in\mathcal H$ can be expanded as
\begin{eqnarray}\label{abexpansion}
|\Psi\rangle=\sum_{mn}c_{mn}|\psi^A_m\rangle\otimes|\psi^B_n\rangle
\end{eqnarray}
where we have $|\psi^{A (B)}_m\rangle$ spanning $\mathcal H_{A (B)}$. Since $|\Psi\rangle$ is a pure state, the density matrix is one-dimensional $\rho=|\Psi\rangle\langle\Psi|$. One can then define a reduced density matrix for the subsystem A
\begin{eqnarray}\label{reducem}
\rho_A=Tr_B\rho=\sum_k\left( I\otimes\langle\psi_k^B|\right)\rho\left(I\otimes|\psi_k^B\rangle\right)
\end{eqnarray}
This is an $N_A\times N_A$ square matrix, where $N_A$ is the dimension of $\mathcal H_A$. Similarly we can also have $\rho_B=Tr_A\rho$, an $N_B\times N_B$ square matrix with $N_B$ the dimension of $\mathcal H_B$. For a local operator $\hat O_{A (B)}$ that acts entirely within $\mathcal H^{A (B)}$, its expectation value is given by
\begin{eqnarray}\label{localexpectation}
\langle\hat O_{A (B)}\rangle=\langle\Psi|\hat O_{A (B)}|\Psi\rangle=Tr_{A (B)}\left(\rho_{A (B)}\hat O_{A (B)}\right)
\end{eqnarray}
While the information about the part which is traced out is completely lost, the reduced density matrix retains all the information about the un-traced part. If the two parts of the Hilbert space are \emph {entangled}, a local experimental measurement on a pure state is equivalent to a measurement of a mixed state represented by the reduced density matrix. This quantum-statistical correspondence is the hallmark of non-locality.

Quantitatively we can define the entanglement entropy, or the Von-Neumann entropy as
\begin{eqnarray}\label{vnentropy}
\mathcal S=-Tr\rho_A\ln\rho_A=-Tr\rho_B\ln\rho_B
\end{eqnarray}
The von Neumann entropy is a unique measure of the bipartite entanglement in the following senses\cite{amico}. 1). $\mathcal S$ is invariant under local unitary operations. 2). $\mathcal S$ is a continuous function of the state in the Hilbert space\cite{virmani}. 3). $\mathcal S$ is additive: \(\mathcal S(|\psi_1\rangle\otimes|\psi_2\rangle) = \mathcal S(|\psi_1\rangle)+\mathcal S(|\psi_2\rangle)\). If A and B are not entangled, the reduced density matrix is calculated from a product state, and $\mathcal S=0$. Eq.(\ref{vnentropy}) is well-defined because the non-zero part of the spectrums of $\rho_A$ and $\rho_B$ are identical. This can be shown with a singular value decomposition (SVD) of the $N_A$ by $N_B$ matrix $(C)_{mn}=c_{mn}$ from coefficients in Eq.(\ref{abexpansion}):
\begin{eqnarray}\label{svd}
C=U\Lambda V^*
\end{eqnarray}
where $U,V$ are unitary square matrices of dimensions $N_A$ and $N_B$ respectively, and $\Lambda$ is an $N_A\times N_B$ diagonal matrix with non-negative real numbers $\lambda_i,i=1,\cdots \min\{N_A,N_B\}$ on the diagonal. Writing $|\psi^A_i\rangle=\sum_m U_{mi}|\psi^A_m\rangle, |\psi^B_i\rangle=\sum_n V^*_{in}|\psi^B_n\rangle$, Eq.(\ref{vnentropy}) can be converted into the diagonal form:
\begin{eqnarray}\label{svdstate}
|\Psi\rangle=\sum_i\lambda_i|\psi^A_i\rangle|\psi^B_i\rangle
\end{eqnarray}
which leads to $\rho_{A (B)}=\sum_i\lambda_i^2|\psi^{A (B)}_i\rangle\langle\psi^{A (B)}_i|$ and $\mathcal S=2\sum_i\lambda_i^2\ln\lambda_i$. Normalization of the state requires $\sum_i\lambda_i^2=1$.

The entanglement entropy measures the minimal amount of the information needed to fully characterize the state. While it is a very important quantity that can contain topological signatures\cite{entropy}, Li and Haldane\cite{lh} pointed out that the entire spectrum of $\rho_{A(B)}$, or the entanglement spectrum, contains additional important information as well. One can rewrite Eq.(\ref{svdstate}) as $|\Psi\rangle=\sum_ie^{-\frac{\epsilon_i}{2}}|\psi_i^A\rangle|\psi_i^B\rangle$, where $\epsilon_i=-2\ln\lambda_i$ is the so-called entanglement energy. In this way, the reduced density matrix resembles the partition function of a quantum system at a ``pseudo-temperature" equal to unity. If any state is missing in the entanglement spectrum, its entanglement energy goes to infinity.

To explore the real space entanglement of a quantum state, a partition of the Hilbert space in real space is usually performed. For the FQHE a real space cut involves both the cyclotron and guiding center coordinates\cite{rezayi}, which is technically more demanding, and not desirable if one wants to explore the entanglement involving only the guiding center degrees of freedom. The alternative is to perform an orbital cut. With spherical geometry and a monopole of strength $2S$ sitting at the center, there are in total $2S+1$ orbitals in the LLL. The Hilbert space for each orbital is two-dimensional (occupied or un-occupied). One can thus separate these orbitals into two groups, the Hilbert space of each is the direct product of the single particle Hilbert space for all orbitals in that group. 

The entanglement entropy calculated from the reduced density matrix of the ground state depends on how strongly correlated these two subgroups are. In \cite{lh}, the cut is made near the equator of the sphere, so the dimension of the reduced density matrix is maximized. Two most important observations can be summarized as follows:

\begin{itemize}
\item{The virtual cut that separates the Hilbert space resembles a physical cut that separates the sphere into two hemispheres with real edges. The low-lying part of the entanglement spectrum has the same counting as the physical edge states of the FQHE, as predicted by the conformal field theory}
\item{The entanglement spectrum of the model wavefunction (e.g. the Laughlin wavefunction or Moore-Read Pfaffian) has significantly fewer states than the dimension of the sub-Hilbert space, i.e. many basis elements in the sub-Hilbert space do not participate in the ground state. When the model Hamiltonian is adiabatically tuned towards the Coulomb interaction, the missing states appear with a small but finite weight. As long as the FQH phase persists, these non-universal basis elements remain \emph{partially} gapped from the universal ones in the entanglement spectrum}
\end{itemize}

\begin{figure*}[h!]
  \begin{minipage}[l]{0.8\linewidth}
\hspace{25pt}
    \psfrag{B}{$\xi$}
    \psfrag{A}{$L_{z}^{A}$}
    \includegraphics[width=\linewidth]{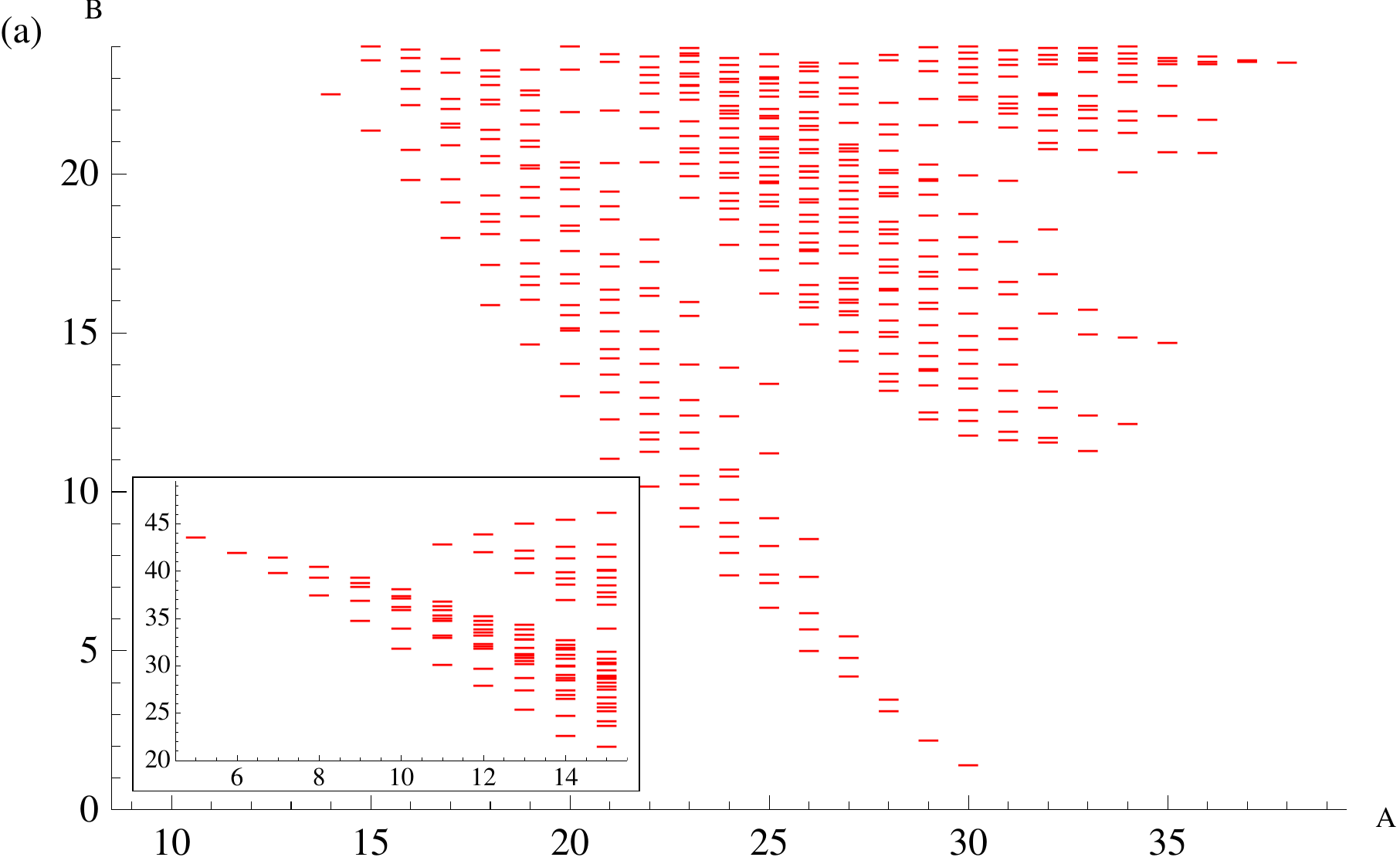}
  \end{minipage}
\\
\hspace{30pt}
  \begin{minipage}[l]{0.8\linewidth}
\hspace{25pt}
    \psfrag{B}{$\xi$}
    \psfrag{A}{$L_{z}^{A}$}
    \includegraphics[width=\linewidth]{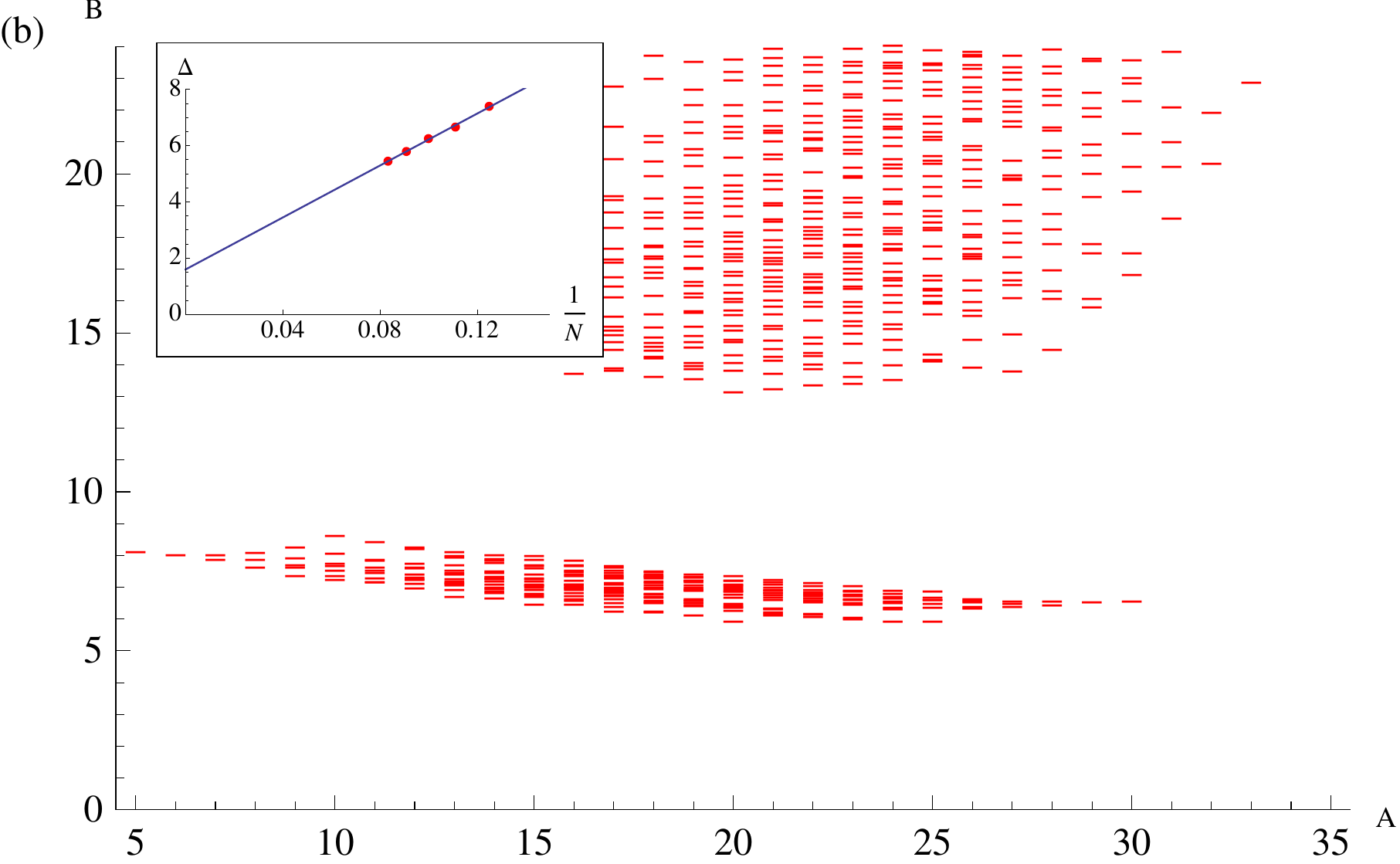}
  \end{minipage}
  \caption{Entanglement spectrum for the $N=11$ bosons, $N_\phi=20$,
    $\nu=1/2$ Coulomb state on the sphere. The cut is such that
    $l_A=10$ orbitals and $N_A=5$ bosons.  (a) Standard normalization
    on the quantum hall sphere. The inset show the remainder part of
    the spectrum where the entanglement levels exceed $\xi=24$. (b)
    Conformal limit (CL) normalization. The CL
    separates a set of universal low-lying energy states,
    which allows an unambiguous definition of the entanglement gap
    over all $L^A_z$ subsectors as
    the minimal difference between the highest energy CFT state and
    lowest generic state. The inset in (b)
    shows the the finite size scaling of the entanglement gap for the
    Coulomb state, which remains finite in the thermodynamic limit.\cite{thomale}}
\label{coubos-ba}
\vspace{0pt}
\end{figure*}

Apart from being used as a diagnostic tool for the topological phases of many-body ground states, the entanglement spectrum also has practical applications for numerical techniques like the density matrix renormalization group (DMRG)\cite{white}. It is clear that the eigenstates in the entanglement spectrum contribute differently to the ground state; the presence of the entanglement gap further indicates that most of the probabilistic weight is carried by the basis elements below the gap (remember the entanglement energy is the negative logarithmic function of the probability amplitude). This is particularly appealing for DMRG, which uses the entanglement spectrum to judiciously truncate away the unimportant part of the Hilbert space. While DMRG does not generally perform well for two-dimensional systems due to the exponential growth of the entanglement entropy with the system size, it can be more useful for systems with gapped topological phases.
\begin{figure}[h!]
\centerline{\includegraphics[width=4.5in]{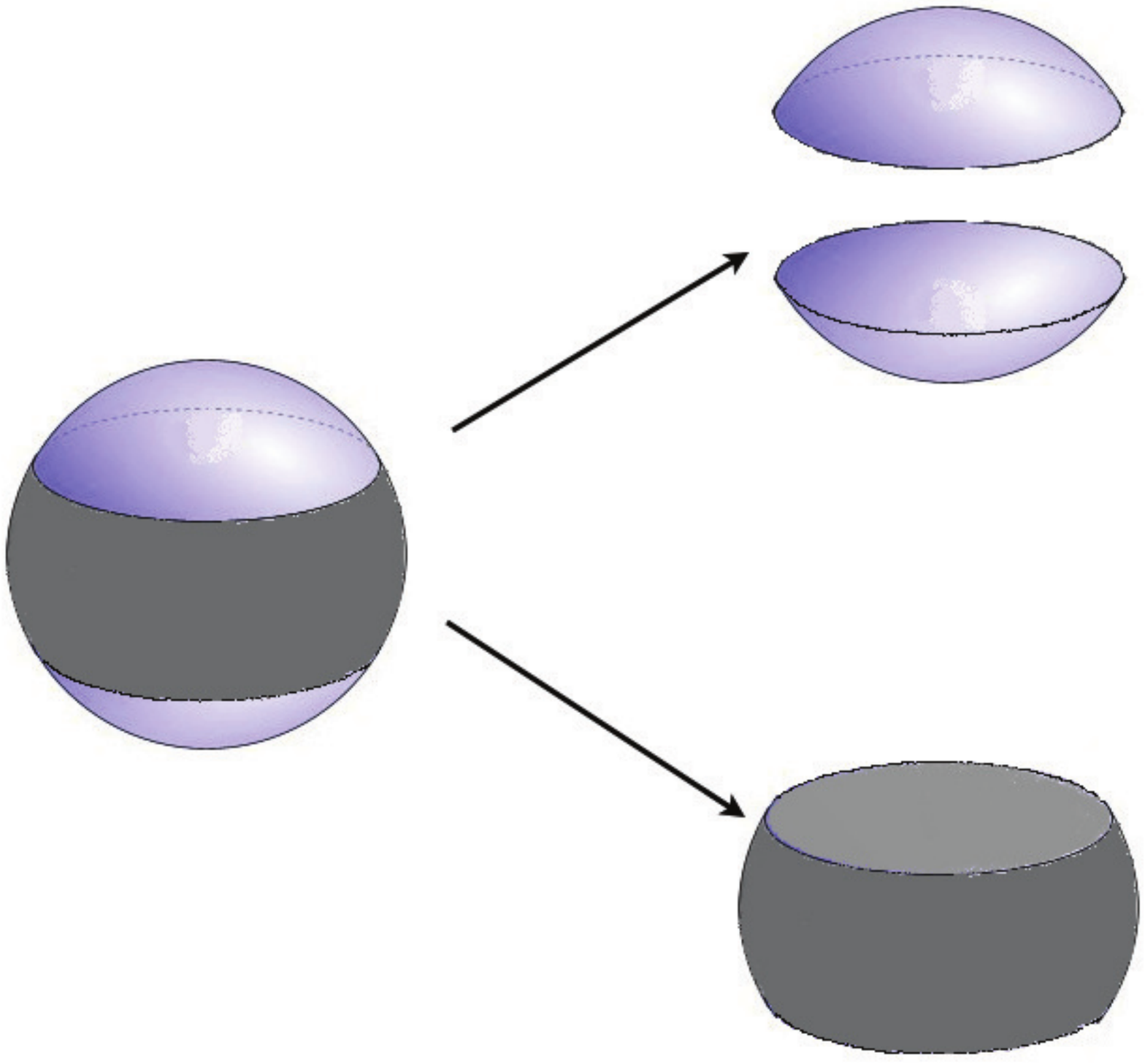}}
\caption{The Hilbert space of the sphere is partitioned into two parts: the top part consists of two caps of the north and the south pole; the bottom part is the rest of the sphere around the equator.}
\label{partition}
\end{figure}
We will now illustrate the idea of the entanglement spectrum for the FQHE ground state with a different partition of the Hilbert space (See Fig.(\ref{partition})), as compared to the work in Fig.(\ref{coubos-ba}). In their work a single cut at the equator is employed, and there does not exist a unique entanglement energy that separates the topological part of the spectrum from the non-universal part. From the DMRG point of view, this makes selection of ``good basis" difficult. Here instead of just one cut, two cuts are performed on the sphere that are parallel and symmetric about the equator. The two resulting subsystems (one including two ``caps" around the north and south pole, the other including the bulk around the equator) have equal number of orbitals. In general, more cuts imply greater entanglement entropy between the two subsystems, which tends to disfavor such partitions; on the other hand, almost all the topological part of the spectrum are below the non-universal part, making the truncation of the unimportant Hilbert space less ambiguous.
\begin{figure}[h!]
\centerline{\includegraphics[width=4.5in]{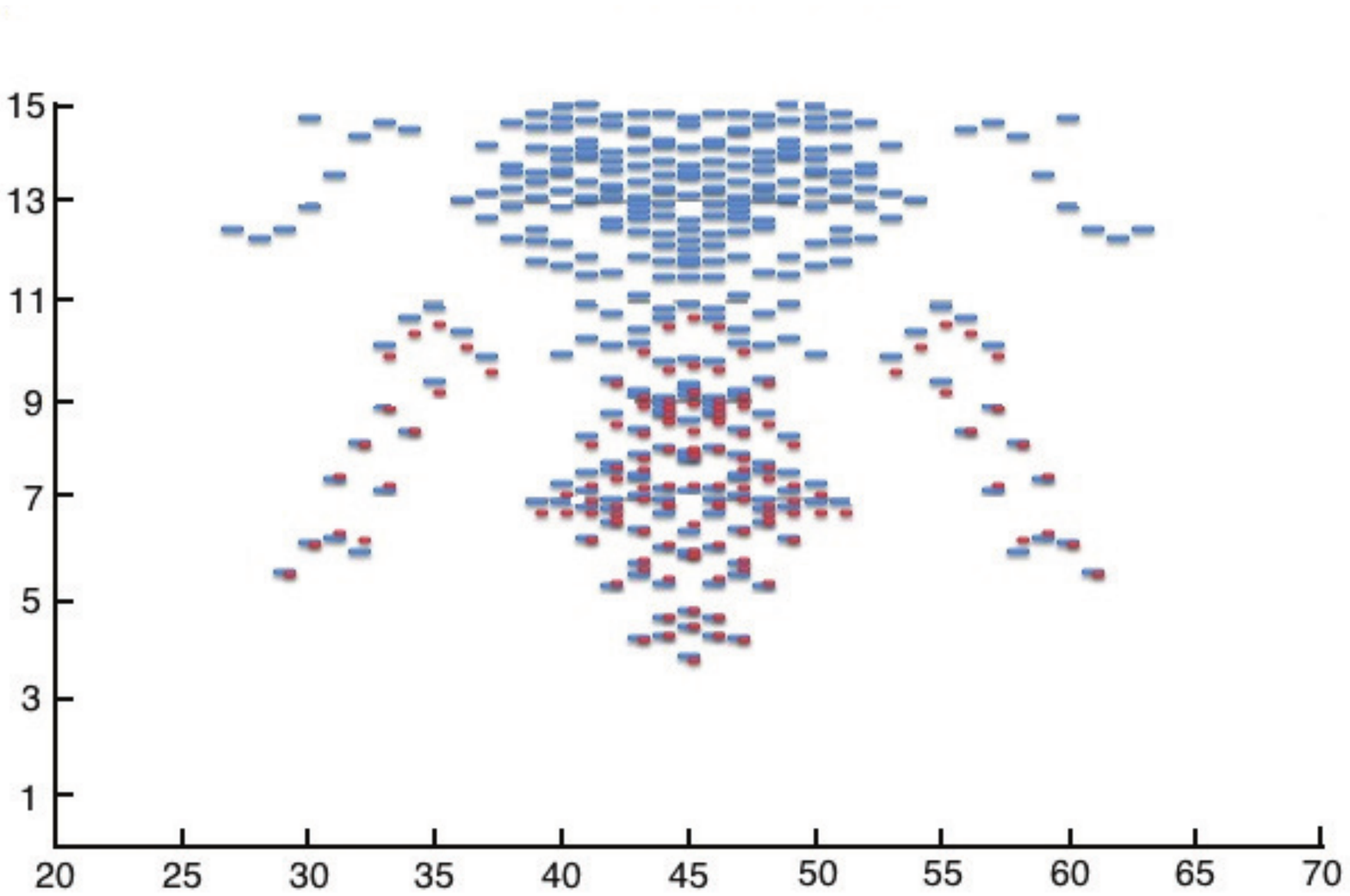}}
\caption{Entanglement spectrum of the FQHE ground state obtained from exact diagonalization with ten particles at filling factor $\nu=\frac{1}{3}$. The blue plots are entanglement spectrum from the Coulomb interaction. The red plots come from the model wavefunction of the same system. As we can see the entanglement spectrum of the model wavefunction is mostly below the non-universal part of the entanglement spectrum obtained from the Coulomb ground state (Compare Fig.(\ref{coubos-ba})).}
\label{twocuts1}
\end{figure}

From Fig.(\ref{twocuts1}) the counting of the low-lying states come from the two branches of the edge excitations. We can form the projection operator from the reduced density matrix of the model wavefunction, and it faithfully projects out the non-universal part of the entanglement spectrum calculated from the coulomb interaction, as shown in Fig.(\ref{twocuts2}):
\begin{figure}[h!]
\centerline{\includegraphics[width=4.5in]{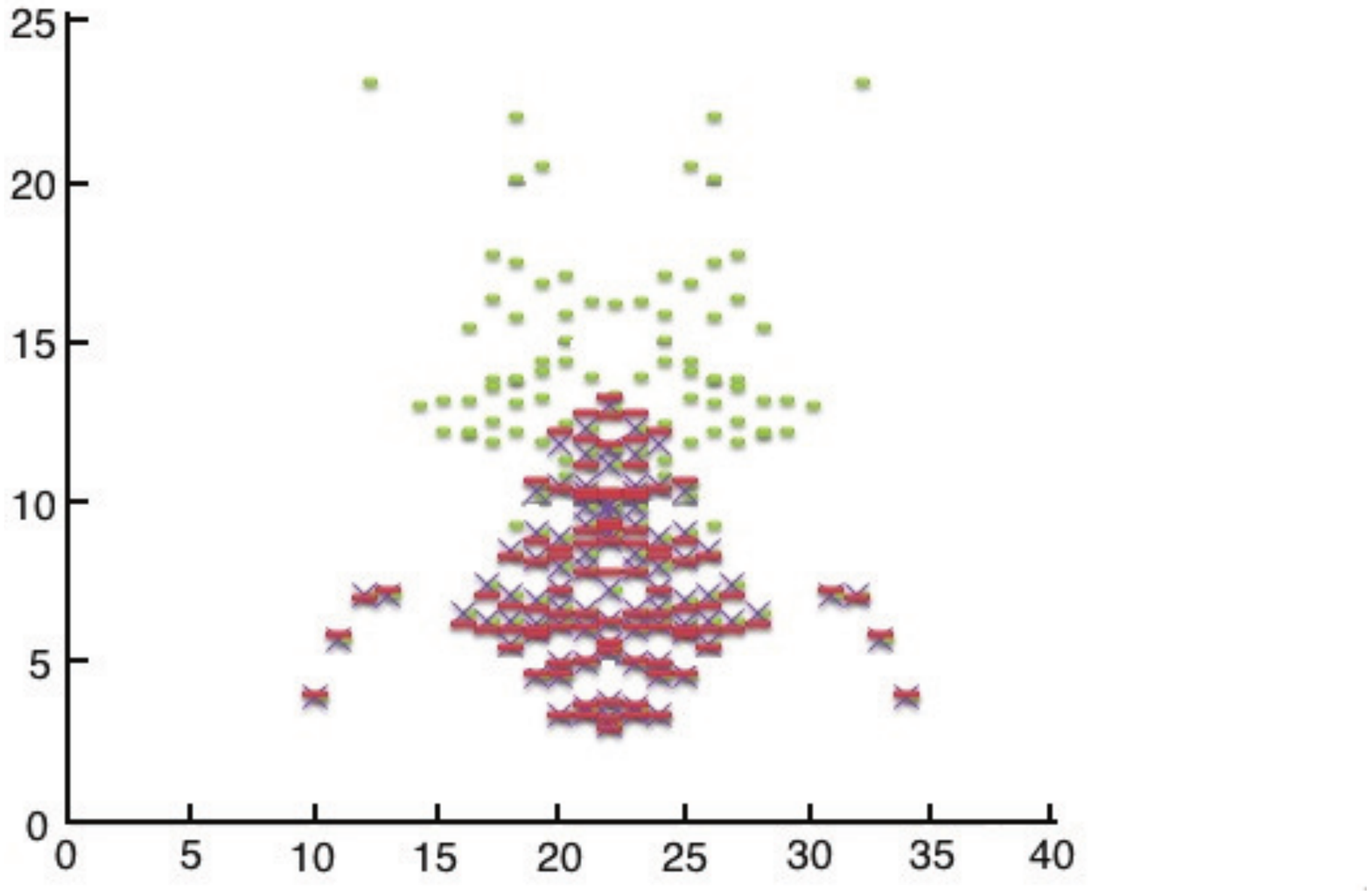}}
\caption{The entanglement eigenstates from the model wavefunction projects out the non-universal part of the entanglement spectrum from the Coulomb interaction. The total system size is 8 electrons at filling factor $\nu=\frac{1}{3}$. The yellow plots are from the entanglement spectrum of the Coulomb ground state; the red plots are from the entanglement spectrum of the Laughlin model wavefunction. The basis elements of the red plots are used to form the projection operator, and the purple plots are the eigenstates of the reduced density matrix of the Coulomb ground state after the application of the projection operator.}
\label{twocuts2}
\end{figure}

The very small entanglement energy overlap between the non-universal states and the ``topological" states is advantageous for DMRG: instead of generating projection operators from diagonalizing intermediate Hamiltonians, one can obtain them directly from Jack polynomials (see Chapter 4). This presents a new approach for DMRG on the FQH states. One should note however be aware that the increase of the entanglement due to the presence of two edges of the partition leads to a larger number of low lying states. It is thus not yet clear which factor outweighs the other.
\begin{figure}[h!]
\centerline{\includegraphics[width=4.5in]{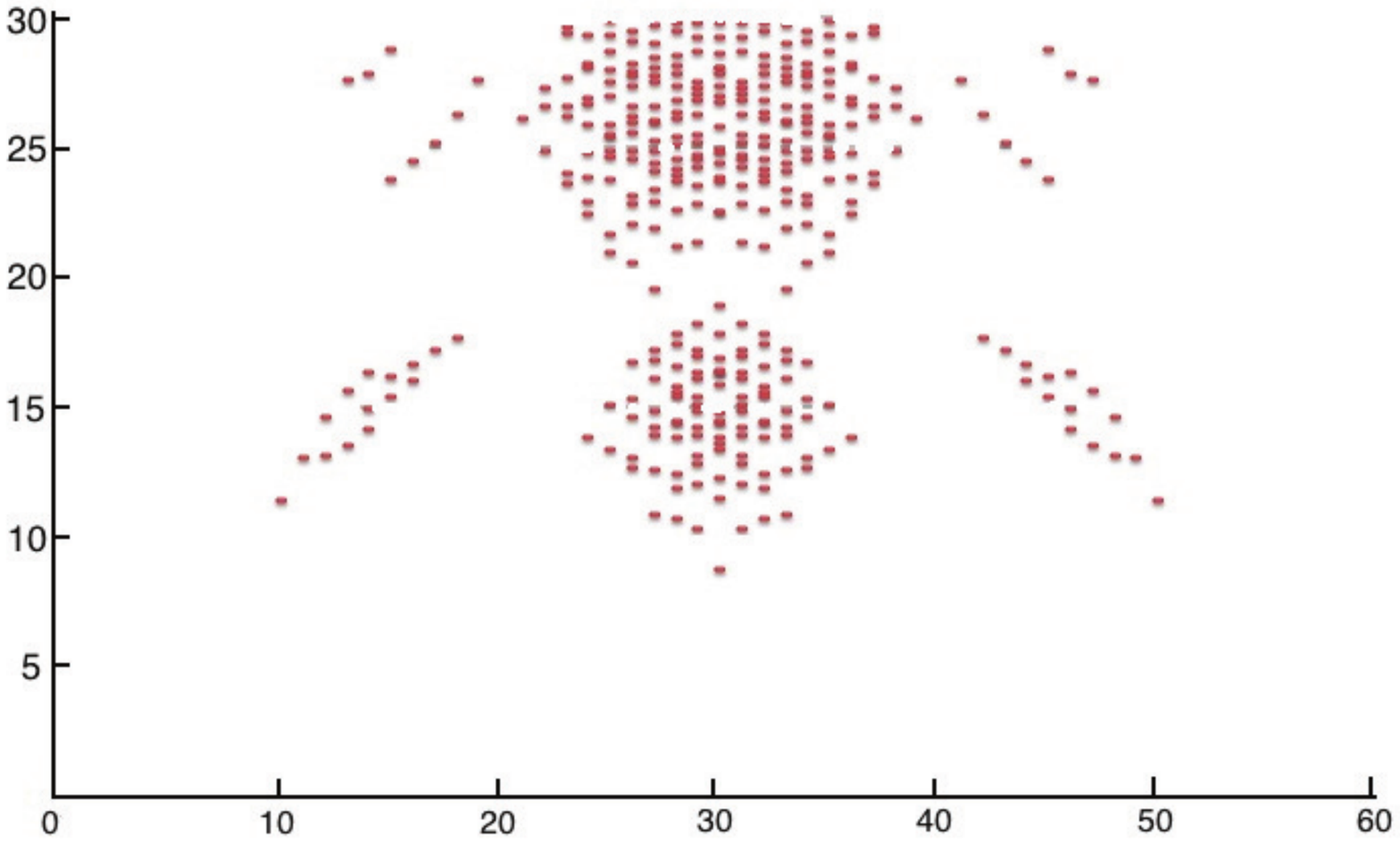}}
\caption{The entanglement spectrum calculated in the conformal limit, with the same system size as the one in Fig.(\ref{twocuts1})}
\end{figure}
Before closing the section let us look at the entanglement spectrum of the partition with two edges in the conformal limit\cite{thomale}(also defined in the caption of Fig.(\ref{coubos-ba})), and compare it with Fig.(\ref{coubos-ba}). The spectrum before and after stripping away the single particle normalization looks pretty much the same, suggesting the new partition is less prone to the finite size effect that tends to introduce environmental errors in the DMRG procedure.

\section{Example B: Guiding Center Metric In FQHE}

It was first pointed out by Haldane that the rotational invariance is not required to protect the topological phases of the FQHE. From previous chapters we know that the rotational invariance only exists if the cyclotron metric and the guiding center metric are congruent: $\widetilde g_{ab}=\bar g_{ab}$. This is not necessarily the case in many physical situations. For Galilean invariant systems, the cyclotron metric is defined by the effective mass tensor. Microscopically the effective mass tensor depends on the band structure of the underlying lattice model, which is anisotropic in materials like ALAs many-valley semiconductors,
or Si in the presence of uniaxial stress\cite{bandanisotropy}. For a Hall surface with finite thickness, we can also tune the effective mass tensor by tilting the magnetic field\cite{yb3,das}.

On the other hand, the interaction part of the Hamiltonian generally contains an independent metric. For Coulomb interaction, this metric is defined by the dielectric tensor, which has the shape of equipotential lines around a charged particle. Explicitly, the Fourier component of the effective interaction is given by
\begin{eqnarray}\label{effectivev}
V_q=\frac{1}{|q|_c}F_N(|q|_m^2)^2
\end{eqnarray}
where $|q|_c=\sqrt{g_c^{ab}q_aq_b}$ and $|q|_m=\sqrt{g_m^{ab}q_aq_b}$ with $g_c^{ab}$ from the dielectric tensor and $g_m^{ab}$ from the effective mass tensor. Once again if $g_m^{ab}=g_c^{ab}$, we have $\bar g^{ab}=\widetilde g^{ab}$, and rotational invariance is preserved. Notice here the definition of rotational invariance allows anisotropy where $\bar g^{ab}=\widetilde g^{ab}\neq \delta^{ab}$. Physically, only the relative difference between different metrics matters. Thus without loss of generality, the dielectric tensor is always taken to be isotropic; the rotational invariance is broken when the effective mass tensor is anisotropic, which is given in the following form:
\begin{eqnarray}\label{gm}
g_m^{ab}\sim\left(\begin{array}{ccc}
\cosh2\theta+\sinh2\theta\cos2\phi &\sinh2\theta\sin2\phi\\
\sinh2\theta\sin2\phi & \cosh2\theta-\sinh2\theta\cos2\phi\end{array}\right)
\end{eqnarray}
A unimodular metric only has two free parameters, where $\theta$ parametrizes squeezing of the metric, and $\phi$ parametrizes the rotation. The anisotropy parameter is given by $\alpha=\cosh 2\theta+\sinh 2\theta$. The point of isotropy is given by $\alpha=1$.

\begin{figure}[ttt]
 \includegraphics[width=0.7\linewidth,angle=270]{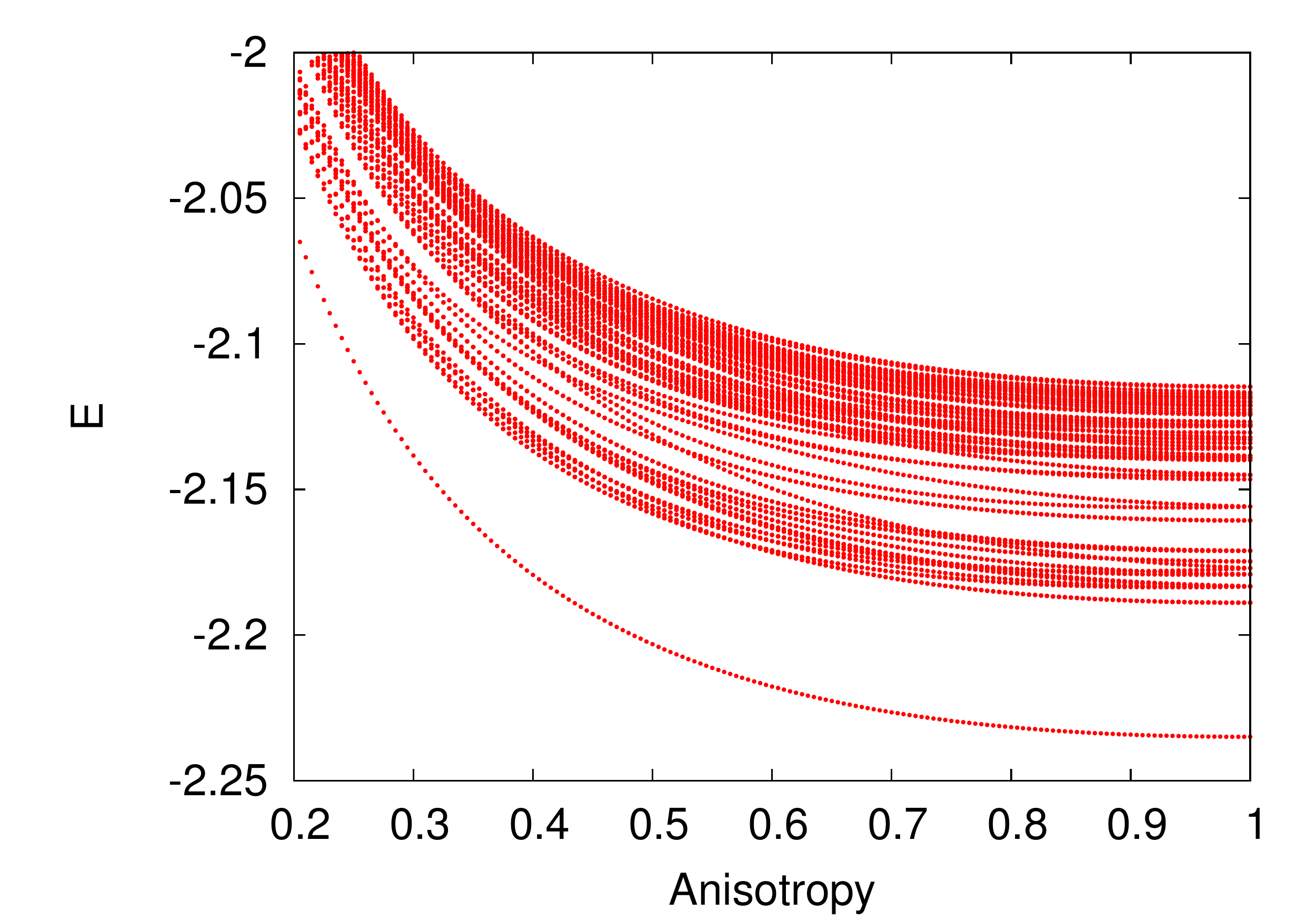}
\caption{Energy spectrum in units of $e^2/\epsilon l_B$ as a function of anisotropy $\alpha$ for the square unit cell and $n=0$ LL Coulomb interaction at $\nu=1/3$. The system is $N_e=7$ electrons and $\phi=0$. Due to the square unit cell, the spectrum is symmetric under $\alpha \rightarrow 1/\alpha$.}
\label{fig_onethirdenergy}
\end{figure}

In the LLL, exact diagonalization shows that the incompressibility is quite robust against anisotropy of $g_m^{ab}$. In Fig.~\ref{fig_onethirdenergy}, the energy spectrum of the Coulomb interaction at $\nu=1/3$ as a function of anisotropy is plotted, and the ground state is always gapped. Level-crossing only occurs among the excited states. On the other hand the isotropic Laughlin wavefunction no longer gives good overlap with the ground state.

Since the FQHE depends on the Landau level form factor, the distortion of the effective mass tensor induces a change of the guiding center metric, a purely ground state property which can be viewed as a hidden variational parameter of the FQHE state. For model wavefunctions, one can define a family of generalized Laughlin wavefunctions parametrized by the guiding center metric $\bar g_{ab}=\bar\omega_a^*\bar\omega_b+\bar\omega_a\bar\omega_b^*$, with the guiding center ladder operators given by $b_i=\bar\omega_aR_i^a,b^
\dagger_i=\bar\omega_a^*R_i^a$. In the plane geometry the generalized Laughlin state is given by 
\begin{equation}\label{glaughlinwf}
|\Psi_{\bar g}^{\nu=1/m} \rangle= \prod_{i<j}(b^\dagger_i-b^\dagger_j)^m|0\rangle.
\end{equation}
where the vacuum is defined as $b_i|0\rangle=0$. The ladder operators are now explicitly metric dependent, and $b_i(g),b^\dagger_i(g)$ with different metrics are related to each other by a Bogoliubov transformation. Equivalently, the wavefunction can be expressed by a unitary transformation $\Psi_L(g)=\exp(-i\xi_{\alpha\beta}\Lambda^{\alpha\beta})\Psi_L^0$, where $\xi_{\alpha\beta}$ is a real symmetric tensor and $\Lambda^{\alpha\beta}=\frac{1}{2}\sum_i \{ R_i^a, R_i^b\}$ is the generator of area-preserving diffeomorphism, and $\Psi_L^0$ is isotropic. The model Hamiltonian of Eq.(\ref{glaughlinwf}) is given by
\begin{eqnarray}\label{ghamiltonian}
\mathcal H(\bar g)=\int \frac{d^2ql_B^2}{(2\pi)^2}\sum_{n=1}^{m-2}\mathcal L_n(|q|^2)e^{-\frac{1}{2}|q|^2}\bar\rho_q\bar\rho_{-q}
\end{eqnarray}
where we have $|q|^2=\bar g^{ab}q_aq_b$. Thus for finite systems Eq.(\ref{glaughlinwf}) can be generated numerically by exact diagonalization. Similarly, the ground state $|\Psi^{\nu=1/m}_{g_m,g_c}\rangle$ without rotational invariance can be obtained numerically by diagonalizing the following Hamiltonian
\begin{eqnarray}\label{rhamiltonian}
\mathcal H=\int \frac{d^2ql_B^2}{(2\pi)^2}\frac{1}{|q|_c}F_N(|q|_m)^2\bar\rho_q\bar\rho_{-q}
\end{eqnarray}
\begin{figure}[ttt]
 \includegraphics[width=0.85\linewidth]{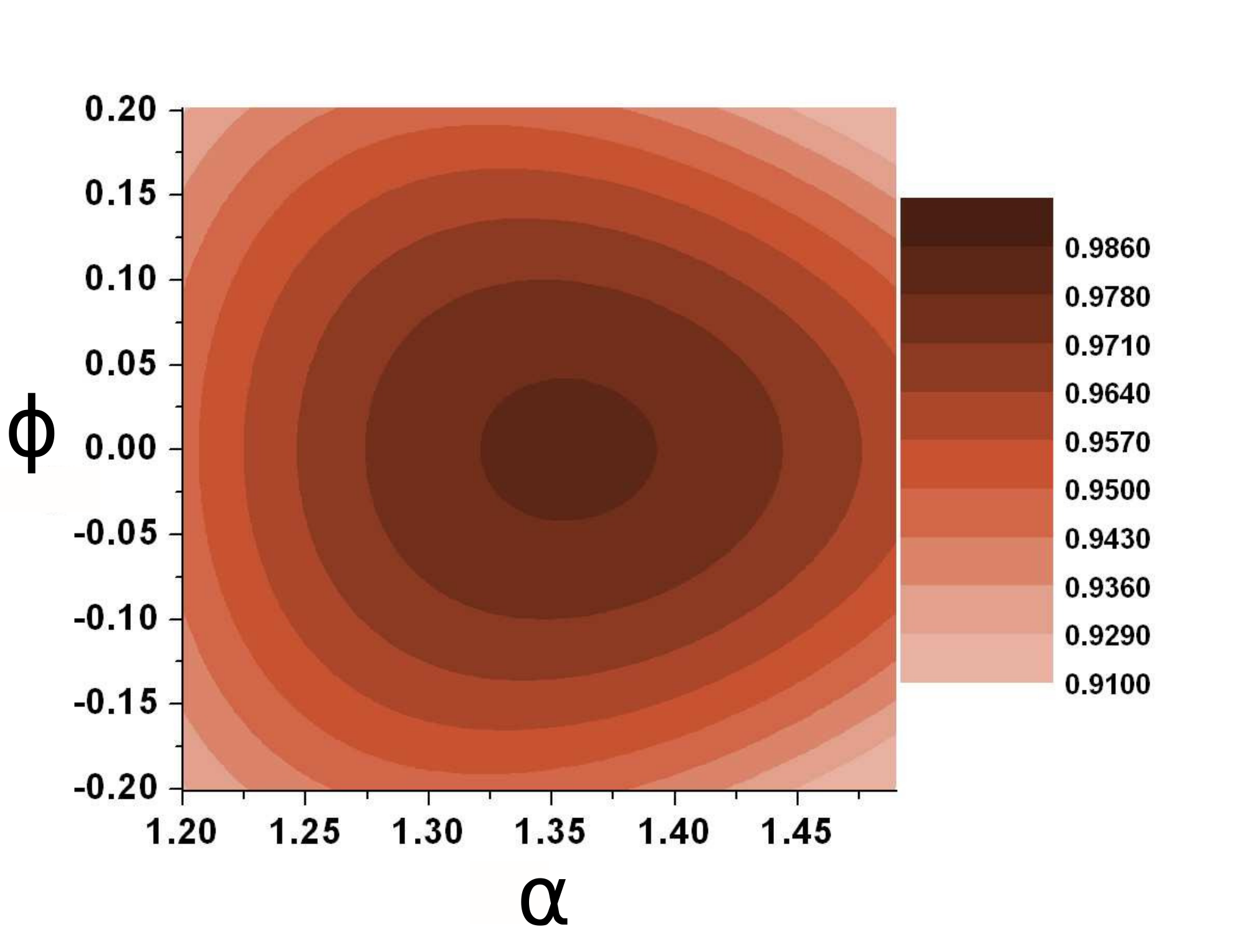}
\caption{Overlap between the Coulomb ground state at $\nu=1/3$ for fixed anisotropy $\alpha_0=2,\phi_0=0$ and the family of Laughlin states parametrized by varying $\alpha$, $\phi$. The system is $N_e=9$ electrons on a hexagonal torus.}
\label{fig_optimize}
\end{figure}
\begin{figure}[ttt]
 \includegraphics[width=0.8\linewidth,angle=0]{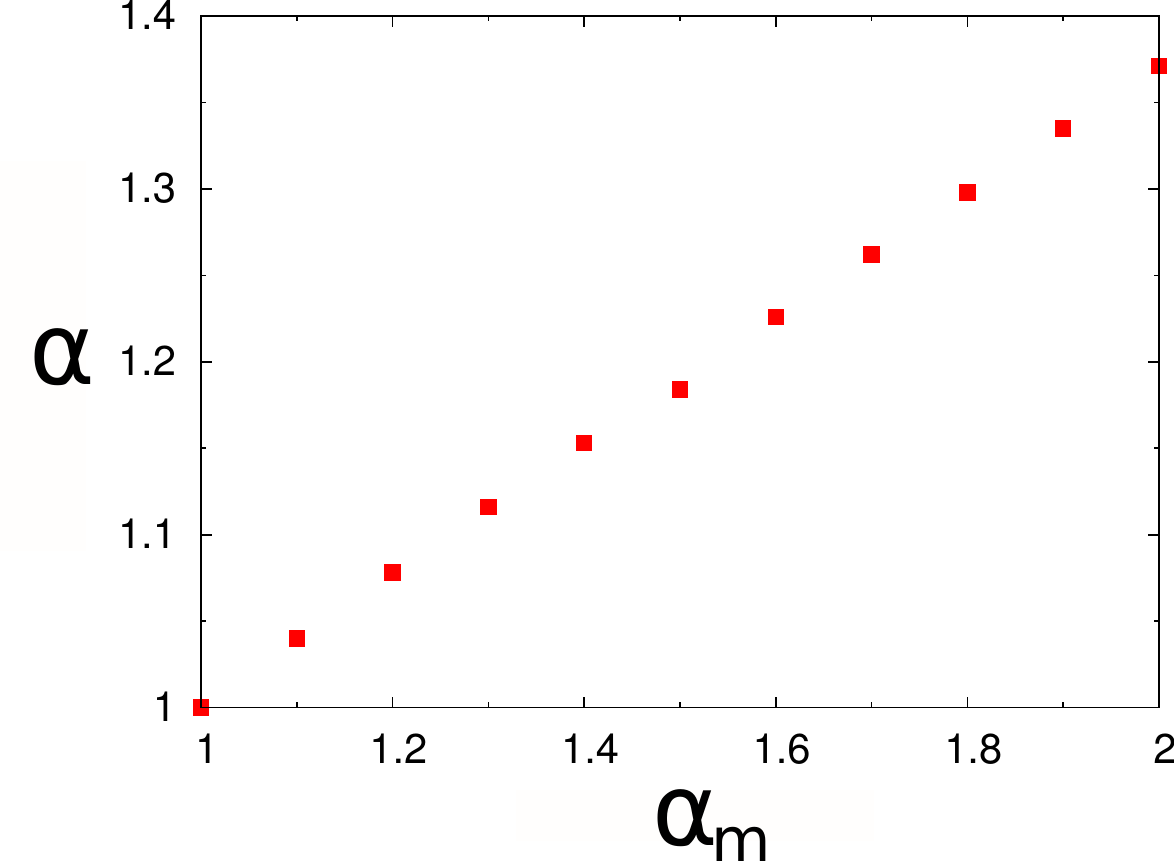}
\caption{Dependence of the intrinsic metric $\alpha$ on the mass metric $\alpha_m$ (Coulomb metric is set to $\delta^{ab}$).}
\label{fig_intrinsic}
\end{figure}
where $V_q$ is given by Eq.(\ref{effectivev}). One can thus define the guiding center metric of $|\Psi^{\nu=1/m}_{g_m,g_c}\rangle$ as the $\bar g^{ab}$ that maximizes the overlap $|\langle\Psi^{\nu=1/m}_{g_m,g_c}|\Psi_{\bar g}^{\nu=1/m} \rangle|^2$, treating $\bar g^{ab}$ as the variational parameter. In Fig.(\ref{fig_optimize}), the exact diagonalization is done on the torus with periodic boundary conditions. The ground state of the Coulomb interaction has fixed mass anisotropy $\alpha_0=2, \phi_0=0$ (the metric of the dielectric tensor is implicitly assumed to be $\alpha=1, \phi=0$), and we evaluate the overlap with a family of Laughlin states generated by varying $\alpha, \phi$. The overlap $|\langle \Psi_L^{\alpha,\phi}| \Psi_C^{\alpha_0=2,\phi_0=0}\rangle|$ is plotted as a function of $\alpha$ and $\phi$. The principal axis of the Laughlin state is aligned with that of the Coulomb state (maximum overlap occurs for $\phi=\phi_0=0$). Notably, the maximum overlap occurs for some value of the anisotropy that is a ``compromise" between the dielectric $\alpha=1$ and a cyclotron one $\alpha=2$. The value of the anisotropy that defines the intrinsic metric depends linearly on the band mass anisotropy (Fig.~\ref{fig_intrinsic}). This result illustrates the ability of the Laughlin state to optimize the shape of its fundamental droplets and maximize the overlap with a given anisotropic ground state of a finite system. 

\begin{figure}[ttt]
 \includegraphics[width=1\linewidth,angle=0]{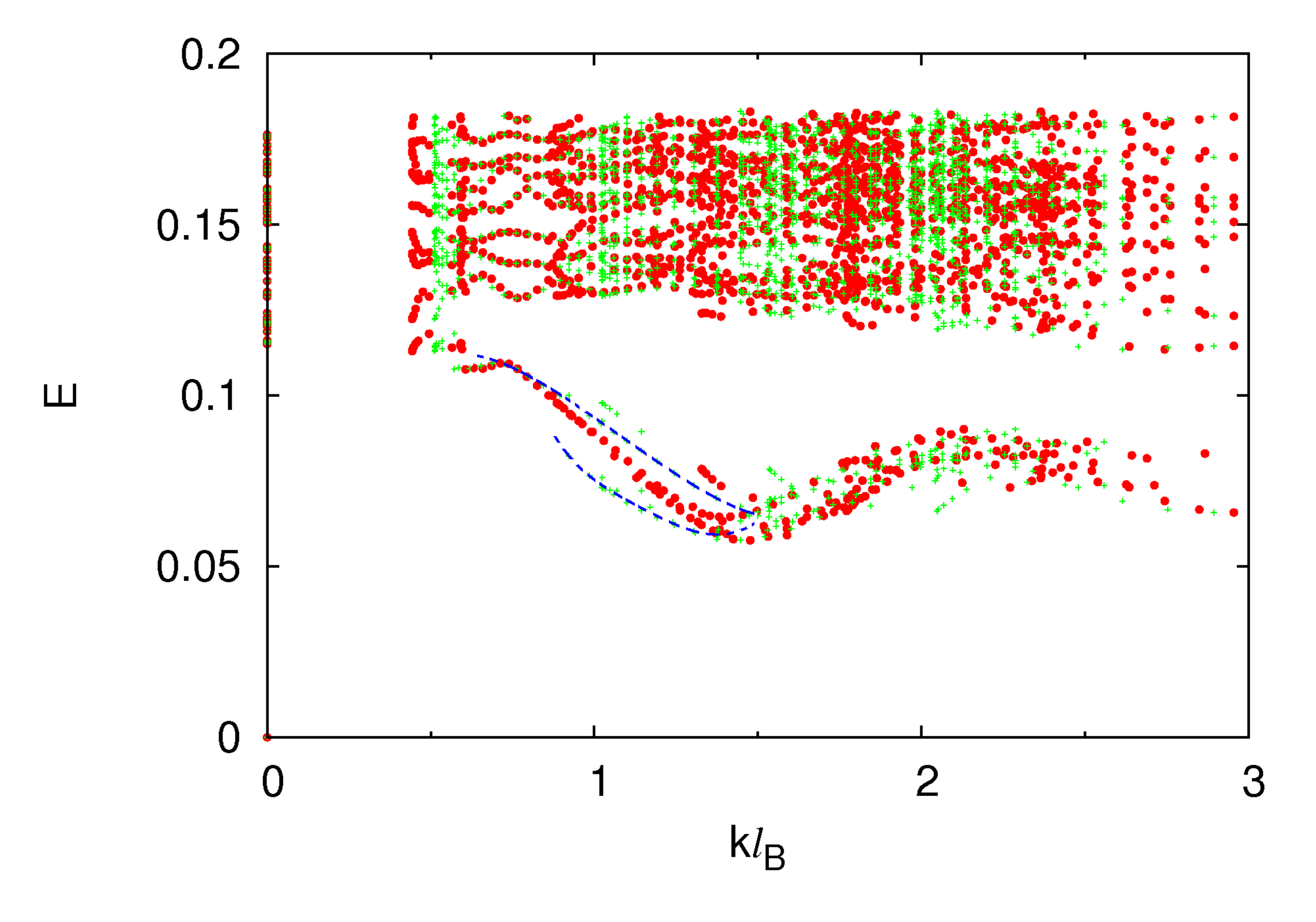}
\caption{Energy spectrum of $N_e=9$ electrons at $\nu=1/3$ with the effective mass anisotropy $\alpha_m=2$  along the $x$-axis. When plotted as a function of $\sqrt{k_x^2+k_y^2}$ (green crosses), two branches of the magneto-roton mode are present (blue dotted lines are guide to the eye). If the spectrum is plotted as a function of $\sqrt{g^{ab}k_a k_b}$, the two branches collapse onto the same curve.}
\label{fig_rotonaniso}
\end{figure}

The guiding center metric obtained by minimizing the wavefunction overlap is purely a ground state property. The structure factor of $|\Psi^{\nu=1/m}_{g_m,g_c}\rangle$ is thus also anisotropic with the same guiding center metric. One would ask if the guiding center metric could be used to characterize the Hamiltonian Eq.(\ref{rhamiltonian}), where excited states in the energy spectrum is involved. The elementary neutral excitations of the FQHE is given by the magneto-roton mode, so an alternative way to obtain the intrinsic metric is to analyze the shape of the two-dimensional momenta of the roton minimum. In a rotationally-invariant case, this mode has a minimum at $|k| \sim \ell_B^{-1}$. In the presence of anisotropy, the minima occur at different $|k|$ in the different directions (Fig.~\ref{fig_rotonaniso}). This leads to an alternative definition of the intrinsic metric based on the shape of the roton minimum in the 2D momentum plane.
In Fig.~\ref{fig_rotonaniso} the energy spectrum of an anisotropic Coulomb interaction at $\nu=1/3$ is plotted as a function of the rescaled momentum $\sqrt{g^{ab}k_a k_b}$, where $g$ is the guiding center metric that maximizes the overlap with the family of Laughlin wavefunctions (Fig.~\ref{fig_intrinsic}). With the usual definition of the momentum $|k|$, the magnitudes of the roton minima are now direction dependent. Different magneto-roton branches collapse onto the same curve if we plot them as a function of $\sqrt{g^{ab}k_ak_b}$. This is reasonable, because the magneto-roton mode is well approximated by the single mode approximation (SMA) up to the roton minimum~\cite{yb1}, and the SMA can be calculated entirely in terms of the properties of the ground state (See Eq.(\ref{smaenergy})). The anisotropy of the peak of the ground state structure factor dictates the position of the roton minima.

\section{Example C: Phase Transition in Second Landau Level}

In the LLL ($\mathcal N=0$, where $\mathcal N$ is the LL index) the incompressible phase is quite robust against anisotropy of the effective mass tensor. In higher LLs, due to a number of nodes in the single-particle wavefunction, the region of the phase diagram where incompressible states occur becomes increasingly narrower, and compressible phases such as stripes and bubbles take over. In this section we briefly present some results on the effects of anisotropy on FQH states in higher LLs, focusing on fillings $\nu=1/3$ and $1/2$. A more detailed analysis of the issue can be found in\cite{yb2}.

\subsection{Stripes and Bubbles in $\mathcal N\ge 2$}

\begin{figure}[ttt]
 \includegraphics[width=0.95\linewidth,angle=0]{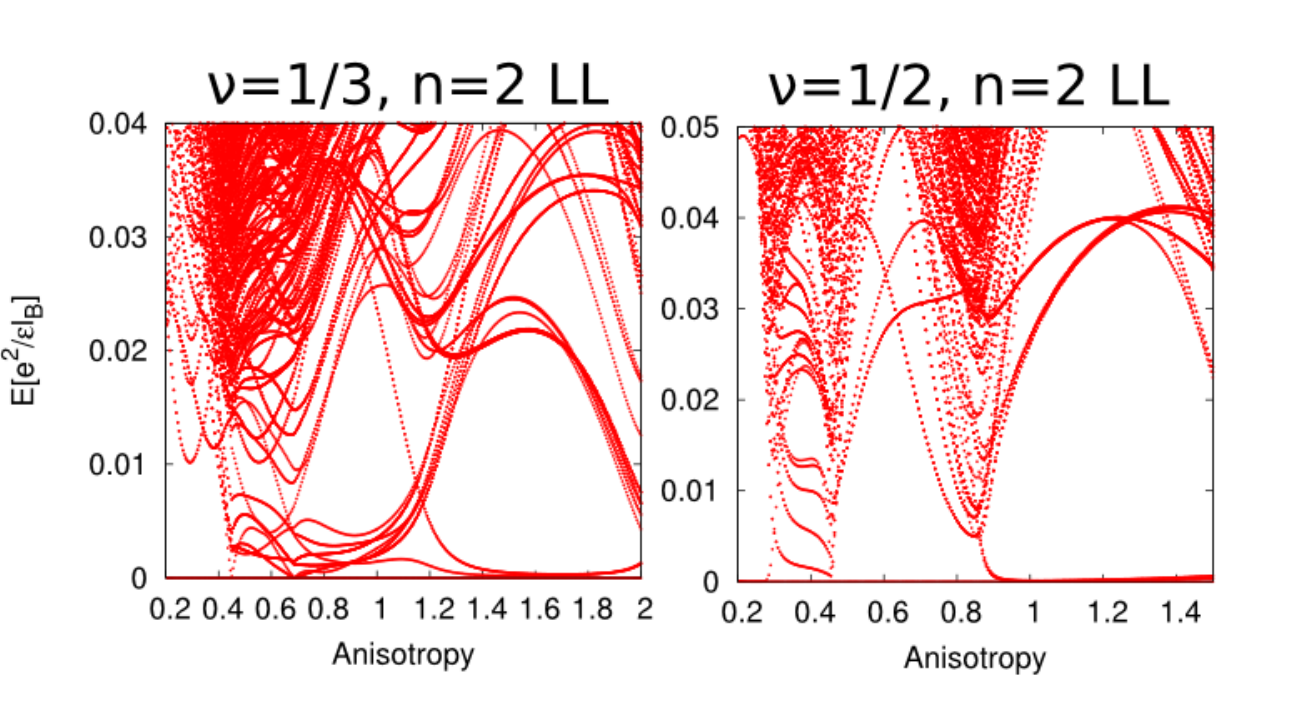}
\caption{Energy spectrum of $\nu=1/3$ (left) and $\nu=1/2$ filled $n=2$ LL (right): mass anisotropy establishes and reinforces the stripe
order.}
\label{fig_ll2}
\end{figure}

For $\mathcal N\ge 2$ the ground state is generally compressible with stripes and bubbles phases, and these phases are enhanced by anisotropic effective mass tensor. In Fig.~\ref{fig_ll2}  the energy spectrum (in units of $e^2/\epsilon\ell_B$) is shown as a function of the anisotropy $\alpha$ (the angle $\phi$ is set to zero). Energies are plotted relative to the ground state at each $\alpha$. As we see on the right panel of Fig.~\ref{fig_ll2}, the increase in $\alpha$ leads to a more pronounced quasi-degeneracy of the ground-state multiplet, and an increase of the gap between this multiplet and the excited states. Level crossing occurs for even larger $\alpha$, but that could be due to finite size effect, which is more pronounced when the effective mass tensor is highly distorted.

In case of $\nu=2+1/3$ state, one expects a two-dimensional CDW order at $\alpha=1$ known as the bubble phase~\cite{edduncanky_bubble}. A bubble differs from a stripe in having a larger degeneracy and a two-dimensional mesh of (quasi)degenerate ground-state wavevectors (as opposed to the one-dimensional array in case of a stripe). The spread of the quasidegenerate levels was also found to be somewhat larger than in the case of stripes. From Fig.\ref{fig_ll2} (left) for $\alpha=1$. The bubble phase remains stable to some extent when $\alpha$ is reduced; for very small $\alpha$ it is eventually destroyed and replaced by a simple CDW. On the other hand, when $\alpha$ is increased, a smaller subset of momenta becomes very closely degenerate with some of the excited levels. This second-order (or weakly first order) transition results in a stripe phase. As for the $\nu=1/2$ case, this stripe becomes enhanced as $\alpha$ is further increased.  Therefore, in $n\geq 2$ LLs mass anisotropy generally produces stripes, even when isotropic ground states have a tendency to form a bubble phase.   

\begin{figure}[ttt]
 \includegraphics[width=0.9\linewidth,angle=0]{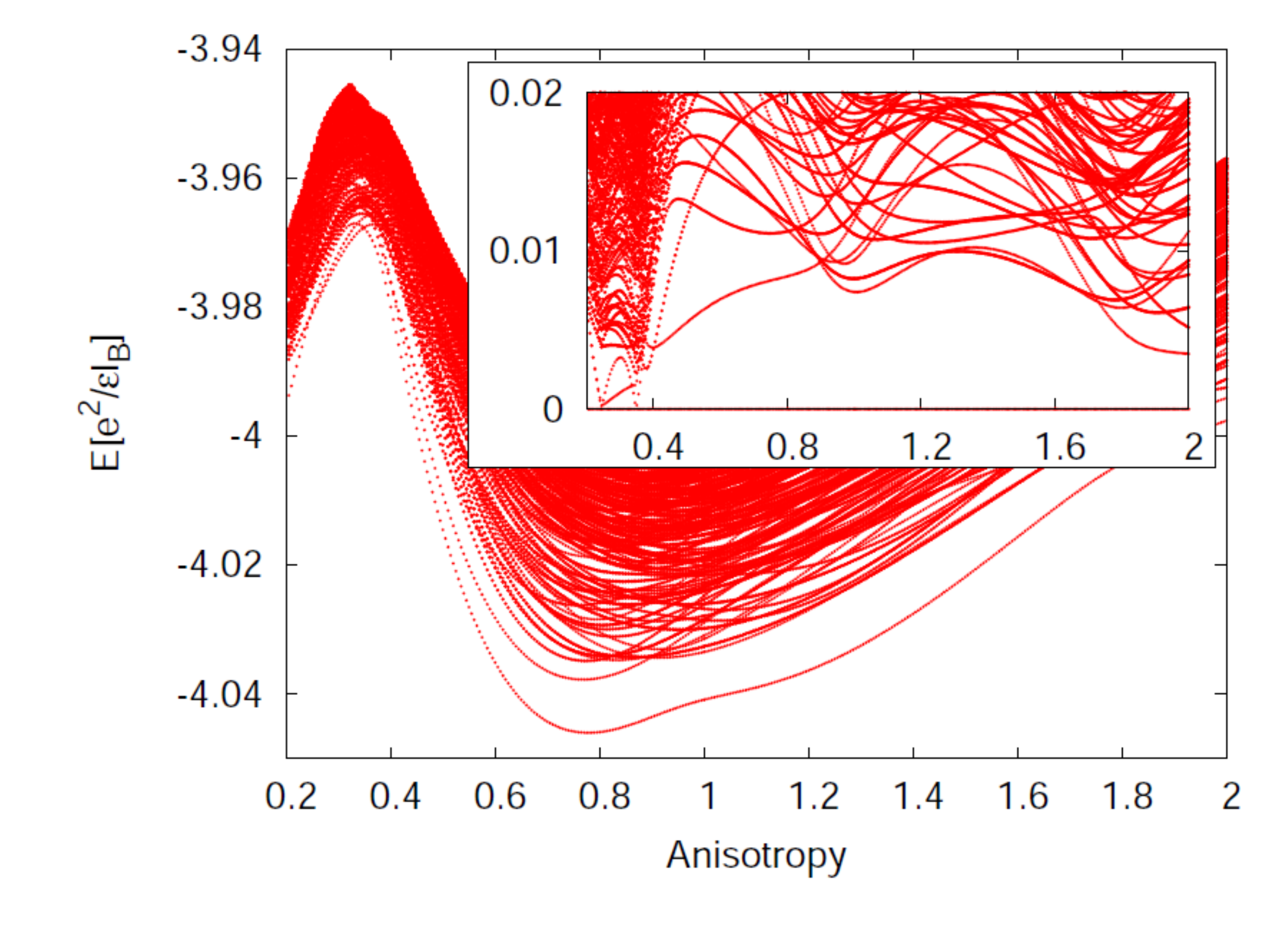}
\caption{Spectrum of $N_e=8$ electrons at $\nu=1+1/3$ with thickness $w=2\ell_B$. Inset: same spectrum plotted relative to the ground state at each $\alpha$. Unit cell has a rectangular shape with aspect ratio $3/4$.}
\label{fig_onethirdanisoll1w2.0}
\end{figure}

\subsection{Incompressible to Compressible Transitions in $\mathcal N=1$}

For pure Coulomb interaction in $\mathcal N=1$, early numerical calculations found the ground state to be at the transition point between compressible and incompressible phases~\cite{haldane_prange}. An experimentally incompressible phase does exist at $\mathcal N=1$, which is believed to be stabilized by the finite thickness of the two-dimensional electron gas, which renormalizes the Coulomb interaction.

\begin{figure}[ttt]
\centerline{\includegraphics[width=0.8\linewidth]{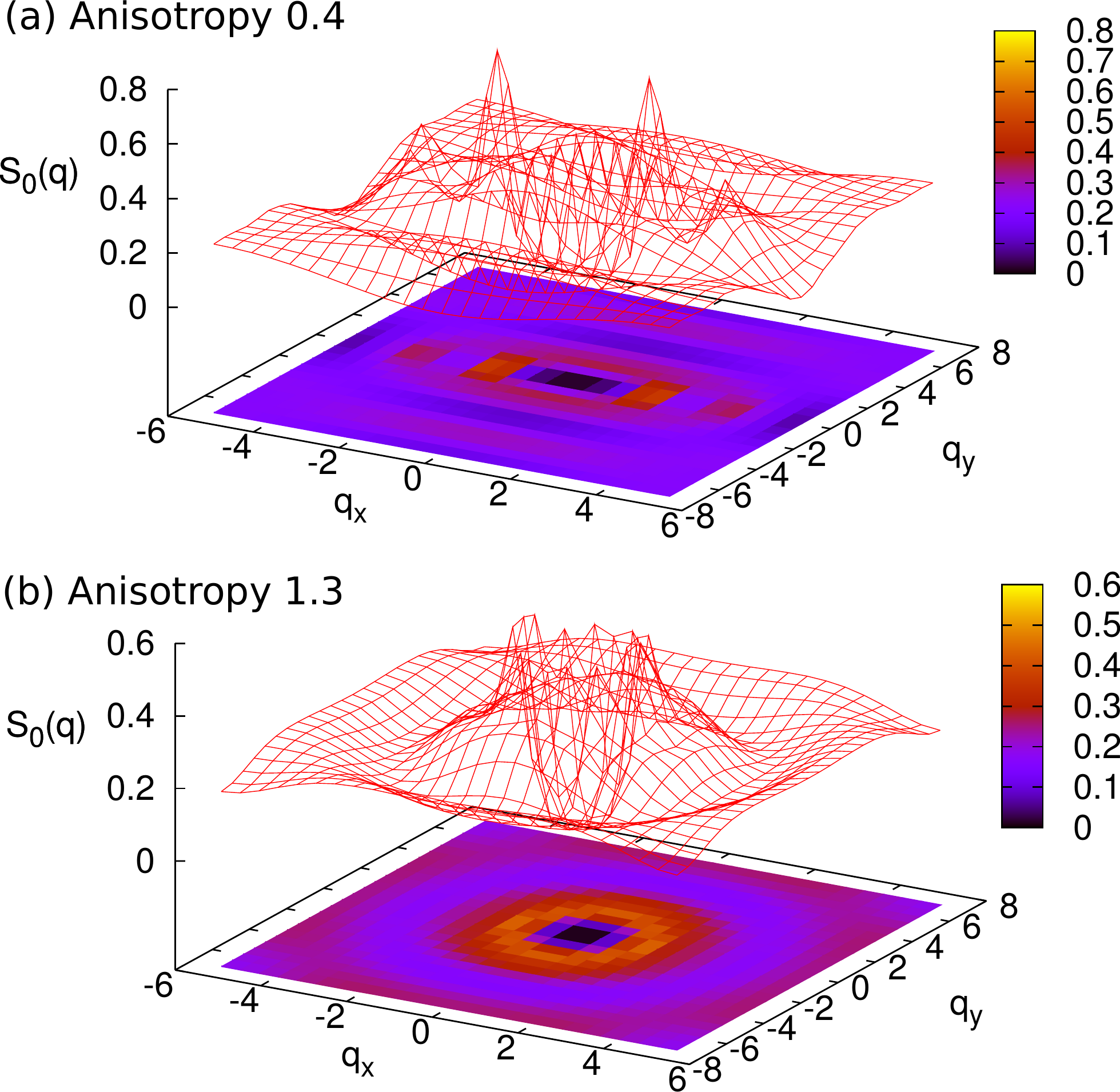}}
\caption{Guiding-center structure factor $S_0(\mathbf{q})$ for $\nu=1/3$ state in $n=1$ LL with thickness $w=2l_B$ and anisotropy $\alpha=0.4$ (a). For comparison, we also show $S_0(\mathbf{q})$ for the state with $\alpha=1.3$ which is in the Laughlin universality class (b). Two peaks in the response function (a) represent the onset of compressibility and CDW ordering.}
\label{fig_s0qonethird}
\end{figure}  

Numerically, varying the $V_1$ pseudopotential leads to the following outcomes: (i) generically, for $\delta V_1<0$, the system is pushed deeper into a compressible phase; (ii) for $\delta V_1>0$, finite-size calculations on systems up to $N_e=9$ electrons permit the existence of two regimes: for $0<\delta V_1^a < \delta V_1 < \delta V_1^b$, the ground state is in the Laughlin universality class, but the lowest excitation is not the magneto-roton; for $\delta V_1 > \delta V_1^b$, the ground state \emph{and} the excitation spectrum is the same as in the LLL. For smaller systems, $\delta V_1^b$ is estimated to be around $0.1e^2/\epsilon \ell_B$, while $\delta V_1^a$ is around $0.04e^2/\epsilon \ell_B$. Larger systems suggest that these two points might merge in the thermodynamic limit, when only a small modification of the interaction might be needed for the Laughlin physics to appear at $\nu=1/3$ in $n=1$ LL. Alternatively, we can consider the Fang-Howard ansatz that mimicks the finite-width effects. In this case, the width of $\ell_B$ or smaller is sufficient to drive a phase transition between the compressible state and the Laughlin-like state, in agreement with results on the sphere and using an alternative finite-width ansatz~\cite{papic_zds}.   

In Fig.~\ref{fig_onethirdanisoll1w2.0} we plot the energy spectrum as a function of anisotropy. One notices that the isotropy point ($\alpha=1$) does not bear any special importance -- indeed, the system appears more stable in the vicinity of it where it can lower its ground state energy or increase the neutral gap. On either side of the isotropy point, however, the system remains in the Laughlin universality class; e.g. at $\alpha=0.8$ and $\alpha=1.3$ the maximum overlap with the Laughlin state is $75\%$ and $80\%$, respectively (these overlaps, although modest compared to the standards of $n=0$ LL, can be adiabatically further increased by tuning the $V_1$ pseudopotential). Note that the quoted maximum overlaps are achieved by the Laughlin state with $\alpha'$ somewhat different from $\alpha$ of the Coulomb state, analogous to Fig.\ref{fig_optimize}. 

The new aspect of Fig.\ref{fig_onethirdanisoll1w2.0} is the transition to a compressible state with CDW ordering for $\alpha\leq 0.4$. In that region of the parameter space, the system is very sensitive to changes in the boundary condition -- the sharp degeneracies seen in the rectangular geometry in Fig.\ref{fig_onethirdanisoll1w2.0} are not obvious in case of higher symmetry, square or hexagonal, unit cell. As an additional diagnostic tool for the compressible states, it is useful to consider a guiding-center structure factor,
\begin{equation}\label{sq}
S_0(\mathbf{q}) = \frac{1}{N_\phi} \sum_{i,j}\langle e^{i\mathbf{q}\cdot \mathbf{R}_i} e^{-i\mathbf{q}\cdot \mathbf{R}_j} \rangle - \langle e^{i\mathbf{q}\cdot \mathbf{R}_i} \rangle \langle e^{-i\mathbf{q}\cdot \mathbf{R}_j} \rangle,
\end{equation}
where the expression for the Fourier components of the guiding-center density, $\rho (\mathbf{q}) = \sum_i^N e^{i\mathbf{q}\cdot \mathbf{R}_i} $,
has been used. Note that $S_0(\mathbf{q})$ is normalized per flux quantum rather than (conventional) per particle~\cite{duncan}. In Fig.\ref{fig_s0qonethird}(a) we show the plot of $S_0(\mathbf{q})$ evaluated for the state with $\alpha=0.4$ in Fig.\ref{fig_onethirdanisoll1w2.0}. Two sharp peaks in the response, similar to those previously identified in $n\geq 2$ LL states~\cite{edduncanky_stripe}, are the hallmark of the CDW order. They are to be contrasted with the smooth response in case of an anisotropic state in the Laughlin universality class for $\alpha=1.3$, Fig.\ref{fig_s0qonethird}(b).

\begin{figure}[ttt]
\centerline{\includegraphics[width=0.78\linewidth,angle=270]{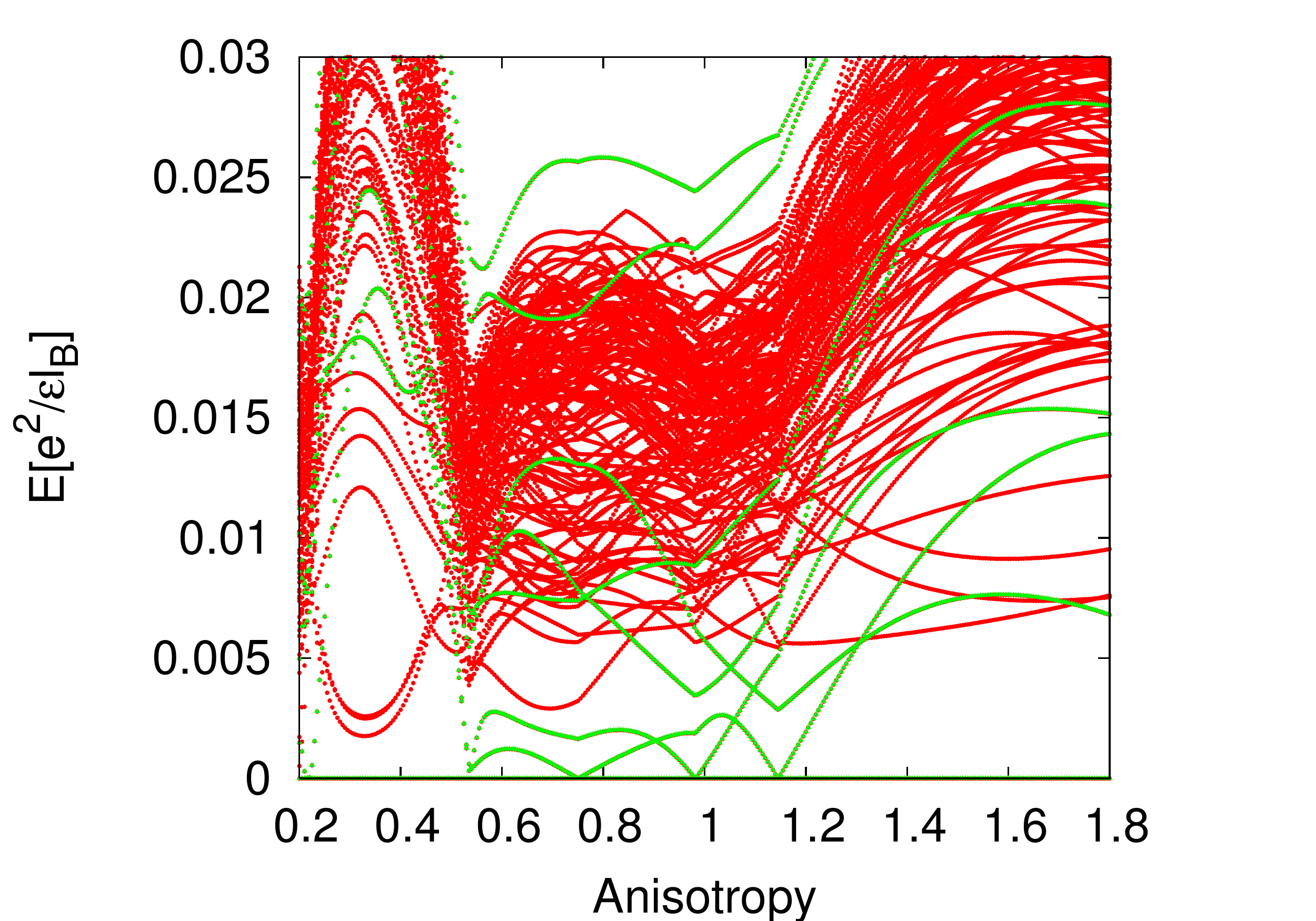}}
\caption{Spectrum of $N_e=14$ electrons at $\nu=1+1/2$ with thickness $w=2l_B$, as a function of anisotropy $\alpha$. Energies are plotted relative to the ground state at each $\alpha$, and the unit cell has a rectangular shape with aspect ratio $3/4$.}
\label{fig_10_20}
\end{figure}

As a second example in $n=1$ LL, we consider half filling where the Moore-Read Pfaffian state~\cite{mr} is believed to be realized in some regions of the phase diagram. This state has a non-Abelian nature, which is reflected in the non-trivial ground state degeneracy~\cite{readgreen} when subjected to periodic boundary conditions. For $\nu=1/2$, the eigenstates of any translationally-invariant interaction possess a twofold center-of-mass degeneracy~\cite{duncan_translations}. On top of this, Moore-Read state has an additional threefold degeneracy. Conventionally, the many-body Brillouin zone is defined for $p=1$,$q=2$ and has a size $N^2$ ($N$ being the GCD of $N_e$ and $N_\phi$), which forces the degenerate groundstates to belong to a Brillouin zone corner $\mathbf{K}=(N/2,N/2)$ and centers of the sides, $\mathbf{K}=(0,N/2);(N/2,0)$.

In Fig.\ref{fig_10_20} we plot the spectrum of the Coulomb interaction as a function of anisotropy (states belonging to $\mathbf{K}$ sectors where the Moore-Read state is realized, are indicated). As earlier, we assume finite width of $w=2\ell_B$ in order to instate the Pfaffian correlations~\cite{rh_pf}. With two-body (Coulomb) interaction, therefore in each finite system the Moore-Read state will mix with its particle-hole conjugate pair, the anti-Pfaffian~\cite{antipf}. The mixing between the two states can be controlled by including higher LLs~\cite{llmix}.  For $0.5 \leq \alpha \leq 1.3$, we find a three-fold quasi-degenerate multiplet, suggesting the presence of the Moore-Read state at the isotropy point and in the neighborhood of it. In finite systems, there is some splitting of the degeneracy that might be reduced upon tuning the $V_1,V_3$ pseudopotentials. Also, upon tuning the anisotropy around $\alpha=1$, there are crossings within the multiplet of degenerate ground states without apparent closing of the gap. 
The region of the Moore-Read state is defined by sharp transitions towards crystal phases. These transitions are likely second order because they do not appear to involve any level crossing, but rather lifting of the degeneracy within a ground-state multiplet.

\chapter{Geometry and Linear Response in FQHE\label{ch:linresponse}}

The interplay between topology and geometry in the FQHE is a fascinating subject that is recently gaining more attention. Wen and Zee first studied the fact that the finite-size QHE realized on spherical geometry does not have the same filling factor as the one in the thermodynamic limit\cite{wen4}. Instead the general relation between the number of fluxes $N_\phi$ and the number of particles $N_e$ at various incompressible phases is given by 
\begin{eqnarray}\label{shift}
N_\phi=\nu^{-1}N_e+\mathcal S
\end{eqnarray}
where $\nu$ is the filling factor in the thermodynamic limit $N_e\rightarrow \infty$, and $\mathcal S$ is the so-called ``shift", an $O(1)$ correction thought to be distinctive for different topological phases. For IQHE with $\nu=N$, the shift $\mathcal S=N$ is positive. For FQHE, on the other hand, the shift is negative. For Laughlin states at $\nu=1/m$ the shift is given by $\mathcal S=1-m$. Indeed, if we ignore $\mathcal S$ in Eq.(\ref{shift}) when performing numerical diagonalizations for the FQHE, the spectrum will no longer be incompressible. For model Hamiltonians the null space (spanned by the zero energy states) is a degenerate manifold of quasihole states.

The shift is topological because it is quantized and stable against perturbations that preserve rotational symmetry, as long as the system remains gapped. Microscopically, a particle with a non-zero spin accumulates a phase when going around a loop in curved real space. This phase is analagous to the Aharonov-Bohm phase of a charged particle moving in a magnetic field. In the QHE the wavefunction of a particle on the sphere can always be viewed as a coherent state with the $SU(2)$ symmetry of a spinor; its orbital angular momentum is effectively its spin. The coupling of the angular momentum (either internal or orbital) to the curvature can be viewed as a result of an additional effective magnetic field, therefore increasing/decreasing the total number of available orbitals. This is the origin of the shift in the QHE. The shift is quantized to integers or half-integers on the sphere because of this observation: a spinning particle going around the equator can pick up the Gaussian curvature of either the northern hemisphere or the southern hemisphere; the phase ambiguity will be $4\pi$ times the shift, which has to be an integer multiple of $2\pi$.

On a flat Hall surface, the quantity closely related to the shift is the Hall viscosity\cite{asz,rr2}. In this chapter,  the microscopic origin of the Hall viscosity will be presented. The Hall viscosity of the FQHE is characteristic of its topological phase, which couples to the geometry of the FQH system resulting from both the quantum fluctuation and the external perturbation. A thorough overview of the linear response to a spatially varying electromagnetic perturbation is given in this chapter, starting with the most general case where neither Galilean invariance nor rotational invariance is assumed. The role of the Hall viscosity is highlighted to illustrate the necessary condition for such a quantity to be measured experimentally.

\section{Hall Viscosity : A Formal Definition}

The Hall viscosity arises in studies of the stress tensor $\sigma_{\alpha\beta}$ in a continuum medium. Any deformation of the continuum media from its equilibrium state generates stress, and for a small deformation the stress tensor is a linear response to the strain $u^{\alpha}_\beta$, and the rate of strain $\dot u^\alpha_\beta$. Denoting the displacement of the elements from their equilibrium position to be $u^\alpha(r,t)$, the strain tensor is given by
\begin{eqnarray}\label{strain}
u^\alpha_\beta=\frac{1}{2}\left(\frac{\partial u^\alpha}{\partial r^\beta}+\frac{\partial u^\beta}{\partial r^\alpha}\right)
\end{eqnarray}
If at equilibrium the local metric is given by $g^{ab}$, the strain tensor gives the local deformation of the metric to the lowest order: $\delta g^{ab}=g^{ac}u^{b}_c$. The linear response of the stress tensor gives
\begin{eqnarray}\label{tensorlr}
\sigma^\alpha_\beta=\lambda^{\alpha\gamma}_{\beta\delta}u^\delta_\gamma-\eta^{\alpha\gamma}_{\beta\delta}\dot u^\delta_\gamma
\end{eqnarray}
Here $\lambda^{\alpha\gamma}_{\beta\delta}$ is the elastic modulus tensor and $\eta^{\alpha\gamma}_{\beta\delta}$ is the viscosity tensor. The elastic modulus tensor is symmetric under the following exchange: $\alpha\leftrightarrow\beta,\gamma\leftrightarrow\delta, \{\alpha\beta\}\leftrightarrow\{\gamma\delta\}$. The viscosity tensor is symmetric under the following exchange: $\alpha\leftrightarrow\beta,\gamma\leftrightarrow\delta$. For exchange between $\{\alpha\beta\}\leftrightarrow\{\gamma\delta\}$ the viscosity tensor can be separated into the symmetric and anti-symmetric part:
\begin{eqnarray}\label{symasym}
\widetilde \eta^{\alpha\gamma}_{\beta\delta}&=&\frac{1}{2}\left(\eta^{\alpha\gamma}_{\beta\delta}+\eta^{\gamma\alpha}_{\delta\beta}\right)\\
\bar \eta^{\alpha\gamma}_{\beta\delta}&=&\frac{1}{2}\left(\eta^{\alpha\gamma}_{\beta\delta}-\eta^{\gamma\alpha}_{\delta\beta}\right)
\end{eqnarray}
We can also separate out the traceless part of the strain tensor by writing $u^{\alpha}_\beta=\left(u^\alpha_\beta-\frac{1}{2}\delta^\alpha_\beta u^\mu_\mu\right)+\frac{1}{2}\delta^\alpha_\beta u^\mu_\mu$. Thus Eq.(\ref{tensorlr}) can be rewritten as 
\begin{eqnarray}
\sigma^\alpha_\beta=\widetilde\lambda^{\alpha\gamma}_{\beta\delta}\left(u^\alpha_\beta-\frac{1}{2}\delta^\alpha_\beta u^\mu_\mu\right)+\frac{1}{2}\bar\lambda^{\alpha\gamma}_{\beta\delta}{2}\delta^\alpha_\beta u^\mu_\mu-\left(\widetilde \eta^{\alpha\gamma}_{\beta\delta}+\bar \eta^{\alpha\gamma}_{\beta\delta}\right)\dot u^\delta_\gamma
\end{eqnarray}
Here, $\widetilde\lambda^{\alpha\gamma}_{\beta\delta}$ is the pure shear modulus, and $\bar\lambda^{\alpha\gamma}_{\beta\delta}$ is the bulk modulus. Classically the shear modulus vanishes for a fluid, which is made of point particles with no internal structure. However as we shall see in Eq.(\ref{apdenergy}) of Chapter 4, a uniform deformation of the guiding center metric of the FQH fluid does cost energy. Phenomenologically the FQH fluid is made of particle-flux composite with internal structures. The shear modulus comes from the deformation of the shape of these composite particles, and is purely a quantum effect. The shear of the cyclotron metric (i.e. the effective mass tensor for Galilean invariant systems) induces LL mixing and also costs energy. In the limit of strong magnetic field the LL mixing can be ignored; it is only in this sense the IQH fluid has vanishing shear modulus.

The bulk modulus for a classical incompressible fluid is infinity. For the case of the QHE, the gapped QH fluid can transmit force by gapless chiral edge modes, but not through the bulk. A spatially varying force (i.e. induced by the gradient of the electric field) induces a locally varying cyclotron/guiding center metric. This will modulate density as will be shown in Eq.(\ref{grealspace}). The internal structure of the coherent state/composite particles will again lead to finite bulk modulus which only vanishes in the long wavelength limit (i.e. with a spatially uniform force).

For a dissipationless fluid, $\widetilde \eta^{\alpha\gamma}_{\beta\delta}$ has to vanish. On the other hand, $\bar \eta^{\alpha\gamma}_{\beta\delta}$ is non-zero only if time-reversal symmetry is broken. This term is the so-called Hall viscosity. For gapped quantum fluids with broken time-reversal symmetry, the Hall viscosity dominates the response at low temperature. 

Quantum mechanically, the total stress tensor over the entire volumn, induced by a uniform strain is given by
\begin{eqnarray}\label{qversion}
\int dV \sigma^\alpha_\beta=\langle\psi|\frac{\partial H}{\partial u^\alpha_\beta}|\psi\rangle=\frac{\partial E}{\partial u^\alpha_\beta}+\text{Im}\langle\partial_{\alpha\beta}\psi|\partial_{\gamma\delta}\psi\rangle\dot u^\gamma_\delta
\end{eqnarray}
where $H$ is the Hamiltonian and $E$ is the energy of the eigenstate $|\psi\rangle$, with $|\partial_{\alpha\beta}\psi\rangle=\frac{\partial}{\partial u^\alpha_\beta}|\psi\rangle$. Comparing Eq.(\ref{qversion}) with Eq.(\ref{tensorlr}), the Hall viscosity is the Berry curvature when the Hamiltonian is adiabatically deformed. Thus Eq.(\ref{qversion}) can be taken as the quantum mechanical definition of the Hall viscosity\cite{asz}.

\section{The Geometry of the Hilbert Space}

Let us look at a physical state $|\psi\rangle$ that depends on a set of external parameters $\{\mu,\nu,\cdots\}$. Notice all states in the Hilbert space has a $U(1)$ gauge invariance so that the physics is not affected by the following transformation 
\begin{eqnarray}\label{u1gauge}
|\psi\rangle\rightarrow e^{i\alpha}|\psi\rangle
\end{eqnarray}
The phase $\alpha$ can now depend on the external parameters, but local gauge invariance in the parameter space should be preserved; the set of $|\psi(\mu,\nu,\cdots)\rangle$ form a $U(1)$ fiber bundle over the manifold of the external parameters. The gauge invariant derivative of $|\psi\rangle$ in the parameter space is given by\cite{duncan}
\begin{eqnarray}\label{derivative}
|D_\mu\psi\rangle=|\partial_\mu\psi\rangle-|\psi\rangle\langle\psi|\partial_\mu\psi\rangle
\end{eqnarray}
This allows us to define the gauge invariant quantum metric $\mathcal{G}_{\mu\nu}$ and the Berry phase $\mathcal{F}_{\mu\nu}$:
\begin{eqnarray}\label{def}
\langle D_\mu\psi|D_\nu\psi\rangle&=&\frac{1}{2}\left(\mathcal{G}_{\mu\nu}+i\mathcal{F}_{\mu\nu}\right)\\
\mathcal{G}_{\mu\nu}&=&\langle\partial_\mu\psi|\partial_\nu\psi\rangle-\langle\partial_\mu\psi|\psi\rangle\langle\psi|\partial_\nu\psi\rangle+\mu\leftrightarrow\nu\label{defl1}\\
i\mathcal{F}_{\mu\nu}&=&\langle\partial_\mu\psi|\partial_\nu\psi\rangle-\mu\leftrightarrow\nu\label{defl2}
\end{eqnarray}
If $|\psi\rangle$ is a non-degenerate eigenstate of a Hermitian operator (which could be the Hamiltonian) $h$, there is useful way to express Eq.(\ref{def}) in terms of $h$:
\begin{eqnarray}\label{def2}
\mathcal{G}_{\mu\nu}&=&\sum_{n}\frac{\langle\psi|\partial_\mu h|\psi_n\rangle\langle\psi_n|\partial_\nu h|\psi\rangle+\mu\leftrightarrow\nu}{\left(\epsilon-\epsilon_n\right)^2}\label{def2l1}\\
i\mathcal{F}_{\mu\nu}&=&\sum_{n}\frac{\langle\psi|\partial_\mu h|\psi_n\rangle\langle\psi_n|\partial_\nu h|\psi\rangle-\mu\leftrightarrow\nu}{\left(\epsilon-\epsilon_n\right)^2}\label{def2l2}
\end{eqnarray}
where the summation is over all states in the Hilbert space defined by $h$ that are \emph{orthogonal} to $|\psi\rangle$, and $\epsilon, \epsilon_n$ are eigenvalues of $|\psi\rangle,|\psi_n\rangle$ respectively. While Eq.(\ref{defl1}) and Eq.(\ref{defl2}) suggest the Berry curvature and the quantum metric are both the properties of a single state, Eq.(\ref{def2l1}) and Eq.(\ref{def2l2}) explicitly shows how the entire Hilbert space is required for the two quantities to be well-defined. This is especially important when the Hilbert space is physically truncated, and Eq.(\ref{def2l1}) and Eq.(\ref{def2l2}) are the proper ways to calculate the Berry curvature and the quantum metric. These two equations are also advantageous for numerical calculations for the quantum metric and the Berry phase, because one does not have to worry about random phases of the ground state obtained from exact diagonalizations at different points in the parameter space.

\subsection{Single-Particle Case: Coherent States}

The coherent states are defined by ladder operators with commutation relations $[a,a^\dagger]=1$. The family of coherent states can be parametrized by the ``center of mass" position in the phase space, i.e. a complex number $k$ such that $a|k\rangle=k|k\rangle$. The definition of a coherent state in the phase space requires a metric, and explicitly let us have $a=\omega_a^*R^a,a^\dagger=\omega_aR^a$, where $\omega_a$ is the usual complex vector with the unimodular metric $g_{ab}=\omega_a^*\omega_b+\omega_a\omega_b^*$ (Eq.(\ref{omega})) and $[R^a,R^b]=-i\epsilon^{ab}$. One can identify $R^a$ as the guiding center coordinates in quantum Hall systems with $l_B=1$, but in general for any one-dimensional harmonic oscillators, $R^a$ is the non-commutative coordinates in the phase space. The auxillary Hamiltonian is given by
\begin{eqnarray}\label{h2}
h=g_{ab}(z)(R^a-k^a(z))(R^b-k^b(z))
\end{eqnarray}
where $z$ denotes the set of parameters upon which the Hamiltonian can vary adiabatically, and $a,b=1,2$. Thus the unique ground state of $h$ is a coherent state located at $k=\omega_ak^a$ in the complex plane with its shape determined by $g_{ab}(z)$. Defining $\Lambda^{ab}=\frac{1}{2}\{R^a,R^b\}$, the Lie algebra of translation and area-preserving deformation is given by
\begin{eqnarray}\label{poincare}
&&[R^a,R^b]=-i\epsilon^{ab}\\
&&[R^a,\Lambda^{bc}]=-i\epsilon^{ab}R^c-i\epsilon^{ac}R^b\\
&&[\Lambda^{ab},\Lambda^{cd}]=-i\epsilon^{ac}\Lambda^{bd}-i\epsilon^{ad}\Lambda^{bc}-i\epsilon^{bd}\Lambda^{ac}-i\epsilon^{bc}\Lambda^{ad}
\end{eqnarray}
To adiabatically drag a state around and deform it at the same time, we define the following unitary operator
\begin{eqnarray}\label{unitary}
U(q_a,\alpha_{ab})=e^{iq_a(z)R^a+i\alpha_{ab}(z)\Lambda^{ab}}
\end{eqnarray}
For infinitesmal parameters $\delta q_a$ and the symmetric tensor $\delta \alpha_{ab}$ we have
\begin{eqnarray}\label{transformation}
UhU^\dagger&=&h+[i\delta q_aR^a+i\delta\alpha_{ab}\Lambda^{ab},h]\nonumber\\
&=&\bar g_{ab}\left(R^a-\bar k^a\right)\left(R^b-\bar k^b\right)\\
\bar g^{ab}&=&g_{ab}+2g_{bd}\epsilon^{cd}\delta\alpha_{ac}+2g_{ad}\epsilon^{cd}\delta\alpha_{bc}\\
\bar k^a&=&k^a+\epsilon^{ab}\delta q_b+2\epsilon^{ab}\delta\alpha_{bc}k^c
\end{eqnarray}
Only linear orders of infinitesmal parameters are kept, and constants are dropped since they are irrelevant. There is no one-to-one relationship between $\delta\alpha_{ab}$ and $\delta g_{ab}$ due to a physically insignificant phase (see Eq.(\ref{repa}) and the comments below), the most convenient choice is to write $\delta\alpha_{ab}=\frac{1}{4}\epsilon_{bc}g^{cd}\delta g_{da}$, where $\delta g_{ab}$ is the infinitesmal change of the metric. The infinitesmal translation of the coherent state is given by
\begin{eqnarray}\label{infinitesmal}
\delta k^a&=&\epsilon^{ab}\delta q_b+2\epsilon^{ab}\delta\alpha_{bc}k^c
\end{eqnarray}
Going back to Eq.(\ref{def2}) and using $\partial_\mu h=[i\partial_\mu q_aR^a+i\partial_\mu\alpha_{ab}\Lambda^{ab},h]=[\mathcal{A}_\mu,h]$ we have
\begin{eqnarray}\label{metricberry}
\mathcal{G}_{\mu\nu}&=&-\langle\psi_0(z)|\{\mathcal{A}_\mu,\mathcal{A}_\nu\}|\psi_0(z)\rangle+2\langle\psi_0(z)|\mathcal{A}_\mu|\psi_0(z)\rangle\langle\psi_0(z)|\mathcal{A}_\nu|\psi_0(z)\rangle\\
i\mathcal{F}_{\mu\nu}&=&-\langle\psi_0(z)|[\mathcal{A}_\mu,\mathcal{A}_\nu]|\psi_0(z)\rangle
\end{eqnarray}
Defining $\widetilde \Lambda^{ab}=\frac{1}{2}\{R^a-k^a,R^b-k^b\}$, we have $\langle\psi_0(z)|\widetilde\Lambda^{ab}|\psi_0(z)\rangle=\langle\widetilde \Lambda^{ab}\rangle_0=sg^{ab}(z)$, where $s=\langle a^\dagger a\rangle+\frac{1}{2}$ is the orbital spin in the phase space, and $\langle R^a\rangle_0=k^a$. Identifying the parameter space with the real space by $q_\mu=\epsilon_{\mu\nu}k^\nu$ and with a bit of algebra one obtains the following results
\begin{eqnarray}\label{gf}
\mathcal{G}_{\mu\nu}&=&2sg_{\mu\nu}+\frac{1}{16}\left(s^2+\frac{3}{4}\right)\left(g^{ac}g^{bd}-\epsilon^{ac}\epsilon^{bd}\right)\partial_\mu g_{ab}\partial_\nu g_{cd}\label{gg}\\
i\mathcal{F}_{\mu\nu}&=&-i\epsilon_{\mu\nu}-\frac{is}{4}\epsilon^{ac}g^{bd}\partial_\mu g_{ab}\partial_\nu g_{cd}\label{ff}
\end{eqnarray}
The orbital spin is related to the Hall viscosity $\eta$ by $\eta=\frac{s}{4}$. Time reversal symmetry in the phase space is always broken, since momentum is odd under time-reversal. For particles with a non-zero orbital spin, the quantum metric is well-defined and proportional to the phase space metric, as shown by the first term of Eq(\ref{gg}). If the phase space is not flat, there is an additional correction coming from the second term of Eq.(\ref{gg}). The first term of Eq.(\ref{ff}) is the Aharanov-Bohm phase of dragging the coherent state around in the phase space with an effective uniform magnetic field. For the QHE where the phase space is mapped to the real space, the Aharanov-Bohm phase results from the external magnetic field. Additional phase from the second term of Eq.(\ref{ff}) comes from the coupling of the single particle orbital spin to the $U(1)$ curvature of the underlying geometry $\mathcal B_f=\frac{1}{8}\epsilon^{\mu\nu}\epsilon^{ac}g^{bd}\partial_\mu g_{ab}\partial_\nu g_{cd}$. Note unlike the shift of a spinor, this $U(1)$ curvature is only one part of the Gaussian curvature given as follows:
\begin{eqnarray}\label{gaussian}
\mathcal K=\mathcal B_f-\frac{1}{2}\partial_a\partial_bg^{ab}
\end{eqnarray}
Eq.(\ref{gaussian}) is only valid for a unimodular metric $g_{ab}$, and is the curl of the spin connection\cite{wen4}. The second term of Eq.(\ref{ff}) gives an additional effective magnetic field when the particle is dragged in a loop with a non-trivial geometry of the phase space. For the QHE, the phase space is mapped to the real space, so the $U(1)$ curvature of the real space will modify the number of orbitals available for the single particle state.

\subsection{FQH State: Thermodynamic Limit and Edge Effect}

The analysis for a single particle state can be generalized to a many-body ground state as long as the ground state is non-degenerate. In this section we focus on the ground state of the FQH Hamiltonian at filling factor $\nu$ on the disk geometry, the most natural geometry from an experimental point of view. To make the ground state unique, one can impose a hard-wall boundary condition, whereby the Hilbert space is restricted to a finite number of orbitals; the alternative is to impose a soft confining potential. The latter is more realistic, but the analysis in the following does not depend on the details of the Hamiltonian $\mathcal H$ for $N_e$ particles. We only need to assume the many-body ground state is non-degenerate. 

The disk geometry has rotational invariance, thus $\mathcal H$ is parametrized by a metric $g_{ab}$. To adiabatically vary the metric of the Hamiltonian, the generator for area-preserving deformation is given by $\Lambda^{ab}=\widetilde\Lambda^{ab}+\bar\Lambda^{ab}$, where $\widetilde\Lambda^{ab}=\sum_i\frac{1}{2}\{\widetilde R_i^a,\widetilde R_i^b\}$ only depends on the cyclotron coordinates and $\bar\Lambda^{ab}=\sum_i\frac{1}{2}\{\bar R_i^a,\bar R_i^b\}$ is the guiding center analog. The two parts can be evaluated separately.

Since the ground state $|\psi_0\rangle$ only involves a single LL (which we label as the $N^{\text{th}}$ LL), the cyclotron part can be easily evaluated. Its Berry curvature and quantum metric contributions are given by
\begin{eqnarray}\label{cyclotronpart}
i\widetilde{\mathcal F}_{\mu\nu}&=&\partial_\mu\alpha_{ab}\partial_\nu\alpha_{cd}\langle\psi_0|[\widetilde\Lambda^{ab},\widetilde\Lambda^{cd}]|\psi_0\rangle\nonumber\\
&=&\frac{iN_e\widetilde s}{4}\epsilon^{ac}g^{bd}\partial_\mu g_{ab}\partial_\nu g_{cd}\label{mbf}\\
\widetilde{\mathcal G}_{\mu\nu}&=&\partial_\mu\alpha_{ab}\partial_\nu\alpha_{cd}\left(\langle\psi_0|\{\widetilde\Lambda^{ab},\widetilde\Lambda^{cd}\}|\psi_0\rangle-2\langle\psi_0|\widetilde\Lambda^{ab}|\psi_0\rangle\langle\psi_0|\widetilde\Lambda^{cd}|\psi_0\rangle\right)\nonumber\\
&=&\frac{N_e}{16}\left(\widetilde s^2+\frac{3}{4}\right)\left(g^{ac}g^{bd}-\epsilon^{ac}\epsilon^{bd}\right)\partial_\mu g_{ab}\partial_\nu g_{cd}\label{mbg}
\end{eqnarray}
Both are extensive quantities obtained by summing up single particle contributions, where $\widetilde s$ is the cyclotron spin. The evaluation of the guiding center contribution requires more care. Replacing $\widetilde\Lambda^{ab}$ with $\bar\Lambda^{ab}$ in Eq.(\ref{mbf}) and Eq.(\ref{mbg}), the quantity of interest we need to evaluate are:
\small
\begin{eqnarray}\label{sas}
\mathcal P^{abcd}_A&=&\langle[\bar\Lambda^{ab},\bar\Lambda^{cd}]\rangle_0=\sum_{n>0}\langle\bar\Lambda^{ab}\rangle_{0n}\langle\bar\Lambda^{cd}\rangle_{n0}-\{ab\}\leftrightarrow \{cd\}\label{pa}\\
\mathcal P^{abcd}_S&=&\frac{1}{2}\langle\{\bar\Lambda^{ab},\bar\Lambda^{cd}\}\rangle_0-\langle\bar\Lambda^{ab}\rangle_0\langle\bar\Lambda^{cd}\rangle_0=\sum_{n>0}\langle\bar\Lambda^{ab}\rangle_{0n}\langle\bar\Lambda^{cd}\rangle_{n0}+\{ab\}\leftrightarrow\{cd\}\label{sas1}
\end{eqnarray}
\normalsize
where $\langle\mathcal O\rangle_{mn}=\langle\psi_m|\mathcal O|\psi_n\rangle$, and $|\psi_n\rangle$ for $n>0$ are excited states of $\mathcal H$. Since $\mathcal H$ is rotationally invariant, we can define $b_i=\omega_a\bar R_i^a,b^\dagger_i=\omega_a^*\bar R_i^a$, and $|\psi_n\rangle$ is an eigenstate of $\sum_ib^\dagger_ib_i$, Eq.(\ref{sas}) and Eq.(\ref{sas1}) can be written in a more illuminating form by defining $\mathcal P^{abcd}_A=\frac{i}{2}\left(g^{ac}\epsilon^{bd}+g^{bd}\epsilon^{ac}\right)\mathcal Q_A, \mathcal P^{abcd}_S=\frac{1}{4}\left(g^{ac}g^{bd}+g^{ad}g^{bc}-g^{ab}g^{cd}\right)\mathcal Q_S$:
\begin{eqnarray}\label{qsas}
\mathcal Q_A&=&\sum_{ij}\sum_{n}\langle b^2_i\rangle_{0n}\langle \left(b^\dagger_j\right)^2\rangle_{n0}-\langle \left(b^\dagger_i\right)^2\rangle_{0n}\langle b_j^2\rangle_{n0}\label{qa}\\
\mathcal Q_S&=&\sum_{ij}\sum_{n}\langle b^2_i\rangle_{0n}\langle \left(b^\dagger_j\right)^2\rangle_{n0}+\langle \left(b^\dagger_i\right)^2\rangle_{0n}\langle b_j^2\rangle_{n0}\label{qs}
\end{eqnarray}
Physically, $\frac{1}{N_e}\mathcal Q_A$ is the guiding center Hall viscosity, and $\frac{1}{N_\phi}\mathcal Q_S$ is the $O(q^4)$ coefficient of the guiding center structure factor defined in Eq.(\ref{sq}). It is instructive to first calculate Eq.(\ref{qa}) and Eq.(\ref{qs}) for a droplet of IQH fluid at filling factor $\nu=1$ with $N_e=N_\phi$. The ground state $|\psi_0\rangle$ is a Slater determinant of all single particle orbitals given by
\begin{eqnarray}\label{slater}
\langle\{z_i\}|\psi_0\rangle=\prod_{i<j}^{N_e}\left(z_i-z_j\right)e^{-\frac{1}{2}\sum_i|z_i|^2}
\end{eqnarray}
The second term of Eq.(\ref{qa}) and Eq.(\ref{qs}) vanishes. A simple calculation shows $\mathcal Q_A=\mathcal Q_S=2N_e^2$, and both are super-extensive. We know that for the bulk of the IQH fluids there is no guiding center degrees of freedom, so both the guiding center spin and the guiding center structure factor should vanish. The super-extensive part is coming from the edge of the droplet, where electrons in the outermost two orbitals can hop into un-occupied orbitals. Note this process is forbidden on a compact geometry like the sphere or the torus. Thus to calculate the bulk contributions to $\mathcal Q_A$ and $\mathcal Q_S$ on the disk, one needs to exclude the edge modes when summing over the excitation spectrum in Eq.(\ref{qa}) and Eq.(\ref{qs}); in particular, one cannot just rewrite Eq.(\ref{qa}) as the ground state expectation value of a commutator, even in the thermodynamic limit $N_e\rightarrow \infty$.

For FQHE at $\nu=1/m$, the ground state is the Laughlin state. Without removing the edge modes, $\mathcal Q_A=\frac{1}{2}mN_e^2+\bar s N_e$, where $\bar s=\frac{1-m}{2}$ is the coefficient of the extensive part. The analytic expression for $\mathcal Q_S$ is unknown, but it also contains both the extensive and super-extensive part. The removal of the edge modes of the FQH fluid on the disk is a tricky issue. The first term of Eq.(\ref{qa}) and Eq.(\ref{qs}) couples the ground state to the excited states in the momentum sector $\delta L=2$, where the ground state is taken as $\delta L=0$. Since $\sum_i(b_i^\dagger)^2|\psi_0\rangle$ is another zero energy state, it only coupled to the two-dimensional manifold of zero energy edge states (the counting of the zero energy edge modes will be explained in Chapter 5). For finite size systems, this will give both an extensive and a super-extensive part.

The second term of Eq.(\ref{qa}) and Eq.(\ref{qs}) couples the ground state to the highest weight excited states in the momentum sector $\delta L=-2$ (by definition, the highest weight state is annihilated by $\sum_ib_i$). In this sector, there are bulk excitations, as well as states containing both bulk and edge excitations (see Fig.\ref{fig_disk}); because of the presence of the latter, this term also contains both extensive and super-extensive part. In the thermodynamic limit, the bulk and edge excitations are independent. Projecting out the gapless edge excitations will remove the super-extensive part of both Eq.(\ref{qa}) and Eq.(\ref{qs}). It is conjectured that for maximally chiral FQHE\cite{haldane6} we have $|\mathcal Q_A|=|\mathcal Q_S|$ and both are given by the guiding center Hall viscosity. This implies in the thermodynamic limit the first term of Eq.(\ref{qa}) and Eq.(\ref{qs}) is purely super-extensive. Numerical calculation on compact torus geometry was performed to show the Hall viscosity is the extensive part of Eq.(\ref{qa}) when the system is large enough\cite{rr2}, both for the Laughlin and Moore-Read state.

From now onwards we will only focus on the bulk properties of FQHE, so gapless edge modes will be projected out whenever Eq.(\ref{qa}) and Eq.(\ref{qs}) are used. In particular, from Eq.(\ref{pa}) we have:
\begin{eqnarray}\label{pacom}
\mathcal P^{abcd}_A=\langle[\bar\Lambda^{ab},\bar\Lambda^{cd}]\rangle_0=\frac{i\bar sN_e}{2}\left(g^{ac}\epsilon^{bd}+g^{bd}\epsilon^{ac}\right)
\end{eqnarray}
which is the regularized result that will be used in the following section.

\section{Hall Viscosity and Long Wavelength Limit of the Linear Response}

In this section, the formalism developed in previous sections will be applied to calculate quantities that can be potentially measured experimentally. The linear response of the quantum Hall fluid to a uniform electric field is known to be a topological index equal to the filling factor. It has been first noticed by Hoyos and Son \cite{Son} that the Hall viscosity constitutes the universal part of the long wavelength corrections to the Hall conductivity, when a spatially varying electric field is present. The result was derived by requiring the effective Chern-Simons (CS) theory for integer and fractional quantum Hall fluids to satisfy diffeomorphism invariance, with the implicit assumption of Galilean and rotational invariance. Later on, a Kubo formula for the Hall viscosity is developed by Bradlyn et al.\cite{bradlyn}, from which the long wavelength limit of the Hall conductivity can be derived for Galilean invariant IQH fluids.

An intriguing fact of the above analysis is that the only FQH contribution to the $O(q^2)$ part of the Hall conductivity is the guiding center Hall viscosity, a topological quantity independent of the details of the interaction. The intra-Landau level dynamics, on the other hand, does not play a significant role, as long as the system is gapped. This is certainly not explicit from the effective Chern-Simons theory, in which the Galilean invariance is assumed even for the FQHE. Galilean invariance generally requires quadratic dispersion of the elementary excitations. Numerical evidences(See Chapter 2) clearly indicate no explicit Galilean invariance of the intra-Landau level dynamics. More importantly, the argument in \cite{Son} does not work if no Galilean invariance is assumed even for the IQH fluids.

In this section, a general account of the various response functions in quantum Hall fluids is presented, in which Galilean invariance and rotational invariance are only treated as special cases. In this way, the roles played by different metrics in the systems are made explicit, and contributions from the intra-Landau level dynamics can also be separated explicitly. The response functions allow us to compute the linear response to a spatially varying external electromagnetic perturbation, which is of experimental interest. On a more formal ground, we can also compute the density response to a spatially varying deformation of the effective mass tensor. Throughout our calculation the limit of strong magnetic field is taken. When the Galilean or the rotational invariance is lacking, the corrections to the transport coefficients are important when comparing the theory to experiments, where situations are almost always less ideal. Conceptually, it is important to understand that even though the Hall conductivity with a uniform electric field is robust from topological arguments (requiring no special symmetry), for the Hall viscosity, the relationship to the experiemental measurement is much less universal. In addition, while rotational invariant perturbations do not alter the Hall viscosity as long as the gap is not closed, the quantization of the Hall viscosity to rational values is generally destroyed once rotational invariance is broken by the perturbation.

\subsection{Model and Algebra}

The full Hamiltonian of the many-body system in this section is given by
\begin{eqnarray}\label{ham}
H&=&h_0+V+\delta V
\end{eqnarray}
where the single particle Hamiltonian $h_0$ only depends on $\widetilde R_i^a$ and satisfies inversion symmetry. The most general form is given by Eq(\ref{generalh0}). $V$ is the \emph{translationally invariant} k-body interaction term for $k\ge 2$. $\delta V$ is the external perturbation. The density operator is given by $J^0_q=\sum_ie^{iq_ar_i^a}$. The cyclotron part of the density operator is given by $\widetilde J^0_q=\sum_ie^{iq_a\widetilde R_i^a}$, while the guiding center part is given by  $\bar J^0_q=\sum_ie^{iq_aR_i^a}$

The gauge invariant current density operator $J^a(x)$ is defined as
\begin{eqnarray}\label{defcurrent}
\delta H=-\int d^2xJ^a(x)\delta A_a(x).
\end{eqnarray}
where $a=1,2$ and $A_a$ is the external vector potential. Up to $O(q^2)$ the Fourier component of the current density operator is given by
\begin{eqnarray}\label{current}
J^a_q&=&\frac{1}{N_\phi}\sum_ie^{iq_aR_i^a}e^{\frac{i}{2}q_a\widetilde R_i^a}\hat v^a_i(q)e^{\frac{i}{2}q_a\widetilde R_i^a}\\
\hat v^a_i(q)&=&i[H,r_i^a]-\frac{iq_bq_c}{24}[[[H,r_i^a],r_i^b],r_i^c]+O(q^4)
\end{eqnarray}
We can divide the current into the longitudinal and transverse part:
\begin{eqnarray}\label{jlt}
J^a_{q}=i\left([h_0, J^a_{q,\parallel}]+\epsilon^{ab}q_bJ_{q,\perp}\right)
\end{eqnarray}
The divergence of the longitudinal part gives the electron density: $q_aJ^a_{q,\parallel}=J^0_q$, while the transverse part is divergenceless. Given that the energy scale of $h_0$ is much larger than the energy scale of $V$, it is also conceptually useful to separate the current density operator into the cyclotron ($\widetilde J^a_q$) and guiding center ($\bar J^a_q$) part by writing $h_0=H_0-V$ and $J^a_q=\widetilde J^a_q+\bar J^a_q$ with
\begin{eqnarray}\label{j2parts}
\widetilde J^a_q&=&i\left([H_0,J^a_{q,\parallel}]+\epsilon^{ab}q_bJ_{q,\perp}\right)\label{jc}\\
\bar J^a_q&=&-i[V,J^a_{q,\parallel}]\label{jg}
\end{eqnarray}
Here $\widetilde J^a_q$ contains the energy scale on the order of LL spacing, and one has to be careful not to ignore the LL mixing induced by many-body interactions. We will also see in later sections that for a rotationally invariant $h_0$, $\bar J^a_q$ is well-defined within the projected Hilbert space, and is a function of only the guiding center coordinates.

Treating $\delta V$ as a small perturbation up to the first order, the ground state (of $H$) expectation value of $J^a_q$ is as follows
\begin{eqnarray}\label{exp}
\left( J^a_q,\delta V_{-q}\right)_0\equiv\sum_{n>0}\frac{\langle J^a_q\rangle_{0n}\langle \delta V_{-q}\rangle_{n0}+\langle \delta V_{-q}\rangle_{0n}\langle J^a_q\rangle_{n0}}{\epsilon_0-\epsilon_n}
\end{eqnarray}
and $\langle A\rangle_{mn}=\langle\psi_m|A|\psi_n\rangle$, where $|\psi_m\rangle$ are eigenstates of $H_0$ with unperturbed eigenvalues $\epsilon_n$. $\delta V_{q}$ is the Fourier component of $\delta V$, with translational invariance imposed. 

One has to include counter-terms to impose gauge invariance if $[r_i^a,\delta V]\neq 0$. This gives an additional term for the current density operator:
\begin{eqnarray}\label{counterterm}
\delta J^a_q&=&\sum_ie^{iq_aR_i^a}e^{\frac{i}{2}q_a\widetilde R_i^a}\hat \delta v^a_i(q)e^{\frac{i}{2}q_a\widetilde R_i^a}\\
\delta\hat v^a_i(q)&=&i[\delta V,r_i^a]-\frac{iq_bq_c}{24}[[[\delta V,r_i^a],r_i^b],r_i^c]+O(q^4)
\end{eqnarray}
While the counter-term is important, one can always ignore it first and deal with it at the end of the calculation, where gauge invariance can be restored by general argument or explicit subtractions. Leaving out the counter-term, Eq.(\ref{exp}) gives:
\begin{eqnarray}\label{exp2}
\left( J^a_q,\delta V_{-q}\right)_0&=&i\langle[J^a_{q,\parallel}, \delta V_{-q}]\rangle_0+i\epsilon^{ab}q_b\left( J_{q,\perp},\delta V_{-q}\right)_0-i\left( [V, J^a_{q,\parallel}], \delta V_{-q}\right)_0
\end{eqnarray}
where $\langle\cdots\rangle_0$ is the expectation value of the ground state of $H_0$. Thus the first term of Eq.(\ref{exp2}) only depends on the ground state. The third term only depends on $V$, so we can just evaluate it with the projected Hilbert space defined in Eq.(\ref{projected}), at the cost of introducing an error of $O(\lambda_0)$, which we can ignore in the limit of large magnetic field. 

It is also useful to define the change of the operator  $O$ with respect to the rescaling of the magnetic field to be $dO$. For single particle Hamiltonian $h_0$ that only depends on the cyclotron coordinates, and the projected interaction Hamiltonian $\bar V$ that only depends on the guiding center coordinates, we have 
\begin{eqnarray}\label{dhdbarv}
dh_0&=&-\frac{l_B^3}{2}\sum_i\partial_{l_B}h_0(l_B^{-1}\widetilde R_i^a)=\frac{i}{4}\epsilon_{ab}\sum_i\{\widetilde R_i^a,[\widetilde R_i^b, h_0]\}\label{d1}\\
d\bar V&=&-\frac{l_B^3}{2}\sum_i\partial_{l_B}\bar V(l_B^{-1} R_i^a)=\frac{i}{4}\epsilon_{ab}\sum_i\{R_i^a,[R_i^b, h_0]\}\label{d2}
\end{eqnarray}
For the un-projected interaction $V$ that depends on $r_i^a$, one can define a dilatation operator $D=i\epsilon_{ab}\sum_iR_i^a\widetilde R_i^b$, and we have:
\begin{eqnarray}\label{dv}
&&d V=-\frac{l_B^3}{2}\sum_i\partial_{l_B}V(l_B^{-1} r_i^a)=\frac{1}{2}[D,V]
\end{eqnarray}

\subsection{Response Functions}

The relevant response functions considered here are charge-charge, charge-current and current-current response functions, defined as $\chi^{\mu\nu}_q=\left(J_q^\mu,J^\nu_{-q}\right)_0$ (following the notations of Eq.(\ref{exp})), where $\mu,\nu=0,1,2$. By gauge invariance we have
\begin{eqnarray}\label{response}
&&\chi^{00}_q=\frac{1}{A}\left(J^0_q,J^0_{-q}\right)_0=O(q^2)\\
&&\chi^{a0}_q=\frac{1}{A}\left(J^a_q,J^0_{-q}\right)_0=i\sigma^H(q)\epsilon^{ab}q_b\label{g1}\\
&&\chi^{ab}_q=\frac{1}{A}\left(J^a_q,J^b_{-q}\right)_0=\chi^m(q)\epsilon^{ac}\epsilon^{bd}q_cq_d\label{g2}
\end{eqnarray}
where $A$ is the area of the Hall surface. Here $\sigma^H$ is the local Hall conductivity which is now $q$ dependent, while $\chi^m$ is the ``magnetic Hall conductivity", defined as the coefficient of the current response to the local curl of the magnetic field. From Maxwell's equation $\chi^m(0)$ is the gradient of the local magnetization density. One would like to evaluate these response functions in the long wavelength limit. The incompressibility of the FQHE ensures\cite{gmp}
\begin{eqnarray}\label{incompressibility}
\langle\bar\psi_0|\bar J^0_q|\bar\psi_n\rangle=-\frac{1}{2}q_aq_b\langle\bar\psi_0|\Lambda^{ab}|\bar\psi_n\rangle+O(q^3)
\end{eqnarray}

On the other hand, the cyclotron coordinates are bounded operators as the excitations to higher LLs are energetically suppressed. Similarly defining $\widetilde\Lambda_i^{a_1a_2\cdots a_n}=\frac{1}{n!}\sum_i\mathcal S_{a_1\cdots a_n}\widetilde R_i^{a_1}\cdots \widetilde R_i^{a_n}$, we have
\begin{eqnarray}\label{expansion}
J^a_{q,\parallel}&=&\sum_ie^{iq_aR_i^a}\left(\widetilde R_i^a+\frac{i}{2}q_b\widetilde\Lambda^{ab}_i-\frac{1}{6}q_bq_c\widetilde \Lambda^{abc}_i\right)+O(q^3)\label{e1}\\
J_{q,\perp}&=&\sum_ie^{iq_aR_i^a}\left(dh_i+\frac{i}{3}q_a\{\widetilde R_i^a,dh_i\}\right)+O(q^3)\label{e2}\\
\bar J^a_{q}&=&-\frac{1}{2}\sum_i[\bar V,q_b\Lambda_i^{ab}]+\frac{i\epsilon^{ab}q_b}{2}[V,D]+O(q^2)\label{e3}
\end{eqnarray}

The average angular momentum per particle can be calculated from the ground state: $\frac{1}{A}\sum_i\langle\widetilde \Lambda_i^{ab}\rangle_0=\frac{n_e\widetilde s}{2}\widetilde g^{ab}, \frac{1}{A}\sum_i\langle\Lambda_i^{ab}\rangle_0=\frac{n_e\bar s}{2}\bar g^{ab}$, where $n_e=N_e/A$ is the electron density and $N_e$ is the number of electrons; $\widetilde g^{ab},\bar g^{ab}$ are the cyclotron and guiding center metrics. $\widetilde s, \bar s$ are the cyclotron and guiding center spin, which are related to the cyclotron and guiding center part of the Hall viscosity by $\widetilde\eta=n_e\widetilde s/4, \bar\eta=n_e\bar s/4$\cite{Son} respectively. From Eq.(\ref{e1}) to Eq.(\ref{e3}) we have
\footnotesize
\begin{eqnarray}
\chi^{a0}_q&=&i\nu\epsilon^{ab}q_b\left(1-q_cq_d\left(\frac{\widetilde \eta}{3n_e}\widetilde g^{cd}+\frac{\bar \eta}{n_e}\bar g^{cd}\right)\right)-\frac{i\epsilon^{ab}q_bq_cq_d}{2A}\left(dh_0,\sum_i\widetilde\Lambda_i^{cd}+\Lambda_i^{cd}\right)_0\nonumber\\
&&+\frac{i\epsilon^{ab}q_bq_cq_d}{3A}\sum_{ij}\left(\{\widetilde R_i^c,dh_0\},\widetilde R_j^d\right)_0-\frac{i\epsilon^{ab}q_bq_cq_d}{4A}\sum_i\left([V, D],\Lambda^{cd}_i\right)_0+O(q^5)\label{cch}\\
\chi^{ab}_q&=&\frac{1}{A}\epsilon^{ac}\epsilon^{bd}q_cq_d\left(\left(dh_0,dh_0\right)_0+\langle d^2h_0+dh_0\rangle_0\right)+O(q^4)\label{cc}
\end{eqnarray}
\normalsize

While the current-current response function is relatively simple due to gauge invariance, the charge-current response requires a bit of explanation. The first line of Eq.(\ref{cch}) is universal, though the Hall viscosities do not have to be quantized in any way when rotational invariance is absent. The second line shows the Hall viscosities, cyclotron metric and guiding center metric are renormalized by $dh_0$. This part actually vanishes when $h_0$ has rotational invariance, as we will see in the next section. The third line is the non-universal part that depends on the LL index, and the last line is the contribution from the intra-Landau level dynamics. Eq.(\ref{cch}) and Eq.(\ref{cc}) are the most general expressions for response functions. 

To make explicit connection to possible transport experiments, a small periodic perturbation of the external electrostatic potential with wavelength $q$ is given by $\delta V_q=\frac{\lambda}{2}\left(J^0_q+J^0_{-q}\right)$, where $\lambda$ is a small parameter. Thus the linear current and density response is given by
\begin{eqnarray}\label{ecurrentres}
\delta J^\mu_q=\frac{\lambda}{2}\chi^{\mu 0}_q 
\end{eqnarray}
For a periodic external perturbation of the magnetic field $\delta B=\epsilon^{ab}\lambda_aq_b\sin q_ar^a$ induced by the change in the vector potential $\delta A_a=\lambda_a\cos q_ar^a$, where $\lambda_a$ are small, we have $\delta V_q=-\frac{e}{4}\lambda_a\left(J^a_q+J^a_{-q}\right)$. The linear current and density response is given by
\begin{eqnarray}\label{bcurrentres}
\delta J^\mu_q=-\frac{e}{4}\lambda_a\chi^{\mu a}_q
\end{eqnarray}
Using Eq.(\ref{g1}) and Eq.(\ref{g2}), the explicit expressions for the coefficients of the Hall conductivity and magnetic Hall conductivity can be extracted from Eq.(\ref{ecurrentres}) and Eq.(\ref{bcurrentres}):
\begin{eqnarray}\label{coefficients}
&&\sigma^H(q)=\nu\left(1-q_cq_d\left(\frac{\widetilde \eta}{3n_e}\widetilde g^{cd}+\frac{\bar \eta}{n_e}\bar g^{cd}\right)\right)-\frac{q_aq_b}{2A}\left(dh_0,\sum_i\widetilde\Lambda_i^{ab}+\Lambda_i^{ab}\right)_0\nonumber\\
&&+\frac{q_aq_b}{3A}\sum_{ij}\left(\{\widetilde R_i^a,dh_0\},\widetilde R_j^b\right)_0-\frac{q_aq_b}{4A}\sum_i\left([V, D],\Lambda^{ab}_i\right)_0+O(q^4)\label{c1}\\
&&\chi^m(q)=\frac{1}{A}\langle d^2h_0+dh_0\rangle_0+\frac{1}{A}\left(dh_0,dh_0\right)_0+O(q^2)\label{c2}
\end{eqnarray}
In this most general case, the Hall conductivity up to $O(q^2)$ \emph{does} depend on the intra-Landau level dynamics, as can be seen in the last term of Eq.(\ref{c1}). On the other hand, in the long wavelength limit the magnetic Hall conductivity is completely independent of the details of the interaction. 

\subsection{Rotationally Invariant $h_0$}

We now specialize to the case where the inversion symmetric $h_0$ contains only one metric. In this case the cyclotron angular momentum operator is given by $\widetilde L_z=\frac{1}{2}g_{ab}\sum_i\widetilde R_i^a\widetilde R_i^b$ and we have$[h_0,\widetilde L_z]=0$.  Such rotationally invariant $h_0$ with metric $g_{ab}$ can be generally expressed as
\begin{eqnarray}\label{rh0}
h_0=\sum_i\sum_{n=1}^\infty \frac{c_n}{(2n)!l_B^{2n}}\left(g_{ab}\widetilde R_i^a\widetilde R_i^b\right)^{2n}
\end{eqnarray}
The energies of LLs depend on the expansion coefficients $c_n$ and without Galilean invariance, they are not evenly spaced. The LL wavefunctions, on the other hand, does not depend on $c_n$. They are well-defined by ladder operators $a_i=l_B^{-1}\omega_a\widetilde R_i^a$ with $[a_i,a^\dagger_j]=\delta_{ij}$, where $\omega_a$ is a complex vector given in Eq.(\ref{omega}). With a little algebra one can show that the guiding center current density operator in Eq.(\ref{jg}) can be written as
\begin{eqnarray}\label{guidingj}
\bar J^a_{q}&=&[\bar V,\bar J^a_{q,\parallel}]+\frac{i}{4}\epsilon^{ab}q_b\bar J_{\perp}\\
\bar J^a_{q,\parallel}&=&-\frac{1}{2}q_b\sum_i\Lambda_i^{ab}\label{jgl}\\
\bar J_{\perp}&=&\sum_ig_{ab}[R_i^a,[R_i^b,\bar V]]+4d\bar V\label{jgt}
\end{eqnarray}
where we define the longitudinal and transverse part of the current operator in Eq.(\ref{jgl}) and Eq.(\ref{jgt}) respectively. Note Eq.(\ref{guidingj}) is defined entirely in terms of the projected operators within the projected Hilbert space.

From Eq.(\ref{rh0}) we also have $[dh_0,h_0]=0$, which greatly simplifies the transverse part of the current. In particular, the magnetic Hall conductivity is given by
\begin{eqnarray}\label{rchi}
\chi^m(q)=\frac{1}{A}\langle d^2h_0+dh_0\rangle_0+O(q^2)
\end{eqnarray}
which is just the spatial derivative of the magnetization density. The expression for the Hall conductivity is a bit more complicated:
\begin{eqnarray}\label{rsigma}
&&\sigma^H(q)=\nu\left(1-q_cq_d\left(\frac{\widetilde \eta}{3n_e}g^{cd}+\frac{\bar \eta}{n_e}\bar g^{cd}+\frac{1}{8N_e}\sum_i\left(\bar J_\perp,\Lambda_i^{cd}\right)_0\right)\right)\nonumber\\
&&-\frac{1}{3A}g^{cd}q_cq_d\sum_i\left(n_i\cdot\frac{(dh)_{n_i}+(dh)_{n_i-1}}{\epsilon_{n_i}-\epsilon_{n_i-1}}+(n_i+1)\cdot\frac{(dh)_{n_i+1}+(dh)_{n_i}}{\epsilon_{n_i}-\epsilon_{n_i+1}}\right)
\end{eqnarray}
Here, $n_i$ is the LL index of the $i^{\text{th}}$ electron, and $(dh)_{n_i}=\langle n_i|dh_0|n_i\rangle$. Without Galilean invariance the contribution of each LL to the Hall conductivity is different.

It is also interesting to look at the linear response to small deformation of the metric $g_{ab}\rightarrow g_{ab}(r)=g_{ab}+\delta g_{ab}(r)$ in $h_0$. A periodic deformation of the metric is given by $\delta g_{ab}(r)=\delta\alpha_{ab}\cos q_ar^a$, and the perturbation is given by $\delta V_q=\frac{i}{4}\delta\alpha_{bc}g_{ad}\epsilon^{ab}\left(\partial_{\widetilde q}^cJ^d_q-\partial_{\widetilde q}^cJ^d_{-q}\right)$, where $\partial_{\widetilde q}^c$ is the $q$ derivative that does not differentiate $e^{iq_aR^a}$, i.e. it only takes the derivative of the cyclotron momentum. The density response is given by
\footnotesize
\begin{eqnarray}\label{3express}
&&\delta J^0_q=\frac{\widetilde \eta}{12n_e}\delta\alpha_{ab}\left(\epsilon^{ac}\epsilon^{bd}-g^{ac}g^{bd}\right)q_cq_d\nonumber+\frac{i}{16A}\delta\alpha_{ab}g_{cd}\epsilon^{ac}q_eq_f\sum_{i}\left(\Lambda_i^{ef},[R_i^b,[R_i^d, V]]\right)_0\nonumber\\
&&-\frac{1}{12A}\delta\alpha_{ab}q_cq_d\left(g^{ab}g^{cd}+g^{ac}g^{bd}\right)\sum_i\left(n_i\cdot\frac{(dh)_{n_i}+(dh)_{n_i-1}}{\epsilon_{n_i}-\epsilon_{n_i-1}}+(n_i+1)\cdot\frac{(dh)_{n_i+1}+(dh)_{n_i}}{\epsilon_{n_i}-\epsilon_{n_i+1}}\right)\nonumber\\
\end{eqnarray}
\normalsize

 Interestingly, if we assume that the deformation of the metric preserves area, e.g. $\det g(r)=1$, then up to lowest order in $\delta g_{ab}(r)$, the second term of Eq.(\ref{3express}), or the guiding center contribution, vanishes. Thus the intra-Landau level dynamics does not contribute, regardless of whether or not the system has full rotational symmetry. It is more useful to look at the response in the real space, which gives us
\footnotesize
\begin{eqnarray}\label{3expressrealspace}
\delta J^0(r)=-\left(\frac{\widetilde \eta}{3n_e}+\frac{1}{6A}\sum_i\left(n_i\cdot\frac{(dh)_{n_i}+(dh)_{n_i-1}}{\epsilon_{n_i}-\epsilon_{n_i-1}}+(n_i+1)\cdot\frac{(dh)_{n_i+1}+(dh)_{n_i}}{\epsilon_{n_i}-\epsilon_{n_i+1}}\right)\right)\partial_a\partial_b\delta g^{ab}(r)\nonumber\\
\end{eqnarray}
\normalsize

It is well-known that the Gaussian curvature of a spatially varying unimodular metric is given by $K=-\frac{1}{2}\partial_a\partial_bg^{ab}(r)$ up to the linear order in metric deformation, thus the induced electron density is locally proportional to the Gaussian curvature for small metric deformation. It is instructive to look at the case where $h_0$ is Galilean invariant, and Eq.(\ref{3expressrealspace}) simplifies to a nice formula
\begin{eqnarray}\label{grealspace}
\delta J^0(r)=\frac{2\widetilde \eta}{n_e}K+O(\delta g^2)
\end{eqnarray}
Galilean invariance plays an important role in the universality of the coefficient in front of the Gaussian curvature, which is the cyclotron Hall viscosity. It is also interesting to see if the guiding center density also has a similar relationship with the Gaussian curvature of the guiding center metric. While analytical calculation seems intractable, the issue can be explored numerically and is part of the on-going research.

\subsection{Galilean Invariant $h_0$}

We can impose Galilean invariance on $h_0$ by keeping just the quadratic term in the expansion of Eq.(\ref{rh0}). In this case we can define the cyclotron frequency as $\omega_c=eB/mc$ with an effective mass $m$. With Galilean metric $g_{ab}$ we have
\begin{eqnarray}\label{gh0}
h_0=\sum_i\frac{1}{2ml_B^2}g_{ab}\widetilde R_i^a\widetilde R_i^b
\end{eqnarray}
In this special case there is a nice relationship between $\sigma^H$ and $\chi^m$:
\begin{eqnarray}\label{relation}
\sigma^H(q)=\nu\left(1-\left(ml_B^2\frac{\chi^m(0)}{\nu}g^{ab}-\frac{\widetilde \eta}{n_e}g^{ab}-\frac{\bar \eta}{n_e}\bar g^{ab}\right)q_aq_b\right)+\frac{1}{4A}\left(\bar J_\perp,\bar J^0_{-q}\right)_0+O(q^4)\nonumber\\
\end{eqnarray}
Note the third term in Eq.(\ref{relation}) only depends on the intra-Landau level dynamics, and without further symmetry is non-vanishing even at the order $O(q^2)$
\begin{eqnarray}\label{thirdterm}
\left(\bar J_\perp,\bar J^0_{-q}\right)_0=-\frac{1}{2}q_aq_b\sum_i\left(\bar J_\perp,\Lambda_i^{ab}\right)+O(q^4)
\end{eqnarray}
The transverse part of the guiding center current operator renormalizes the guiding center spin in a non-universal way depending on the details of the interaction. However if we impose full rotational symmetry so that $\bar g_{ab}=g_{ab}$, the interaction Hamiltonian $\bar V$ commutes with the guiding center angular momentum operator $\bar L_z=\frac{1}{2}g_{ab}\sum_i\Lambda_i^{ab}=\sum_i\left(b_i^\dagger b_i+\frac{1}{2}\right)$, where $b_i=l_B^{-1}\omega_a^*R_i^a$. From Eq.(\ref{jgt}) it is easy to see that $[\bar J_\perp, \bar L_z]=0$, so $\bar J_\perp$ only connects the ground state to excited states in the same $\bar L_z$ sector. On the other hand we have
\begin{eqnarray}\label{gspin}
 \sum_i\Lambda_i^{ab}=g^{ab}\bar L_z+\sum_i\left(\omega^a\omega^bb_ib_i+\omega^{a*}\omega^{b*}b_i^\dagger b_i^\dagger\right)
\end{eqnarray}
As the ground state is an eigenstate of $\bar L_z$, $\sum_i\Lambda_i^{ab}$ only connects the ground state to excite states $|\psi_n\rangle$ of a different $\bar L_z$ sector (with $\langle\psi_n|\bar L_z|\psi_n\rangle_n=\pm 2$). Thus with full rotational invariance, Eq.(\ref{thirdterm}) vanishes up to $O(q^2)$ and Eq.(\ref{relation}) is reduced to the same expression first obtained by Hoyos and Son\cite{Son}. 

Numerical tests \cite{rr2} from Laughlin model wavefunctions have confirmed that the guiding center Hall viscosity at filling factor $\nu=1/m$ is $\bar\eta=1$. The same paper presents a few arguments to show that rotational invariant perturbation of the Hamiltonian does not change the Hall viscosity. To complement those arguments, we can easily see from Eq.(\ref{gspin}) that as long as the perturbation $\delta V$ does not break rotational invariance, i.e. $[\bar L_z,\delta V]=0$, it will not change the ground state expectation value of $\sum_i\Lambda_i^{ab}$, up to any order of perturbation. This argument only works if the perturbation theory is applicable, i.e. the perburtation does not close the gap of the FQH state. In particular for Laughlin states, the guiding center spin will not change when we adiabatically move from the model Hamiltonian with pseudopotential interactions to a physically realistic Coulomb interaction, if rotational invariance is preserved.

\section{Summary}

In this chapter, the geometry aspect of the FQHE is illustrated with both the Berry phase and the quantum metric induced by area-preserving deformation of the metric characterizing the quantum fluid. Both single particle wavefunctions and strongly correlated many-body wavefunctions are considered. For many-body wavefunctions that describe a droplet of FQH fluids, the edge contribution has to be carefully removed in order to obtain the bulk contributions. The bulk contribution to the Berry phase leads to an important quantity called the `` guiding center Hall viscosity", which for rotationally invariant systems is a topological invariant constituting the universal part of various eletromagnetic linear responses of the QH fluids in the long wavelength limit. 

The geometric aspect of the FQHE will also be reflected in the next two Chapters, where we will focus on the collective neutral excitations. For the topological phase of the FQHE to be stabilized, the system has to be incompressible. We shall see the bulk neutral gap in the long wavelength limit, or the so-called ``quadrupole gap", is inversely proportional to the guiding center Hall viscosity, and is proportional to the stiffness of the FQH droplet against the guiding center metric deformation. In particular as a sanity check, when the guiding center Hall viscosity approaches zero (i.e. for IQHE), the intra-Landau level dynamics is frozen by Landau level projections, and the gap of the neutral excitations goes to infinity.

\chapter{Model Wavefunctions for the Neutral Excitations\label{ch:cmwave}}

The neutral excitations, or the collective modes of the FQHE defines the incompressibility of the bulk of the topological phase. Experimentally these neutral excitations were explored by several groups\cite{west1,foxon,cheng,west2}). They can be probed numerically by exact diagonalizations on geometries without boundary (e.g. sphere or torus geometry, see Fig.(\ref{fig_torus}) and Fig.(\ref{fig_sphere})). For the Laughlin state, there is only one branch of the low-lying neutral excitations: the magneto-roton mode, first studied with the single mode approximation (SMA) within the projected Hilbert space. Starting from the Laughlin ground state $|\psi_0\rangle$, the SMA model wavefunctions are constructed as density wave excitations:
\begin{eqnarray}\label{densitywave}
\psi_q\rangle=\delta\bar\rho_q|\psi_0\rangle
\end{eqnarray}
where $\delta\bar\rho_q=\bar\rho_q-\langle\rho_q\rangle_0$ is the regularized guiding center density operator satisfying the GMP algebra in Eq.(\ref{gdensity}). In this way, the model wavefunctions are orthogonal to the translationally invariant ground state $\langle\psi_0|\psi_q\rangle=0$, and the variational energy is given by
\begin{eqnarray}\label{smaenergy}
\epsilon_q&=&\frac{\langle\psi_0|\delta\bar\rho_{-q}\mathcal H\delta\bar\rho_q|\psi_0\rangle}{\langle\psi_0|\delta\bar\rho_{-q}\delta\bar\rho_q|\psi_0\rangle}=\frac{\int_0^\infty d\omega\omega S(\omega,q)}{\int_0^\infty d\omega S(\omega,q)}\nonumber\\
&=&\frac{1}{S(q)}\int \frac{d^2q'l_B^2}{(2\pi)^2}V_{q'}\left(S(q'+q)+S(q'-q)-2S(q')\right)\left(2\sin\frac{q\times q'l_B^2}{2}\right)^2
\end{eqnarray}
where in the first line the variational energy is given in terms of the ground state dynamical structure factor $S(\omega,q)$, thus $\epsilon_q$ can be thought of as the average energy of the excitations that couple to the ground state via density fluctuations. In the second line, the two-body interaction $V_q$ is shown explicitly, and $S(q)$ is the ground state static structure factor. Thus $\epsilon_q$ depends on $S(q')$ for the entire range of $q'$, even in the limit of $q\rightarrow 0$. Details of the SMA can be found in \cite{gmp}.

The SMA model wavefunctions give an upper bound for the energies of the neutral excitations. However, they are only reasonably good up to the roton-minimum, at the momentum which is of order $l_B^{-1}$ (see Fig.(\ref{fig_torus})). For the rest of the section we will try to understand why SMA fails, and present a numerical scheme that constructs the model wavefunctions of the neutral excitations for the entire range of the momenta, based on the formalism of Jack polynomials and clustering properties of many-body wavefunctions.

\section{Jack Polynomials and Clustering Properties}

We first review some basic properties of the Jack polynomials, which are members of the vector space spanned by symmetric monomials. Each symmetric monomial is characterized by two numbers: number of variables \(N_e\), and the total degree N, which is the sum of the powers of all variables in the monomial.

The two numbers do not uniquely determine the monomial; we also need to specify how the total degree is distributed among different variables. There are two schemes to represent each monomial with a string of non-negative integers. In the first scheme, the monomial is represented by \([\lambda_1,\lambda_2\cdots ], 0\leq\lambda_i, \sum_i\lambda_i=N_e\), which is called the ``occupation basis". The string is a partition of \(N_e\); \(\lambda_i\) gives the number of variables with power $(i-1)$ in the monomial. In relation to FQH wavefunctions on the disk, where the variables are $z_j$ with $j=1,2,\cdots N_e$, the subscript of $\lambda_i$ is the orbital index. Thus the wavefunction for the $i^{\text{th}}$ orbital is given by $z^{i-1}$. For example, [2,0,0,1] gives \(z_1^3+z_2^3+z_3^3\), [1,1,0,1] gives \(z_1(z_2^3+z_3^3)+z_2(z_1^3+z_3^3)+z_3(z_1^3+z_2^3)\). We can thus label each monomial by its corresponding partition \(\lambda\) so that all monomials are denoted as \(m^s_{\lambda}\).

In the second scheme, the monomial is represented by $[n_1,n_2\cdots n_{N_e}]=\mathcal S\left(z_1^{n_1}z_2^{n_2}\cdots z_{N_e}^{n_{N_e}}\right)$, and now the subscript of $n_i$ is the particle index, same as the subscript of $z_i$; but the symmetrization $\mathcal S$ is only over the particle indices of $z_i$. In this chapter, the monomials will be represented in the \emph{first} scheme unless otherwise stated.

Within the vector space of symmetric monomials there is a non-hermitian Laplace-Beltrami operator:
\begin{eqnarray}\label{lb}
H_{LB} &=& K+V\nonumber\\
K&=&\sum_i(z_i\frac{\partial}{\partial z_i})^2\nonumber\\
V&=& \frac{1}{\alpha}\sum_{i<j}\frac{z_i+z_j}{z_i-z_j}(z_i\frac{\partial}{\partial z_i}-z_j\frac{\partial}{\partial z_j})\nonumber
\end{eqnarray}
The operator conserves \(N_e\) and N. The kinetic term K is diagonal in the monomial basis, while \(Vm^s_{\lambda}=\sum_{\mu}c^{\alpha}_{\lambda\mu}m^s_{\mu}\). Partitions \(\lambda,\mu\) have the same total degree, but \(\mu\) is squeezed from, or dominated by \(\lambda\). Explicitly by squeezing we mean
\begin{eqnarray}\label{squeeze}
\mu=[\mu_1,\mu_2,\cdots]=[\lambda_1,\cdots,\lambda_i-1,\cdots, \lambda_{i+n}+1,\cdots,\lambda_{j-n}+1,\cdots,\lambda_j-1]
\end{eqnarray}
for $2n<j-i$ and an integer $n>0$. We write \(\mu\preceq\lambda\) and mathematically there is a dominance rule that \(\mu\preceq\lambda\) if and only if \(\sum_i \mu_i\geq \sum_i \lambda_i\) for all i. The set of basis \(m_{\mu}\) with \(\mu\preceq\lambda\) is a partially ordered set. The property of K and V allows us to recursively construct the eigenvectors of \(H_{LB}\) in the monomial basis. These eigenvectors are called Jack polynomials (or Jacks). We denote such polynomials as \(J^{\alpha}_{\lambda}(z_1,\cdots z_{N_e})\), and its expansion in the monomial basis is given by $J^\alpha_\lambda=m^s_\lambda+\sum_{\kappa<\lambda}c^{\alpha}_{\lambda\kappa}m^s_{\kappa}$, where the coefficient of $m^s_\lambda$ is normalized to \emph {one}, and all $m^s_\kappa$ are dominated by $m^s_\lambda$: $\kappa\preceq\lambda$. The coefficients of expansion are given by 
\begin{eqnarray}\label{jalpha}
c_{\lambda\kappa}^{\alpha}=\frac{2/\alpha}{\rho_{\lambda}(\alpha)-\rho_{\kappa}(\alpha)}\sum_{\kappa\leq\mu<\lambda}((\kappa_i+t)-(\kappa_j-t))c_{\mu\kappa}^{\alpha}\nonumber
\end{eqnarray}
where \(\kappa=[\kappa_1,\cdots \kappa_i,\cdots, \kappa_j,\cdots], \mu=[\mu_1,\cdots \mu_i+t,\cdots, \mu_j-t,\cdots]\), and \[\rho_{\lambda}(\alpha)=\sum_i\lambda_i(\lambda_i-1-\frac{2}{\alpha}(i-1))\]
Thus after fixing $N_e$ and $N$, the Jack polynomial is characterized by a root configuration $\lambda$, and it consists of only basis squeezed from $\lambda$, with coefficients determined by $\alpha$ through Eq.(\ref{jalpha}). In the limit $\alpha\rightarrow 0$ the Jack polynomials reduce to the monomials of the root configuration: $\lim_{\alpha\rightarrow 0}J^\alpha_\lambda=m^s_\lambda$, which are orthogonal to each other. When $\alpha$ is non-zero, the Jack polynomials deform into the squeezed basis, and they are in general not orthogonal anymore (Note Eq.(\ref{lb}) is not Hermitian).

The coefficients $c^\alpha_{\lambda\kappa}$ are highly structured. It can be shown from the recursion relation that it has a product rule\cite{ronny}. While in principle $\alpha$ can be complex, here we only consider the case when $\alpha$ is real. When \(\alpha\) is positive, the coefficients of Jacks are well-defined. On the other hand, the wavefunctions for FQHE are all constructed with negative $\alpha$. When \(\alpha\) is negative, it is not always true that all of the coefficients of expansions are well-defined. 

It was shown in\cite{haldane5} that for negative rational \(\alpha=-\frac{k+1}{r-1}\), with k, r integers and $\left(k+1, r-1\right)$ coprime, \(J^{\alpha}_{\lambda}\) is well-defined if \(\lambda\) is $(k,r,N)$ admissible: there are no more than $k$ ``particles" in r consecutive ``orbitals" in the root configuration. These Jack polynomials form a basis for the space of symmetric polynomials with the clustering property that it vanishes when $k+1$ variables coincide.

The criterion shown above does not exhaust all well-defined Jack polynomials for negative \(\alpha\)\cite{bernevigjack}. Define a partition \([n_00^{s(r-1)}n(\lambda_{k,r})]\) to be (k,r,s,N) admissible when \(n_0=(k+1)s-1\) and \(n(\lambda_{k,r})\) is \((k,r,N-n_0)\) admissible. Denote the partition by \(\lambda_{k,r,s}\) and \(J^{\alpha}_{\lambda_{k,r,s}}\) is well-defined. Moreover, \(J^{\alpha}_{\lambda_{k,r,s}}\) satisfies HW condition when  \(\lambda_{k,r}=[k0^{r-1}k0^{r-1}k\cdots]\).

This new set of Jack polynomials corresponds to symmetric polynomials with new clustering properties reflected in the root configuration (the result is not proven but checked numerically):

1) The polynomial vanishes when s clusters of $k+1$ particles are formed, but remains finite when only $s-1$ or fewer clusters of $k+1$ particles are formed.

2) The polynomial is finite when $(k+1)s-1$ particles are at the same point (but vanishes for more particles because of the first clustering property). Letting \(z_1=z_2\cdots z_{(k+1)s-1}=Z\), we have \(J^{\alpha}_{\lambda_{k,r,s}} \sim \prod_{i=s(k+1)}^N (Z-z_i)^{(r-1)s+1}\).

Unlike (k,r,N) admissible states, (k,r,s,N) states do not span the space of symmetric polynomials with the above clustering properties.

Before ending this short introduction of Jack polynomials, one should note that the Jack polynomials are symmetric with the particle indices, which is appropriate for bosonic FQHE wavefunctions. To obtain fermionic FQHE wavefunctions, one uses the fact that for any symmetric monomial $m^s_\lambda(z_1,\cdots, z_n)$, there is a one-to-one mapping to an antisymmetric monomial $m^a_\lambda(z_1,\cdots, z_n)=\prod_{i<j}^n(z_i-z_j)m^s_\lambda(z_1,\cdots,z_n)$: the multiplication of the Vandermonde (given by $\prod_{i<j}(z_i-z_j)$) is all that is needed. Thus throughout this chapter, an ``antisymmetric Jack polynomial'' is always obtained from multiplying the Vandermonde determinant to the symmetric Jack polynomial, parametrized by the same $\alpha$. 

\section{Numerical Construction of Neutral Excitations}

In this section, we only treat fermionic FQHE wavefunctions, while the corresponding bosonic FQHE wavefunctions can be trivially obtained by dividing the factor of the Vandermonde determinant. The wavefunctions will be expanded in terms of the occupation basis. A typical occupation basis is a string of binary numbers like $[1001001001]$, where $\lambda_i$ now can only be either $1$ (occupied) or $0$ (unoccupied). The total degree $N$ now corresponds to the total angular momentum of the wavefunction (where the vaccum is set to have zero angular momentum). For example, a state $|1001001001\rangle$ with four particles has a total angular momentum $0+3+6+9=18$. An example of the mapping between the monomial wavefunctions and occupation basis is shown as follows:
\begin{eqnarray}\label{mtooc}
z_i^3-z_2^3\sim |1001\rangle,\quad z_1^2z_2-z_1z_2^2\sim |0110\rangle
\end{eqnarray}
For the ground state and quasihole states of the Read-Rezayi series (including the Laughlin and Moore-Read states), the model wavefunctions are Jack polynomials. Quasielectron states, on the other hand, are more complicated~\cite{bernevigjack} because they contain local defects where electrons are forced to get closer to each other than allowed in the ground state. The same difficulty arises in neutral excitations which consist of quasielectron-quasihole pairs. However, we do assume the excitations are local defects of the ground state root configuration, and each model wavefunction for the neutral excitations can be expanded into its root configuration and the corresponding squeezed basis. This defines the Hilbert space of each model wavefunction; the next step is to determine the coefficients of these squeezed basis.

\subsection{Magneto-roton Mode at $\nu=1/3$}

The explicit set of root configurations for the magneto-roton mode is shown as follows: 
\begin{eqnarray}\label{laughlinwfs}
&&\d{1}\d{1}0\textsubring{0}\textsubring{0}01001001001001\cdots    L=2\nonumber\\
&&\d{1}\d{1}0\textsubring{0}010\textsubring{0}01001001001\cdots    L=3\nonumber\\
&&\d{1}\d{1}0\textsubring{0}010010\textsubring{0}01001001\cdots    L=4\nonumber\\
&&\d{1}\d{1}0\textsubring{0}010010010\textsubring{0}01001\cdots    L=5\nonumber\\
&&\d{1}\d{1}0\textsubring{0}010010010010\textsubring{0}01\cdots    L=6\nonumber\\
&&\vdots
\end{eqnarray} 

The states are labeled by their total angular momentum $L$ on the sphere, though once the single particle normalization is removed the wavefunctions are suitable for any genus-0 manifold (e.g. sphere/disk/cylinder\cite{qhecylinder}). In Eq.(\ref{laughlinwfs}) the black dot schematically indicates the position of a quasielectron, while the white dot that of a quasihole. To determine the position of a quasiparticle, one can look at any three consecutive orbitals in the root configurations above, and count the number of electrons to see if it violates the ground state clustering property. In this particular case, if there is more (less) than one electron in every three consecutive orbitals, we have a quasielectron (quasihole), which is located right below the middle of the three consecutive orbitals. 

From the root configuration we can see that in the long wavelength limit, i.e. when $L$ is small, the quasihole merges with the quasielectrons at the north pole (to the left of the root configuration), creating a quadrupole excitation. Note the ground state is in the sector of $L=0$, and the neutral excitations start at $L=2$, thus in the long wavelength limit the quadrupole excitation can be viewed as a ``spin-2 graviton".

Due to rotational invariance on the sphere, we impose the highest weight condition on the wavefunctions $|\psi_\lambda^L\rangle$ to single out from the degenerate states the one with quasiparticles piled up at the north pole. On the disk this also means picking out a state with no center-of-mass rotation:
\begin{eqnarray}\label{hwt}
\nonumber  L^+|\psi^L_\lambda\rangle=0,\\
|\psi^L_\lambda\rangle=\sum_{\mu\preceq\lambda}a_\mu m_\mu
\end{eqnarray}
where $m_\mu$ is the monomial with partition $\mu$~\cite{haldane5}. The summation is over all partitions $\mu$ that can be squeezed from the root configuration $\lambda$. The constraints in Eq.(\ref{hwt}) substantially reduce the Hilbert space dimension (e.g., the basis dimension is less than 20 for 10 particles. A formal explanation of this issue will be presented in Chapter 5). The $L=1$ state actually vanishes, when we impose the constraints to generate the coefficients of expansions. For now we take it as a numerical observation; more insight will be shed on this issue in Chapter 5. The resulting lowest-energy eigenstates of the Hamiltonian, restricted to this Hilbert space, are very good approximations to the exact magneto-roton mode.

Instead of exact diagonalization within the restricted Hilbert space, we follow the spirit of Jack polynomials and impose the following constraint:
\begin{eqnarray}\label{hwt2}
\hat{V}_1c_1c_2|\psi_\lambda^L\rangle=0.
\end{eqnarray}
Here $\hat{V}_1$ is the operator corresponding to the first Haldane pseudopotential of which the Laughlin state is the exact zero energy state, and $c_i$ annihilates an electron at the $i$th orbital. This additional constraint renders $|\psi^L_\lambda\rangle$ unique by enforcing the following clustering property: the wavefunction is vanishing only when two or more clusters of two particles coincides in the real space.

The resulting implementation is numerically much less expensive, with variational energies only slightly above the lowest energy state obtained in the Hilbert space defined by only constraints Eq.(\ref{hwt}), and improving with the increase in system size. Two features of the resulting wavefunction is worth noting: firstly, when the geometric normalization factors of the single particle orbitals on the sphere are removed, the coefficients of the decomposition in the Fock space are integers, with the coefficient of the root configuration normalized to one; secondly, there is a ``product rule"~\cite{productrule,ronny} if the first five orbitals are treated as one ``big" orbital, which allows us to generate a large subset of coefficients recursively. An approximation to $|\psi_\lambda^L\rangle$ can be built from the product rules; the overlap between the approximate and exact model wavefunctions is high and increasing with system size (see Table I). The approximate state is thus used as the seed state, or the initial trial state for the Lanczos procedure that imposes the highest weight condition. The use of the approximate state as the seed state can reduce the computing time by a factor of four.

\begin{table}\label{overlaps}
\caption{ The overlap of the approximate model wavefunctions constructed from product rules and the true model wavefunctions.}
\begin{tabular}{|l|l|l|l|l|}
\hline
No. of electrons & 9 &10 & 11 & 12 \\ \hline
L=2 & 89.83\% & 90.13\% & 90.31\% & 90.42\% \\ \hline
L=3 & 86.42\% & 86.99\% & 87.37\% & 87.63\% \\ \hline
L=4 & 83.63\% & 84.59\% & 85.23\% & 85.69\% \\ 
\hline
\end{tabular}
\end{table}

\begin{figure}
\centerline{\includegraphics[width=15cm,height=12cm]{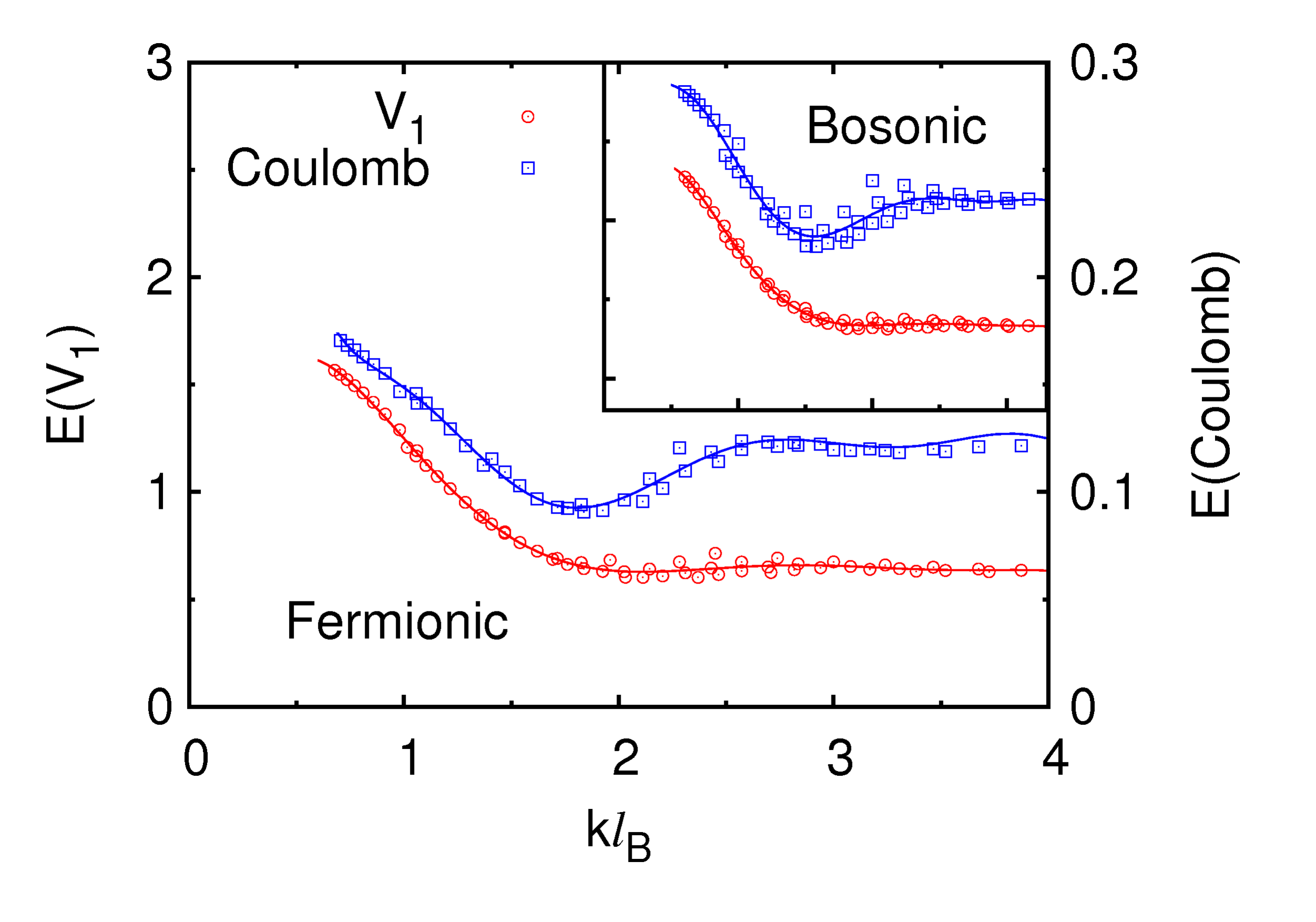}}
\caption{The variational energy of the model wavefunctions defined by Eqn (2) and (3), against $V_1$ (left axis, arbitrary units) and Coulomb Hamiltonian (right axis, in units of $e^2/\epsilon \ell_B$), plotted as a function of momentum. The data is generated from system sizes ranging from 6 to 12 electrons (the inset shows the same plot for the bosonic Laughlin state).}
\label{fig:laughlin}
\end{figure} 

To check how good our model wavefunctions are, we can evaluate their variational energies and compare them with exact diagonalization. In Fig.~\ref{fig:laughlin}, the variational energies are plotted versus momentum $k=L/\sqrt{S}$, where $N_{\rm orb}=2S+1$ is the number of orbitals in the LLL. This is how the linear momentum is obtained from the angular momentum on the sphere. We include the data for a number of system sizes and rescale the magnetic length $\ell_B$ by a factor $\sqrt{S/N_{\rm orb}}$ to minimize the finite size effects. For the model $V_1$ Hamiltonian and Coulomb Hamiltonian, the dispersion obtained using the model wavefunction is in excellent agreement with the results from exact diagonalization, both in small $k$ and large $k$ regime. The model wavefunctions compare favorably with the exact diagonalization eigenstates, with 99\% overlap for 10 electrons. 

\subsection{Neutral Excitations at $\nu=1/2$}

The same approach can be used to construct the neutral excitation wavefunctions for the entire Read-Rezayi series, once the root configuration is identified. For the magneto-roton mode of the Moore-Read state at filling factor $\nu=1/2$, the root configurations are given by

\begin{figure}[H]
\centerline{\includegraphics[width=6cm,height=3cm]{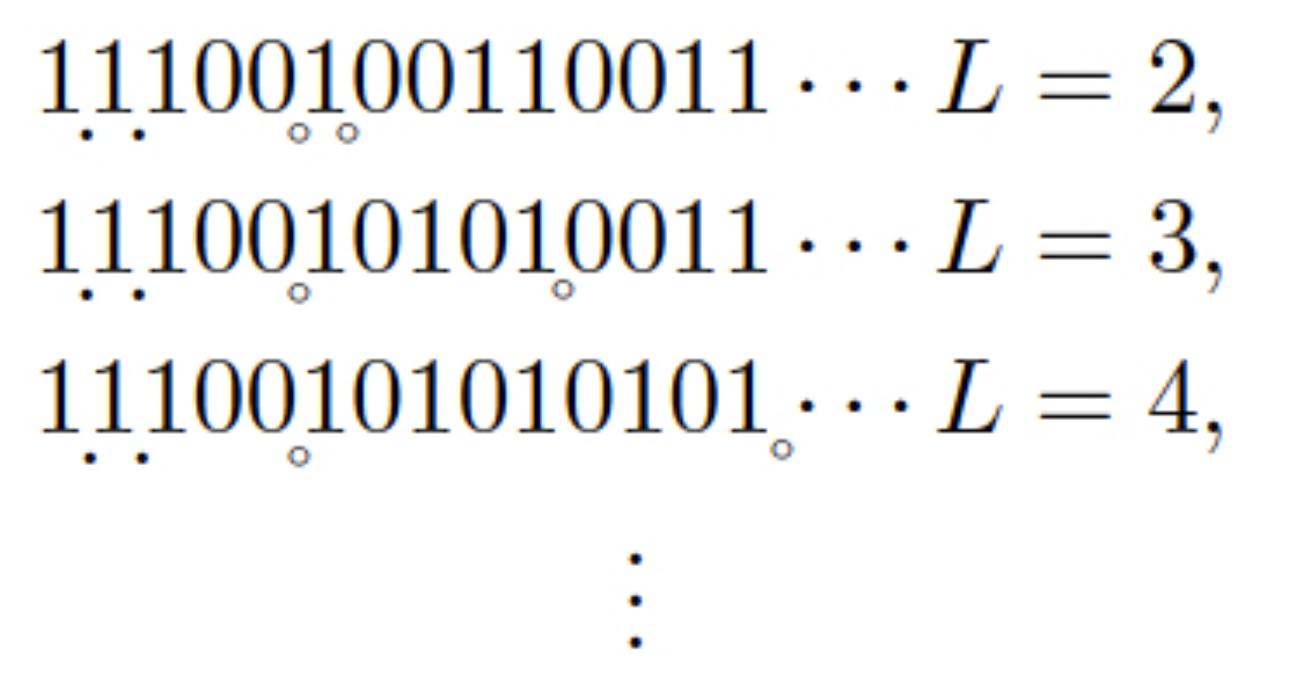}}
\label{fig:mr}
\end{figure} 

In addition to the magneto-roton mode, there is another branch of neutral excitations for MR state due to its non-abelian nature. This is the so called neutral fermion mode\cite{readgreen,moller}, which can be physically interpreted as the breaking of paired particles in the ground state. The root configurations of the neutral fermion mode thus contain an odd number of particles, and are shown as follows:

\begin{figure}[H]
\centerline{\includegraphics[width=6cm,height=3cm]{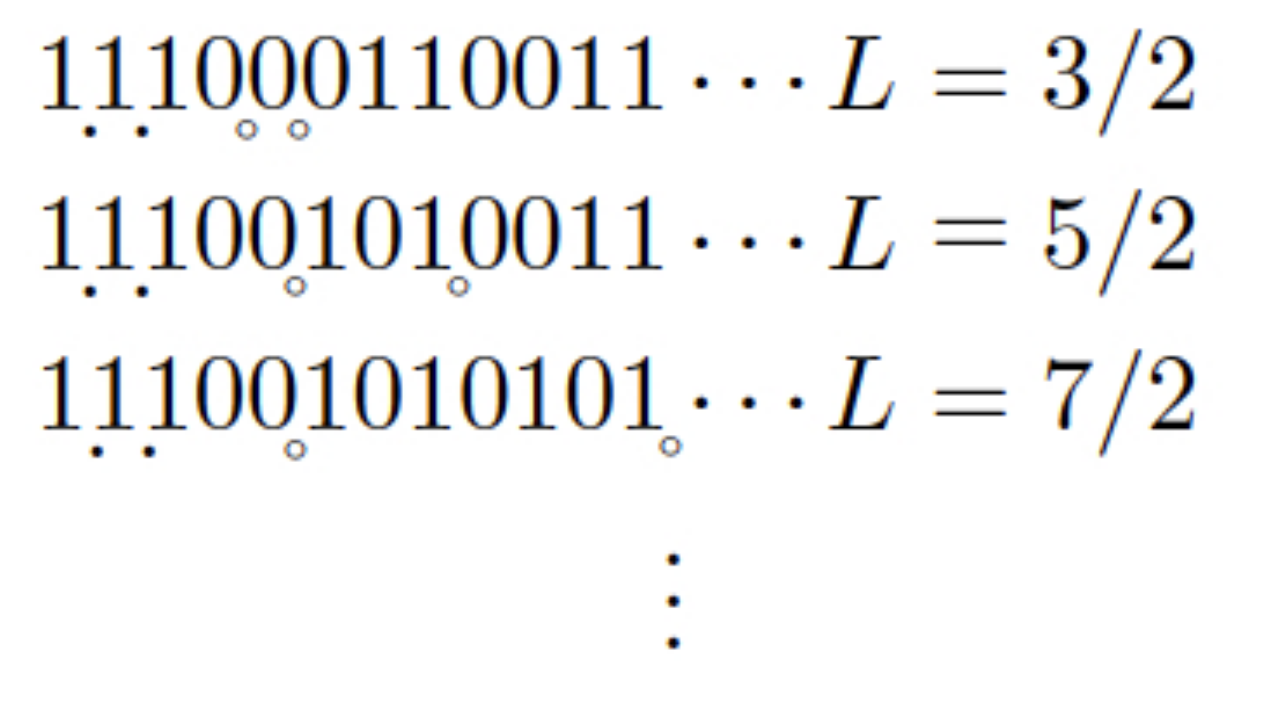}}
\label{fig:nf}
\end{figure}

The Moore-Read ground-state root configuration is given by 2 electrons in 4 consecutive orbitals~\cite{haldane5}. Similarly to the Laughlin state, any deviation from the uniform background density yields the position of the quasihole/quasielectron. Unique model wavefunctions can be constructed by imposing the constraint Eq.(\ref{hwt}), and in addition a modified constraint Eq.(\ref{hwt2}) that reads $H_{\rm 3b}c_1c_2c_3|\psi_\lambda^{\rm MR}\rangle=0$, where $H_{\rm 3b}$ is the Moore-Read three-body model Hamiltonian. Their variational energies are plotted in Fig.~\ref{fig:mr}.

\begin{figure}
\centerline{\includegraphics[width=15cm,height=12cm]{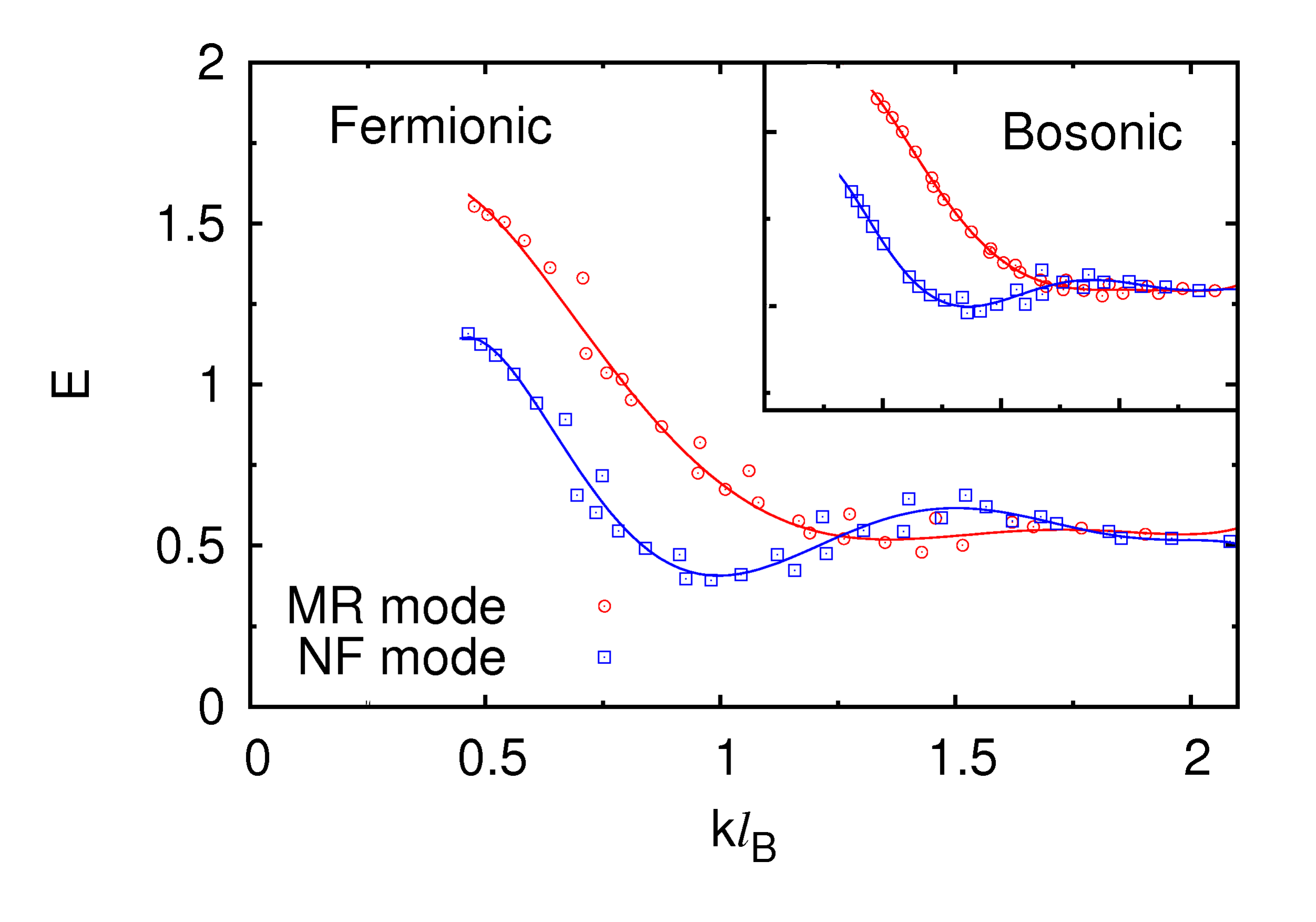}}
\caption{The variational energy of the model wavefunctions for the magneto-roton (MR) mode and the neutral fermion (NF) mode, evaluated against the 3-body Hamiltonian. The data is generated from system sizes ranging from 5 to 17 electrons, where the odd number of electrons contribute to the NF mode, and the even number of electrons contribute to the magneto-roton mode. (The inset shows the same plot for the bosonic Moore-Read state)}
\label{fig:mr}
\end{figure}

It should be emphasized that for both the Laughlin and Moore-Read case, the neutral excitations enter the multi-roton continuum in the long wavelength limit. The continuum starts at the energy that is double the energy gap of the roton minimum. While this makes exact diagonalization ambiguous, the root configurations give clear physical interpretations for the modes for the entire momentum range. Again, the $L=1$ state (and $L=1/2$ state for the neutral fermion mode) vanishes with the set of constraints we impose. Thus in the long wavelength limit, the NF mode can be identified as a spin-$\frac{3}{2}$ ``gravitino", or the ``supersymmetric partner" of the ``graviton" in the magneto-roton mode. 

\section{Validity of SMA}

\begin{figure}
\centerline{\includegraphics[width=15cm,height=12cm]{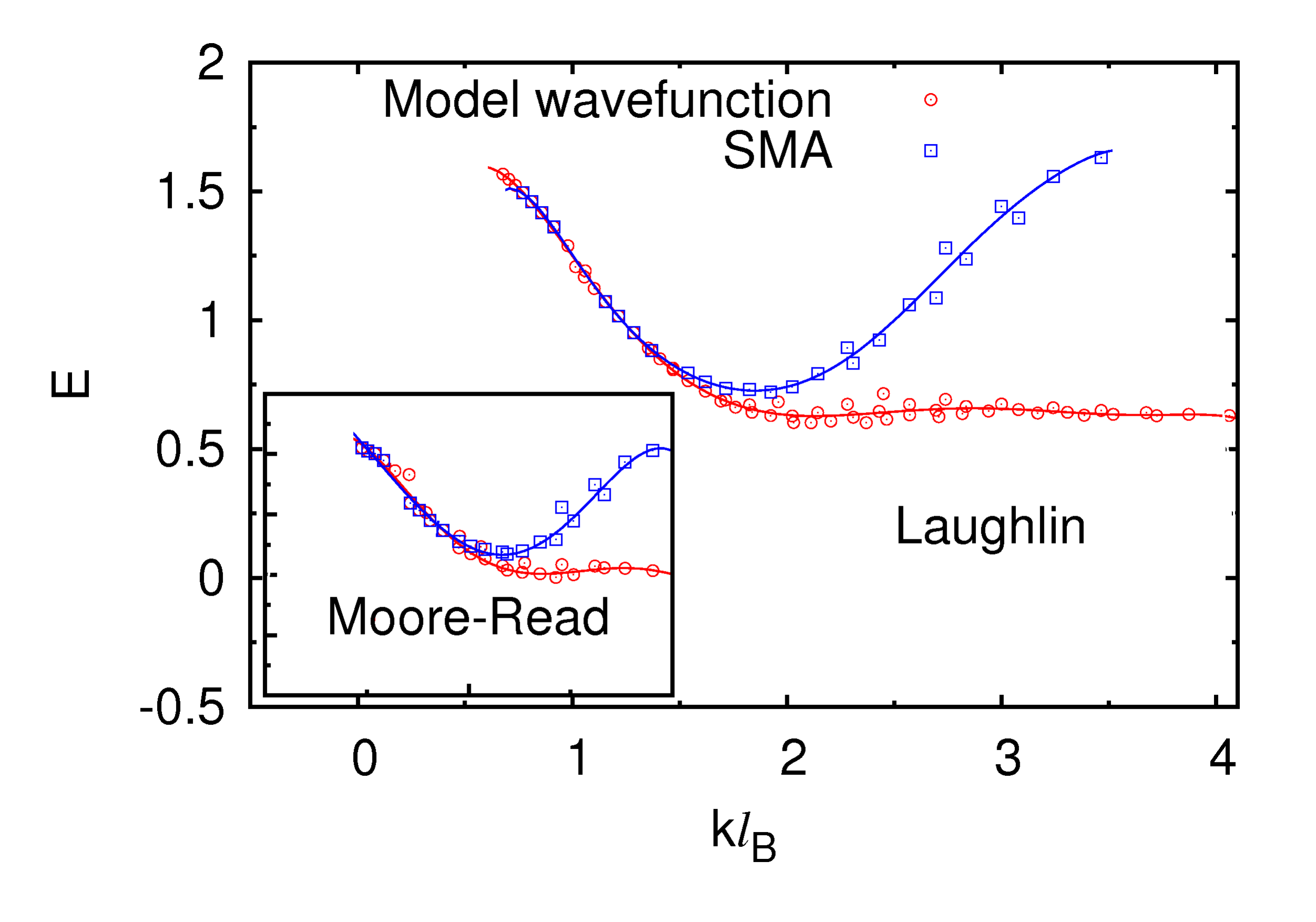}}
\caption{(Color online). The variational energies for the SMA model wavefunctions compared to our model wavefunctions for Laughlin state at $\nu=1/3$ filling. (The inset shows the same comparison for the magneto-roton mode of the Moore-Read state)}
\label{fig:sma}
\end{figure}

It is very instructive to compare the model wavefunctions gerenated in the previous section with SMA wavefunctions obtained from the ground state $|\psi_0\rangle$ by the guiding center density modulation in Eq.(\ref{densitywave}). The SMA yields excitation energies manifestly depends on the guiding center structure factor the ground state. On the sphere, the ground state has the total angular momentum $L=0$, and the SMA wavefunction with total angular momentum $L$ is obtained by boosting one electron with orbital angular momentum $L$. The projection into the LLL is equivalent to the projection of the boosted single-particle state into the sub-Hilbert space of the total spin $S$. Formally we have
\begin{eqnarray}\label{smawf}
|\psi^{\rm SMA}_{LM}\rangle=\sum_i \hat{C}^{S,L,S}_{m_i+M,M,m_i}|\psi_0\rangle,
\end{eqnarray}
where $i$ is the electron index, and $\hat{C}^{S,L,S}_{m',M,m}$ is defined by its action on the single electron state $\hat{C}^{SLS}_{m'Mm}|m\rangle=C^{SLS}_{m'Mm}|m'\rangle$, where  $C^{SLS}_{m'Mm}=\langle m'|\hat{Y}^{LM}|m\rangle$ are the Clebsh-Gordon coefficients, and $\hat{Y}^{LM}$ are the spherical harmonics. This is a result of the Wigner-Eckart Theorem, and due to rotational invariance we can set $M=L$ in Eq.(\ref{smawf}). The dispersion of the SMA wavefunctions is plotted in Fig.~\ref{fig:sma} along with that of our model wavefunctions. 

For small momenta the variational energies of the two classes of wavefunctions agree very well, while the SMA mode evidently becomes invalid for momenta larger than the magneto-roton minimum. Note that at $L=2,3$ the SMA wavefunctions only involve the elements of the basis squeezed from the same root configuration that defines our model wavefunctions. Taking the Laughlin $1/3$ as an example, we now prove the SMA wavefunctions are actually identical to the model wavefunctions at $L=2,3$. By the product rule of the Jack polynomial, we can write
\begin{eqnarray}\label{product}
 |\psi_0\rangle\sim J^\alpha_{\lambda_1}\otimes J^\alpha_{\lambda_2}+J^\alpha_{\lambda_3}\otimes J^\alpha_{\lambda_4}+|\bar{\psi}_0\rangle,
\end{eqnarray}
where the relative coefficients between different terms are ignored because they are unimportant for the proof. The partitions $\lambda_1=[10010],\lambda_2=[01001001\cdots],\lambda_3=[10001],\lambda_4=[10001001\cdots]$, and $|\bar{\psi}_0\rangle$ involves the rest of the squeezed basis. It is easy to check that $c_1c_2\sum_i\hat{C}^{S,L,S}_{m_i+L,L,m_i}|\bar{\psi}_0\rangle=0$. We thus have
\begin{eqnarray}\label{v1product}
\hat{V}_1c_1c_2|\psi^{\rm SMA}_{LL}\rangle&\sim&\hat{V}_1[0000]\otimes (J^\alpha_{\lambda_2}+tJ^\alpha_{\lambda_4})=0
\end{eqnarray}

Again, the coefficients are suppressed in Eq.(\ref{v1product}), and $t=0$ for $L=2$. Thus the SMA wavefunctions satisfy exactly the same constraints as the model wavefunctions, which makes them identical. Note that for $L>3$ the SMA wavefunctions contain unsqueezed basis components with respect to the root configurations used in our model wavefunctions, and the proof breaks down.  

It is also clear from the construction of the model wavefunctions why SMA ceases to give good upper-bound for the neutral excitations at large momenta. The nature of the neutral excitations at large momenta is characterized by a dipole excitation of a quasihole-quasielectron pair. The separation of the quasihole and quasielectron is proportional to the momentum. Heuristically, this is because the Lorentz force on the opposite charges is proportional to the momentum and tends to pull them apart. In SMA the momenta of the neutral excitations are given by the boost of a single particle. The construction in the previous section shows the neutral excitations are actually many-body excitations, whereby the momentum of the state is shared by particles between the quasihole-quasielectron pair. At large momenta the number of particles involved in the neutral excitations is also large, making SMA an increasingly bad approximation.

\section{Geometric Interpretation in the Long Wavelength Limit}

From Eq.(\ref{smaenergy}) the variational energy of the SMA in the long wavelength limit is given by
\begin{eqnarray}\label{smalenergy}
\epsilon_{q\rightarrow 0}&=&\frac{G^{abcd}q_aq_bq_cq_d}{\Gamma^{abcd}q_aq_bq_cq_d}\\
G^{abcd}&=&-\int\frac{d^2q'd^2q''}{(2\pi)^4}V(q')(S(q'')-S_\infty)e^{iq'\times q''}\epsilon^{ag}\epsilon^{bh}\epsilon^{ce}\epsilon^{df}q'_eq'_fq''_gq''_h\\
\Gamma^{abcd}&=&N_\phi^{-1}\left( \left<\{\Lambda^{ab},\Lambda^{cd}\}\right>-2\left<\Lambda^{ab}\right>\left<\Lambda^{cd}\right>\right)\label{guidingcenterspin})
\end{eqnarray}
Here $N_\phi$ is the total number of flux quanta. Eq.(\ref{guidingcenterspin}) is the $O(q^4)$ coefficient of the guiding center structure factor\cite{haldane8}, which we know is the guiding center spin. Thus the guiding center spin controls the gap of the neutral excitations in the long wavelength limit. In particular, the guiding center spin of the IQHE is zero, and the gap goes to infinity. This should be the case, since the guiding center dynamics are frozen in the IQH.

The area-preserving diffeomorphism generator $\Lambda^{ab}$ defined in Eq.(\ref{apd}) can also be used to deform the guiding center metric with a unitary operator $U(\alpha)$ as shown in Eq.(\ref{deformation}). In the limit of small deformation, we have $\lambda^c_d=\delta^c_d+\epsilon^{ac}\alpha_{ad}$, and the variational energy is:
\begin{eqnarray}\label{apdenergy}
\langle\psi_\alpha|H|\psi_\alpha\rangle=\frac{1}{2}G^{abcd}\epsilon^e_a\epsilon^f_d\alpha_{ec}\alpha_{fd}
\end{eqnarray}
This is precisely the energy gap as $q\rightarrow 0$, when Eq.(\ref{apdenergy}) is given by an infinitesmal uniform deformation of the guiding center metric (which defines the shape of the FQH droplets). The four tensor $G^{abcd}$ can thus be interpreted as the  ``guiding center shear modulus ". Since we know the SMA describes neutral excitations exactly in the long wavelength limit, one can now understand the quadrupole excitation as the energy cost of deforming the guiding center metric. It is gapped from the ground state because of the zero-point quantum fluctation from the non-commutivity of the guiding center coordinates. This reinforces the idea that the quadrupole excitation can be identified as a ``spin-2" graviton, with its dynamics controled by the geometry of the FQH droplets.

\section{Summary}

This Chapter presents a numerical scheme to generate the model wavefunctions for the neutral bulk excitations in both the Laughlin and Moore-Read state. The high overlap between the model wavefunctions with those obtained with exact diagonalization suggests the physics of the neutral excitations both for the magneto-roton mode and the neutral fermion mode is well captured by the root configurations of the model wavefunctions. The comparison of the model wavefunctions with the single mode approximation (SMA) shows that SMA gives the correct model wavefunctions only in the long wavelength limit. With that insight the quadrupole gap can be shown to be related to the energy cost of a uniform deformation of the guiding center metric.

Note the model wavefunctions generated in the chapter are not exact eigenstates of any known model Hamiltonians (in particular, they are not the exact eigenstates of the model Hamiltonians of the Read-Rezayi series). One should, however, think of these model wavefunctions as the prefered basis in describing the dynamics of the FQHE, since physical interactions in a two-dimensional system is non-universal and dependent on many experimental conditions (unlike in high energy physics where the presence of Lorentz invariance strongly constrains the physical systems, leading to ``much cleaner" theories as compared to condensed matter physics). The magneto-roton mode and the neutral fermion mode can be treated as the ``elementary excitations" of the FQHE that in the long wavelength limit merges into the multi-roton continuum. The interaction Hamiltonian (either Coulomb interaction or the more artificial pseudopotential interactions) will cause these elementary excitations to scatter and decay, but it is conjectured based on both numerical and analytic (see next chapter) evidences that they have appreciable lifetime in the thermodynamic limit for physically realistic systems, as long as the FQHE incompressible phase persists.

\chapter{Analytical Wavefunctions for the Collective Modes\label{ch:cman}}

The model wavefunctions numerically generated in the previous chapter lead to several interesting observations. Firstly, the model wavefunctions for both the magneto-roton mode and the neutral fermion mode seem to agree exactly with those generated in \cite{jain1,jain2,rodriguez}. In the latter, the neutral excitations of the FQHE are mapped to the excitons of the IQHE made of composite fermions (CF). The many-body wavefunction of an exciton in the IQH is given by
\begin{eqnarray}\label{iexciton}
\phi(z_1,z_2,\cdots,z_N,z_1^*,\cdots z_N^*)=\mathcal A[z_1^0z_2^1\cdots z_l^{l-1}\left(z_{l+1}^*z_{l+1}^{m+l+1}\right)z_{l+2}^{l+1}\cdots z_N^{N-1}]
\end{eqnarray}
where we have $N-1$ particles in the LLL and one particle in the first Landau level (1LL), and the exciton has total angular momentum $L=m$ relative to the ground state. To form the composite fermion wavefunction at $\nu=1/k$, $k-1$ fluxes are attached to each particle:
\begin{eqnarray}\label{cf}
\phi_{\text{CF}}=\phi\prod_{i<j}(z_i-z_j)^{k-1}
\end{eqnarray}
The model wavefunctions for the magneto-roton mode are obtained by projection into the LLL:
\begin{eqnarray}\label{cfmagnetoroton}
\phi_{L=m}=\mathcal P_{LLL}[\phi_{\text{CF}}]
\end{eqnarray}
For the mangeto-roton mode and neutral fermion mode of the MR state, a bipartite CF picture is employed\cite{jain2}, but the scheme is essentially the same. The resultant wavefunctions are numerically found to be identical to those generated in the previous chapter. This is an intriguing fact, as the underlying physical pictures for these two approaches are quite different.

The numerical comparison between the two sets of model wavefunctions are made easy because of the second interesting fact about these model wavefunctions: if we strip away the single particle normalizations of the many-body wavefunctions expanded in the occupation basis, followed by normalizing the coefficient of the root configuration to unity, all the coefficients of the neutral excitations are integers for the Laughlin states, and integers or rational numbers for the MR states. This is generally characteristic of Jack polynomials, where all coefficients are essentially generated via combinatorics. However, as is emphasized in the previous chapter, model wavefunctions for the neutral excitations are not Jack polynomials.

The last interesting observation is the existence of a product rule very similar to Jack polynomials, as shown in the previous chapter. One would then conjecture these wavefunctions are very similar to the Jack polynomials. Just as the ground states and charged excitations of the FQHE, there are natural ways of writing down wavefunctions for the neutral excitations requiring no variational parameters whatsoever. This is indeed the case, as will be illustrated in the next few sections.

\section{Algebraic Structures of the Edge Neutral Excitations}

It is instructive to first explore the algebraic properties of the edge neutral excitations. The edge neutral excitations correspond to the bulk charged quasihole excitations; both are obtained by inserting fluxes into the ground state of the quantum Hall fluid. Labeling the edge excitations by $\delta L$, with the ground state taken as $\delta L=0$, the number of edge excitations in each momentum sector is given by the number of ways of inserting quasiholes in the bulk. Thus the counting of the edge states at $\delta L=N$ is given by the partition number $P(N)$. For finite systems, the counting is only valid for $N\le N_e$, where $N_e$ is the number of particles. Let us take the Laughlin state as an example. For model Hamiltonians, the subspace of the edge modes in each momentum sector is spanned by Jack polynomials satisfying the admission conditions\cite{haldane5}. The root configurations for a few momentum sectors are listed as follows:

\begin{eqnarray}\label{edgejacks}
\delta L=1:\quad&&\cdots 10010010010010001\nonumber\\
\delta L=2:\quad&&\cdots 100100100100100001\nonumber\\
&&\cdots 100100100100010010\nonumber\\
\delta L=3:\quad&&\cdots 1001001001001000001\nonumber\\
&&\cdots 1001001001000100010\nonumber\\
&&\cdots 1001001000100100100
\end{eqnarray}

The counting of Laughlin edge states is $1,1,2,3,5,7,11\cdots$ and all Jack polynomials in Eq.(\ref{edgejacks}) are in the null space of the model Hamiltonian. They are therefore gapless excitations at the edge. The counting also matches that of the Virasoro algebra of $U(1)$ bosons in the conformal field theory (CFT). This is not merely a coincidence. The $W_\infty$ algebra in Eq.(\ref{winfinity}) has a Virasoro sub-algebra
\begin{eqnarray}\label{vsub}
[W_{m,0},W_{n,0}]=(n-m)W_{m+n,0}
\end{eqnarray}
One should note, though, that the negative modes of the Virasoro algebra is missing (so Eq.(\ref{vsub}) is also called the Witt algebra). Nevertheless, it is easy to show for the model Hamiltonian $\mathcal H$, if $\mathcal H\psi_0=0$, then $\mathcal HW_{m,0}\psi_0=0$. This is because in real space $b_i^\dagger\sim z_i, b_i\sim \partial_{z_i}$. Taking $\psi_0$ as the Laughlin ground state, the edge states with $\delta L=N$ are given by $\prod_{i}W_{n_i,0}^{s_i}\psi_0$ with $\sum_i(n_i-1)s_i=N$, each of them has zero energy and is a linear combination of the appropriate Jack polynomials (which themselves are not orthogonal).

The edge states generated by acting ``Virasoro operators" $W_{m,0}$ on the ground state is complete. This is the microscopic connection of the chiral edge modes to the chiral CFT. To see the connection of the edge mode to the bulk mode, one notice that $W_{m,-1}$ also generates the complete edge modes given by $W_{m,-1}\psi_0$. This is related to the Kac-Moody algebra that describes the edge states\cite{drr,edge1,edge2}. With translational invariance we have $[\mathcal H,W_{0,-1}]=[\mathcal H,W_{-1,0}]=0$, thus each state has an infinite degeneracy associated with the center-of-mass rotation. If this ``trivial degeneracy" is removed, the space of edge states at $\delta L=N$ is spanned by
\begin{eqnarray}\label{ntedge}
\psi^e_{n_1,n_2,\cdots,s_1,s_2,\cdots}=\prod_iW_{n_i,-1}^{s_i}\psi_0,\quad \sum_is_i(n_i-1)=N,n_i>1
\end{eqnarray}
The states in Eq.(\ref{ntedge}) are linearly independent but not orthogonal. The edge mode space for each $\delta L=N$ is spanned only by \emph{the highest weight states}, once the center of mass degeneracy is removed. The counting of such edge modes for $N\ge 1$ is given by $0,1,1,2,2,4,4,7,\cdots$

The canonical conjugate of $W_{m,-1}$ is given by $W_{-1,m}$. Since the ground state $\psi_0$ is the highest weight state, we have $W_{-1,0}\psi_0=0$, while $W_{-1,m}\psi_0$ are also the highest weight state. In analogy to the edge modes, at $\delta L=-N$ there is a subspace of highest weight states spanned by
\begin{eqnarray}\label{bhighestweight}
\psi^b_{n_1,n_2,\cdots,s_1,s_2,\cdots}=\prod_iW_{-1,n_i}^{s_i}\psi_0,\quad \sum_is_i(n_i-1)=N,n_i>1
\end{eqnarray}
One can diagonalize the Hamiltonian within the highest weight subspace spanned by states in Eq.(\ref{bhighestweight}) in $\delta L=-N$ sector. The ground state in each momentum sector is a good trial wavefunction for the magneto-roton mode. In particular at $N=1$, the highest weight subspace is empty, so the $L=1$ elementary neutral excitation does not exist.

One should note the edge modes generated by Virasoro operators or Kac-Moody operators are the $U(1)$ charge sector of the edge excitations, which in CFT is obtained by inserting the $U(1)$ current into the ground state correlator. For the Laughlin state only the charge sector is present, so the scheme above generates all the edge excitations. For the non-abelian MR state, there is an additional statistical sector from the fermionic majorana mode, and the counting of the edge modes are thus different. The details of this subtelty can be found in \cite{drr} and the references therein, and will not be pursued further in this thesis.

\section{Magneto-roton modes in Laughlin State}

Let us start by presenting the wavefunctions of the neutral excitations for the fermionic Laughlin state at filling factor $\nu=1/m$ in the LLL, where $m$ is odd. On the sphere the ground state is the Laughlin wavefunction in total angular momentum $L=0$ sector, or the fermionic Jack polynomial  $J^{-m+1}_{1001001\cdots}$\cite{bernevig}. By stripping away the single particle normalization factor, the Laughlin ground state is given by Eq.(\ref{laughlin}). The model Hamiltonian in Eq.(\ref{2body}) can be written as $V=\sum_{i<j}V_{ij}$, with
\begin{eqnarray}\label{vij}
V_{ij}=\int \frac{d^2ql_B^2}{2\pi}\sum_{n=0}^{m-1}L_n(q^2l_B^2)e^{-\frac{1}{2}q^2l_B^2}e^{i\vec q\cdot (\vec R_i-\vec R_j)}
\end{eqnarray}
To make comparison with the wavefunctions in the previous chapter, the family of neutral excitations at $\delta L=-N$ sector (we omit the exponential part of the wavefunction, which is irrelavent in the LLL)  is labeled by the corresponding total angular momentum $L=N$ on the sphere:

\footnotesize
\begin{eqnarray}\label{collect}
&&\mathcal A[(z_1-z_2)^{m-2}\prod'_{i<j}(z_i-z_j)^m]\qquad L=2\nonumber\\
&&\mathcal A[(z_1-z_2)^{m-2}(z_1-z_3)^{m-1}\prod'_{i<j}(z_i-z_j)^m]\qquad L=3\nonumber\\
&&\mathcal A[(z_1-z_2)^{m-2}(z_1-z_3)^{m-1}(z_1-z_4)^{m-1}\prod'_{i<j}(z_i-z_j)^m]\qquad L=4\nonumber\\
\vdots
\end{eqnarray}
\normalsize

Here $\mathcal A$ indicates antisymmetrization over all particle indices, and $\prod'_{i<j}$ means products of only pairs $\{ij\}$ that do not appear in the prefactors to the left of it. Thus the $L=2$ state, which is the quadrupole excitation in the thermodynamic limit\cite{yb1}, is obtained from the ground state by reducing the power of one pair of particles (which we can choose arbitrarily as particle $1$ and $2$ due to the antisymmetrization) by \emph {two}, followed by antisymmetrizing over all particles. This scheme naturally forbids an $L=1$ state by pair excitation, since if we reduce the power of one pair of particles by \emph {one}, antisymmetrization kills the state. 

The $L=3$ state is generated by pairing particle $1$ with another particle (which we arbitrarily label as particle $3$) and reducing their pair power by one. It is now clear how the modes in other momentum sectors are generated. Naturally for a total of $N_e$ particles, the family of the neutral excitations ends at $L=N_e$, agreeing with the numerical scheme in the previous chapter. Indeed all wavefunctions here satisfy the highest weight condition, and the states relax to the ground state far away from the excited pairs; these are exactly the conditions we used to numerically generate the unique model wavefunction in each momentum sector.

\section{Magneto-roton Mode and Neutral Fermion Mode in MR State}

The same scheme applies to the MR state. It is instructive to first see how the MR ground state is obtained. The Laughlin wavefunction at half filling is given by $J^{-2}_{1010101\cdots}(z_i)=\prod_{i<j}(z_i-z_j)^2$. For fermions this is not a valid state; instead the ground state was constructed by a pairing mechanism\cite{mr}, which is also a Jack polynomial $J^{-3}_{1100110011\cdots}$. The pairing reduces the power of each pair of particles by \emph{one}.  For $2n$ particles, the antisymmetrization reproduces the Pfaffian up to a constant as follows:
\footnotesize
\begin{eqnarray}\label{pairing}
&&\prod_{i<j}(z_i-z_j)^2\rightarrow\mathcal A[(z_1-z_2)(z_3-z_4)\cdots(z_{2n-1}-z_{2n})\prod'_{i<j}(z_i-z_j)^2]\nonumber\\
&&=\text{Pf}\left(\frac{1}{z_i-z_j}\right)\prod_{i<j}(z_i-z_j)^2
\end{eqnarray}
\normalsize
where the last line of Eq.(\ref{pairing}) is the familiar Pfaffian for the MR ground state. The explicit use of antisymmetrization instead of the Pfaffian allows us to naturally extend to the case with an odd number of particles: starting from the Bosonic Laughlin wavefunction at half filling, every two particles form a pair except for just one particle. Naturally the ``ground state" of the neutral fermion mode is given by
\small
\begin{eqnarray}\label{opairing}
&&\prod_{i<j}(z_i-z_j)^2\rightarrow\mathcal A[(z_1-z_2)(z_3-z_4)\cdots(z_{2n-1}-z_{2n})\nonumber\\
&&\prod_{k<2n+1}(z_{2n+1}-z_k)^2\prod'_{i<j}(z_i-z_j)^2]
\end{eqnarray}
\normalsize
Note both $i,j$ in $\prod'_{i<j}$ runs from $1$ up to $2n+1$, with pairs appearing before  $\prod'_{i<j}$ excluded. Though we can no longer represent Eq.(\ref{opairing}) as a Pfaffian, comparing the antisymmetrized products we can see Eq.(\ref{opairing}) is really the same as that of Eq.(\ref{pairing}), only with an odd number of particles. For the model three-body Hamiltonian, this is a zero-energy abelian quasihole state $J^{-3}_{1100110011\cdots 0011001}$ in the angular momentum sector $L=\frac{1}{2}(N_e-1)$. The magneto-roton mode and the neutral fermion mode are obtained from Eq.(\ref{pairing}) and Eq.(\ref{opairing}) respectively by reducing the powers in the Jastrow factor the same way as what is done for the Laughlin state.

\section{A Generalized Formal Scheme}

To write down the analytic wavefunctions in a more formal way, we define $\mathcal P_{ij}=\frac{1}{z_i-z_j}$. Notice the Pfaffian for $2n$ particles can be written as $\text{Pf}\left(\frac{1}{z_i-z_j}\right)\sim\mathcal A[\mathcal P^{(2n)}]$, where $\mathcal P^{(2n)}=\mathcal P_{12}\mathcal P_{34}\cdots\mathcal P_{2n-1,2n}$. The magneto-roton mode for the Laughlin state is given by
\begin{eqnarray}\label{laughlinm}
\psi_l^{L=k+2}=\prod_{i<j}^{N_e}(z_i-z_j)^m\mathcal S[\mathcal P_{12}^2\mathcal P_{13}\cdots\mathcal P_{1,2+k}]
\end{eqnarray}
where $\mathcal S$ is the symmetrization over all particle indices. From the bosonic Laughlin wavefunction at filling factor $1/2$ we can impose pairing to obtain
\begin{eqnarray}\label{mrg}
\psi_{\text{mr}}=\prod_{i<j}^{N_e}(z_i-z_j)^2\mathcal A[\mathcal P^{(2n)}]
\end{eqnarray}
For an even number of electrons we have $N_e=2n$ and Eq.(\ref{mrg}) is the MR ground state. The magneto-roton modes are given by
\small
\begin{eqnarray}\label{mrm}
\psi_{\text{mr}}^{L=k+2}=\prod_{i<j}^{N_e}(z_i-z_j)^2\mathcal A[\mathcal P^{(2n)}\mathcal P_{13}^2\mathcal P_{15}\cdots \mathcal P_{1,3+2k}]
\end{eqnarray}
\normalsize
For an odd number of electrons we have $N_e=2n+1$ and Eq.(\ref{mrg}) is the MR quasihole state of Eq.(\ref{opairing}). The neutral fermion modes are given by
\footnotesize
\begin{eqnarray}\label{mrn}
\psi_{\text{mr}}^{L=\frac{3}{2}+k}=\prod_{i<j}^{N_e}(z_i-z_j)^2\mathcal A[\mathcal P^{(2n)}\mathcal P_{N_e,1}^2\mathcal P_{N_e,3}\cdots\mathcal P_{N_e,1+2k}]
\end{eqnarray}
\normalsize

In this way, the class of ground state model wavefunctions are generalized to include neutral excitations under a universal scheme.

\section{A Lattice Diagram Representation}

\begin{figure}[h!]
\centerline{\includegraphics[width=8cm,height=4cm]{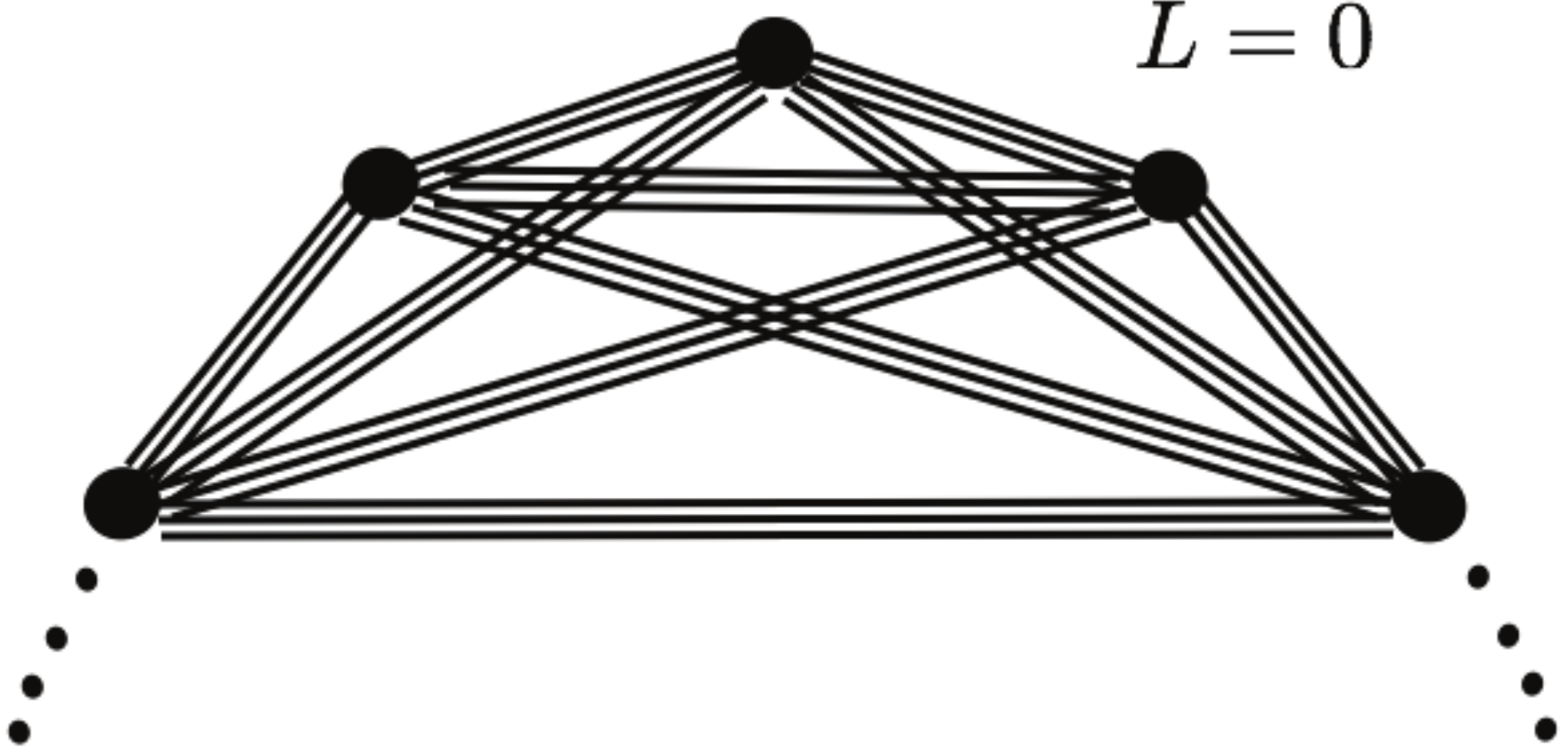}}
\caption{For $N_e$ particles, the lattice can be viewed as an $N_e$-gon, with three bonds connecting every pair of vertices}
\label{fig:laughlin0}
\end{figure}

An intuitive way to visualize the family of the neutral excitations is to map the particles onto a lattice, where each lattice site represents a particle. Since for the FQHE we have a quantum fluid instead of a solid, every two lattice sites interact with each other. The number of bonds between each pair of lattice sites equal to the power of the pair of particles in the wavefunction. As an example we consider the simpliest Laughlin state at $\nu=1/3$, so for the ground state every two lattice sites are connected by three bonds, as shown in Fig.~\ref{fig:laughlin0}.

\begin{figure}[h!]
\centerline{\includegraphics[width=10cm,height=6cm]{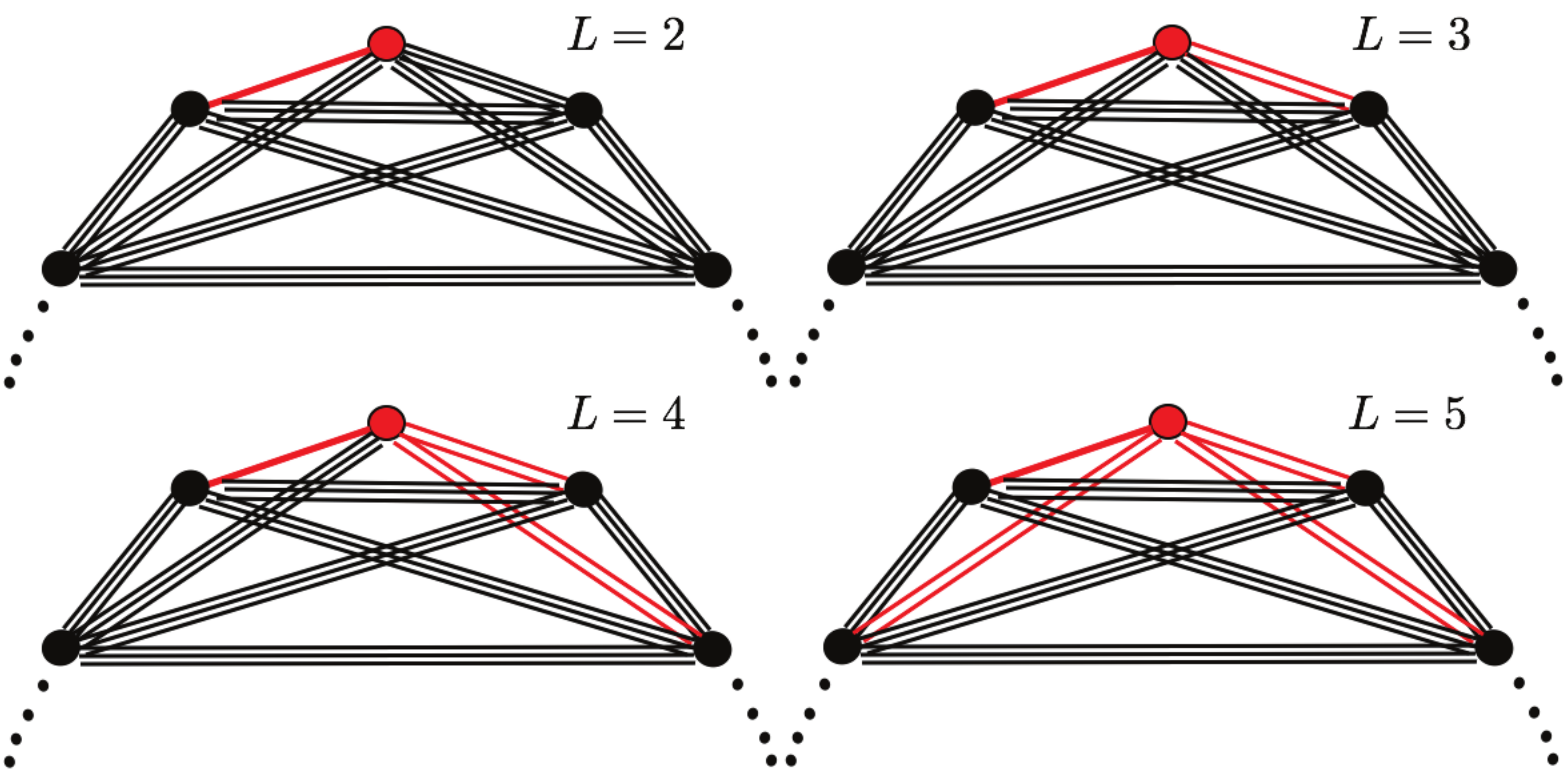}}
\caption{Neutral excitations from $L=2$ to $L=5$, where the change of bonds are highlighted with red color.}
\label{fig:laughlin25}
\end{figure}

The neutral excitations are obtained by breaking the bonds between lattice sites, as shown in Fig.~\ref{fig:laughlin25}. We can view the entire family of the neutral excitations as elementary excitations centered around a single red lattice site. Note the lattice pattern uniquely defines the many-body wavefunction, and different types of ``elementary excitations" can be identified with different patterns of bond-breaking around a single lattice site.

\begin{figure}
\centerline{\includegraphics[width=10cm,height=4cm]{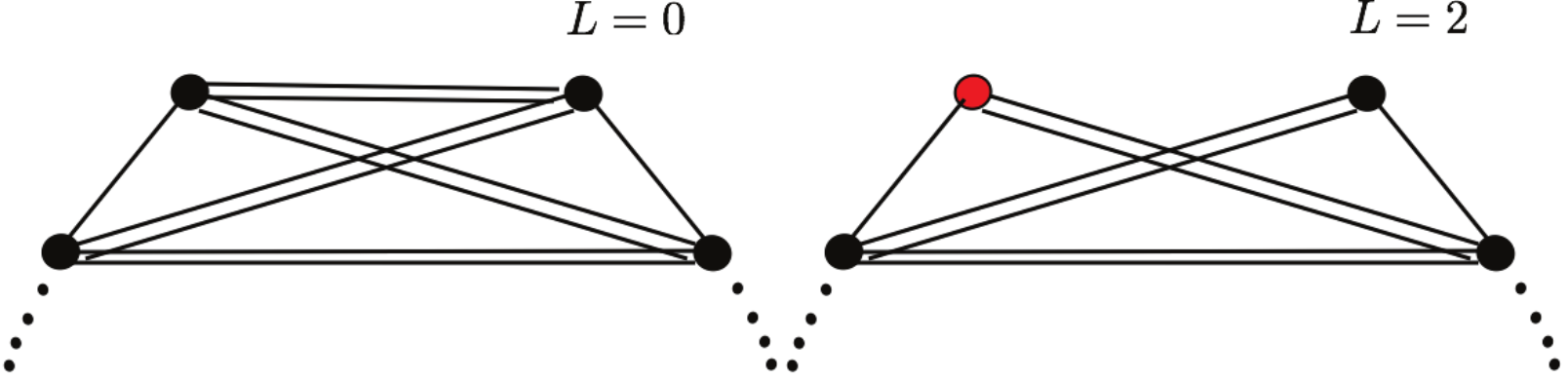}}
\caption{The lattice configuration of the ground state $L=0$ and the first neutral excitation at $L=2$. Consecutive neutral modes can be obtained by breaking one of the \emph {double bonds} connecting the red lattice site to some other site.}
\label{fig:mr02}
\end{figure}

\begin{figure}[h!]
\centerline{\includegraphics[width=8cm,height=10cm]{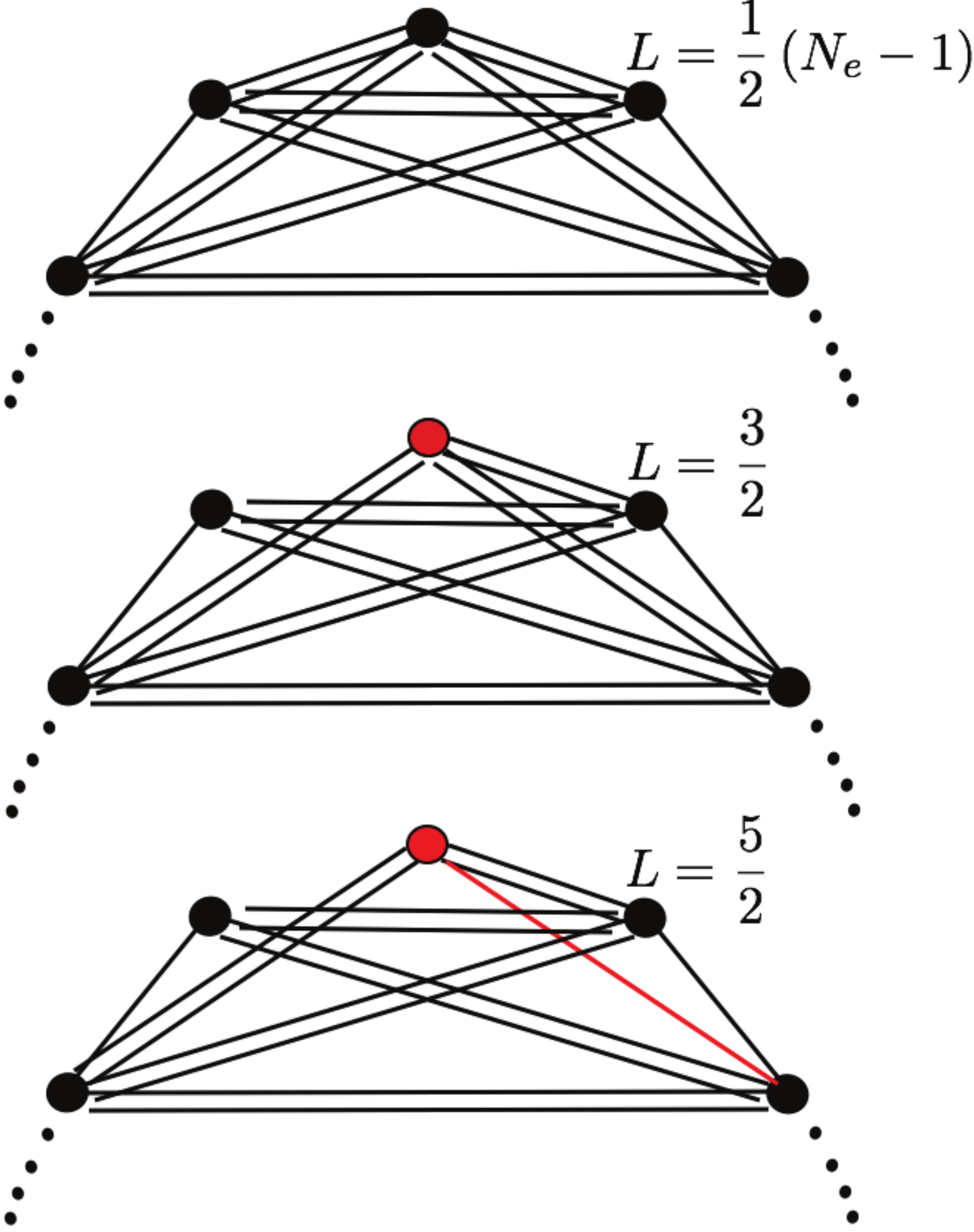}}
\caption{The lattice configuration of the zero mode quasihole state $L=\frac{1}{2}\left(N_e-1\right)$ and the first two neutral fermion modes at $L=\frac{3}{2}\text{ and }L=\frac{5}{2}$. Consecutive neutral fermion modes can be obtained by breaking one of the \emph {double bonds} connecting the red lattice site to some other site.}
\label{fig:mrn02}
\end{figure}

This suggests lattice representation of the MR state and its magneto-roton mode with the same scheme, as shown in Fig.~\ref{fig:mr02}. The representation of the MR quasihole state and those of the neutral fermion modes are given in Fig.~\ref{fig:mrn02}.

\section{Quadrupole Gap in the Thermodynamic Limit}

The analytic wavefunction is useful in calculating the magneto-roton mode energy gap in the long wavelength limit. For the Laughlin state, the energy gap is given by
\begin{eqnarray}\label{energy}
\epsilon_{q\rightarrow 0}=\lim_{N_e\rightarrow\infty}\frac{\langle\psi^{L=2}_l|V|\psi^{L=2}_l\rangle}{\langle\psi^{L=2}_l|\psi^{L=2}_l\rangle}
\end{eqnarray}

We already know from \cite{yb1} that in $L=2$ and $L=3$ sector, SMA is exact for the magneto-roton model wavefunctions. Writing the Laughlin wavefunction as $\psi_l$ and using the guiding center ladder operators $b^\dagger_i=z_i,b_i=\partial_{z_i}$, we have
\begin{eqnarray}\label{sma}
\psi^{L=2}_l&=&\frac{1}{2m(m-1)}\sum_i(b_i)^2\psi_l
\end{eqnarray}
In the thermodynamic limit, the normalization constant of Eq.(\ref{sma}) is thus related to the long wavelength expansion of the ground state guiding center structure factor, as shown in Eq.(\ref{qs}), which leads to

\begin{eqnarray}\label{denominator}
\langle\psi^{L=2}_l|\psi^{L=2}_l\rangle=-\frac{\bar sN_e}{2m^2(m-1)^2}
\end{eqnarray}
The numerator of Eq.(\ref{energy}) can be calculated using the plasma analogy. Note in Eq.(\ref{collect}), before antisymmetrization the term only has one pair of particles with relative angular momentum smaller than $m$. We thus have
\small
\begin{eqnarray}\label{onepair}
\langle\psi^{L=2}_l|V|\psi^{L=2}_l\rangle&=&\frac{N_e(N_e-1)}{2\mathcal N^2}\langle\bar\psi_l|\mathcal P_{12}^2V_{12}\mathcal P^2_{12}|\bar\psi_l\rangle
\end{eqnarray}
\normalsize
where $\mathcal N$ is the normalization constant of the Laughlin state. We note that $V_{12}$ projects out states with relative angular momentum $(z_1-z_2)^{m-2}$, which can be integrated over. The numerator is thus equivalent to evaluating the norm of the following wavefunction:
\small
\begin{eqnarray}\label{numerator}
\bar\psi=\prod_{i=2}^{N_e-1}\left(\frac{1}{\sqrt{2}}z_1-z_i\right)^{2m}\prod_{1<i<j<N_e-1}(z_i-z_j)^m
\end{eqnarray}
\normalsize
which can be evaluated as the free energy of a two-dimensional one-component plasma (OCP) on the disk with radius $R^2=\frac{mN_e}{2}$ and elementary charge $e=2\sqrt{\pi mk_BT}$, where particle $1$ interacts with the rest of the particles with charge $2e$. We thus obtain
\begin{eqnarray}\label{energyr}
\epsilon_{q\rightarrow 0}=-\frac{2^mm(m-1)^2}{\pi\bar s}e^{-\frac{\mathcal F_2-\mathcal F}{k_BT}}
\end{eqnarray}
Both $\mathcal F_2\text{ and }\mathcal F$ are free energies of OCP in the thermodynamic limit ($N_e\rightarrow\infty$), where $\mathcal F$ is for $N_e$ particles, each with charge $e$ and interacting with lthe ogarithmic two-body interaction together with a neutralizing background of radius $R$; for $\mathcal F_2$, we have the same neutralizing background with $N_e-2$ particles of charge $e$, and exactly one particle with charge $2e$. Thus $\mathcal F_2-\mathcal F$ is the free energy cost of fusing two particles of charge $e$ to create a particle of charge $2e$, which is an $O(1)$ effect.

Similar calculations can be carried out for the magneto-roton mode in the MR state. Analogous to Eq.(\ref{sma}) we have $\psi^{L=2}_{\text{mr}}=\frac{1}{24}\sum_ib_i^2\psi_{mr}$,  and in the long wavelength limit we have
\begin{eqnarray}\label{energymr}
\epsilon^{\text{mr}}_{q\rightarrow 0}=-\frac{24}{\pi\bar s_{\text{mr}}}e^{-\frac{\mathcal F_3-\mathcal F_{\text{II}}}{k_BT}}
\end{eqnarray}
where  $\bar s_{\text{mr}}=-2$ is the guiding center spin for the MR state, and $\mathcal F_{\text{II}}$ is the standard two-component plasma free energy for the MR ground state\cite{bonderson}. The charge for the interaction between the two components is given by $Q_1=\pm \sqrt{3k_BT}$, while the charge for the interaction between one of the components and the neutralizing background is given by $Q_2=2\sqrt{k_BT}$. $\mathcal F_3-\mathcal F_{\text{II}}$ is the free energy cost of fusing \emph {three particles} to create one particle for each component with charge $3Q_2$ but with the same $\pm Q_1$. 

The evaluation of the long wavelength gap of the neutral fermion mode is less transparent. The difficulty lies with evaluating the normalization constant of $\psi^{L=\frac{3}{2}}_{\text mr}$. There is no SMA for the neutral fermion mode, and it is not known if in the thermodynamic limit the gap should be inversely proportional to the guiding center spin. On the other hand $\langle\psi^{L=\frac{3}{2}}_{\text{mr}}|V_{\text{3bdy}}|\psi^{L=\frac{3}{2}}_{\text{mr}}\rangle$ can be mapped to a two-component plasma as well, and we obtain
\begin{eqnarray}\label{energymrnf}
\bar\epsilon^{mr}_{q\rightarrow 0}\sim e^{-\frac{\bar{\mathcal F}_3-\bar{\mathcal F}_{\text{II}}}{k_BT}}
\end{eqnarray}

Here $\bar{\mathcal F}_{\text{II}}$ is the free energy of the 2-component plasma similar to that of $\mathcal F_{\text{II}}$ with only one difference: there is exactly one \emph {more} particle carrying charge $Q_2$ that interacts with the neutralizing background, and its $Q_1$ charge is zero. This is how an unpaired fermion in the MR state is interpreted in the plasma analogy. Furthermore, $\bar{\mathcal F}_3-\bar{\mathcal F}_{\text{II}}$ is the energy cost of fusing the unpaired fermion with one pair of two other fermions, creating a particle with charge $Q_2=6\sqrt{k_BT}$ but again with \emph zero $Q_1$. 

\section{Summary}

The numerical results from the previous chapter help to identify the analytic wavefunctions constructed in this chapter. From a practical point of view, these compact analytic forms are useful, because now the energy gap of the quadrupole excitation in the thermodynamic limit can be related to the free energy cost of the fusion of charges in the plasma energy, and is inversely proportional to the guiding center spin which characterizes its topological order. This is the first time that the plasma analogy is extended to the neutral excitations of the FQHE, and the analogy not only applies to the wavefunctions, but also to the energy spectrum as well. Since the neutral excitations in the long wavelength limit is buried in the multi-roton continuum, it is important to calculate the decay rate of these neutral modes. Numerical calculation has been performed to show that even in the continuum the decay rate of the neutral modes is very small\cite{note2}. This opens up the possibility of experimental detection of these modes. A more detailed analysis of the decay rate of the neutral modes is currently research in progress.

The neutral excitations in the single component FQHE can now be understood in several coherent frameworks, at least for the Laughlin and Moore-Read states, with possible generalization to the entire Read-Rezayi series. The composite fermion picture maps the FQHE to the IQHE of the particle-vortex composite, and in this framework the neutral excitations are excitons of composite fermions. The Jack polynomial formalism enables us to describe the wavefunctions of the ground states, the quasihole and quasiparticle states, as well as the neutral excitations in a unified way with root configurations and squeezed basis constrained by clustering properties. It is now satisfactory to see that compact analytic real space wavefunctions in electron coordinates, which initiated the theoretical understandings of the FQHE, can now be extended from ground states and charged excitations to include neutral bulk excitations. One could still ask if the neutral excitations proposed so far completely describes the energy spectrum of the FQHE. Experimental measurements on the Laughlin state\cite{west3} suggest a splitting of the neutral modes in the long wavelength limit, with theoretical explanations proposed from a hydrodynamic point of view\cite{vignale}, and the composite fermion point of view\cite{jain2mode}. It would be interesting to see if the lattice diagram can be generalized to produce suitable analytic wavefunctions that describes the multi-roton excitations and the splitting of the neutral modes as well.

It is well-known in the literature that the wavefunctions of the gapless edge excitations on the disk can be obtained by multiplying the ground state with symmetric polynomials. With model Hamiltonians these are the zero energy states in the positive $\delta L_z$ angular momentum sectors\cite{edge1,edge2}. For the Moore-Read state, in addition to the charge sector generated by the symmetric polynomials, there are also edge excitations obtained from the statistical sectors via inserting Majorana fermions\cite{rezayi}. The analytic wavefunctions of these states are known explicitly. One can also generate wavefunctions by similar operations not only on the ground state, but also on the bulk neutral excitations obtained in this paper. These wavefunctions describe states such that each contains both bulk and edge excitations. We call these roton-edge excitations, which explain the gapped low-lying multitude of states below the multi-roton gap in disk geometry. Recent studies show\cite{boyang} that for the Laughlin state, each bulk neutral excitation generates a branch of quasi-degenerate roton-edge excitations with the same Virasoro counting as the zero-energy edge states (See Fig.(\ref{fig_disk})). For the Moore-Read state, however, the counting of the roton-edge states seem different because of the lack of the linear independence between states in the same momentum sector, possibly due to the non-abelian nature of the FQH fluid.

\chapter{Summary and Outlook\label{ch:conclusion}}

In this thesis, I explored the physics of the fractional quantum Hall effect from a geometric point of view, where the dynamics are governed by the guiding center coordinates with non-commutative spatial components. The recognition of the geometric aspect of the FQHE leads to a better understanding of the guiding center Hall viscosity, which is a topological index that defines the energy gap of the neutral excitations in the long wavelength limit. The guiding center Hall viscosity also captures the universal part of the transport coefficients, when the FQH fluid is perburbed by a spatially varying electromagnetic field. It is also shown that the experimental measurement of the guiding center Hall viscosity as proposed by Son and collaborators requires both Galilean and rotational invariance. The microscopic calculation presented in this thesis include all general corrections when those symmetries are absent, with different geometric dependence made explicit.

The neutral excitations of the FQHE in the Laughlin state and the Moore-Read state are presented both from a numerical perspective and a more general analytic construction. Using the formalism of Jack polynomials, the model wavefunctions for the neutral excitations are constructed for the entire range of momenta, improving earlier attempts with the single mode approximation. The numerical construction with the root configurations allows us to identify the long wavelength limit of the neutral excitations as ``spin-2 gravitons" (with ``spin $3/2$ gravitino" as the super-partner in the Moore-Read case). It is also proven that the single mode approximation gives exact ``graviton" wavefunctions, which allows us to understand the neutral gap in the long wavelength limit as the energy cost of area-preserving deformation of the ground state guiding center metric, and is inversely proportional to the guiding center Hall viscosity as a result of zero-point quantum fluctuation of the non-commuting coordinates in the projected Hilbert space.

The analytic wavefunctions for both the magneto-roton modes and the neutral fermion modes presented in this thesis unify previous numerical constructions of the neutral excitation model wavefunctions, including the one presented in the thesis, as well as the other scheme from the perspective of the composite fermions. The analytic wavefunctions have simple representations in terms of lattice diagrams, and allow analytic computation of the dynamics of the neutral excitations in the long wavelength limit. It was shown that the usual plasma picture for the Laughlin and Moore-Read state, which was previously only applicable to the ground state wavefunctions, has interesting connections to the FQHE neutral excitations. The energy cost from area-preserving deformation of the guiding center metric can be viewed as the free energy cost of fusing charged particles with a neutralizing background in the plasma analogy.

It would be interesting to see how to experimentally measure the guiding center Hall viscosity, which is a result of strong correlation between electrons and is sensitive to rotational invariance as well as the edge effects. Additional theoretical work is needed to fully understand the quantization of the guiding center Hall viscosity when the interaction moves away from the model Hamiltonians where the Laughlin and Moore-Read model wavefunctions are exact. Numerically, one good way of probing the guiding center Hall viscosity is to locally deform the guiding center metric and measure the guiding center density response, and there are ongoing works on cylinder geometry with DMRG technique\cite{sonika}. This calculation is complementary to the work on torus, where the Hall viscosity is measured by deforming the periodic boundary condition. In addition to that, it can also shed light on the formulation of the geometric effective field theory of the FQHE, where the guiding center dynamics are determined by the coupling of the composite boson guiding center spin to the fluctuation of the guiding center metric.

For neutral excitations in the FQHE, there are unanswered questions on the stability and life-time of the neutral excitations, especially in the long wavelength limit where the quadrupole excitation merges into the continuum of multi-roton excitations. With a better understanding of the entire branch of the neutral excitations, one can investigate the tunability of the energy of the mode from a dynamic point of view. For the Moore-Read state with both the magneto-roton mode and the neutral fermion mode, it is interesting to see if in the long wavelength limit this pair of ``super-partners" converge to the same energy. There are interesting questions on the transition of the FQHE from an incompressible phase to a fermi liquid or nemetic phase\cite{sondhi}; the thermodynamic gap can be closed either by the roton minimum going soft, or the quadrupole excitation going soft. For the latter, it is also interesting to see if one can bring the quadrupole excitations below the multi-roton minimum by tuning the Hamiltonian, so it can be probed experimentally.

\appendix 

\singlespacing
\bibliographystyle{plain}

\cleardoublepage
\ifdefined\phantomsection
  \phantomsection  
\else
\fi
\addcontentsline{toc}{chapter}{References}


\begin{thebibliography}{99}
\bibitem{note2}
This is still a part of an on-going research. The claim here is based on the numerical calculation of the first moment of the energy  (or the standard deviation) based on the model wavefunctions: $\langle H^2\rangle-\langle H\rangle^2$, which is a rough estimate of the life time of the state.
\end{thebibliography}

\end{document}